\documentclass[12pt]{article}
\usepackage{amsmath,graphicx,amssymb,amsfonts,amsthm}
\usepackage{verbatim,enumitem,hyperref,lipsum,float}
\usepackage{color,epstopdf,epsfig,lscape,longtable,mathabx,natbib}
\usepackage{xr}
\usepackage{mathrsfs}
\usepackage{authblk}
\usepackage{multirow,soul,enumitem}
\usepackage[utf8]{inputenc}

\newtheorem{Th}{\underline{\bf Theorem}}

\newtheorem{Pro}{Proposition}
\newtheorem{Lem}{\underline{\bf Lemma}}

\newtheorem{Assump}{\underline{Assumption}}

\def\boxit#1{\vbox{\hrule\hbox{\vrule\kern6pt  \vbox{\kern6pt#1\kern6pt}\kern6pt\vrule}\hrule}}
\def\bse{\begin{eqnarray*}}
	\def\ese{\end{eqnarray*}}
\def\be{\begin{eqnarray}}
\def\ee{\end{eqnarray}}
\def\bsq{\begin{equation*}}
\def\esq{\end{equation*}}
\def\bq{\begin{equation}}
\def\eq{\end{equation}}
\def\th{^{th}}
\def\var{\hbox{var}}
\def\cov{\hbox{cov}}

\def\wh{\widehat}
\def\wt{\widetilde}

\def\th{^{\rm th}}

\def\mR{\mathbb{R}}
\def\n{\nonumber}
\def\bias{\mbox{bias}}

\def\cov{\mbox{cov}}

\def\sumi{\sum_{i=1}^n}
\def\sumj{\sum_{j=1}^n}
\def\trans{^{\rm T}}

\def\ba{{\boldsymbol\alpha}}

\def\bb{{\boldsymbol\beta}}
\def\bphi{{\boldsymbol\phi}}
\def\bpsi{{\boldsymbol\psi}}

\def\bg{{\boldsymbol\gamma}}
\def\bmu{\boldsymbol\mu}

\def\a{{\bf a}}
\def\B{{\bf B}}

\def\D{{\bf D}}
\def\V{{\bf V}}

\def\I{{\bf I}}

\def\N{\mbox{ $\mathcal{N}$}}

\def\U{{\bf U}}

\def\W{{\bf W}}

\def\X{{\bf X}}

\def\x{{\bf x}}

\def\Uniform{\hbox{Uniform}}

\def\log{\hbox{log}}
\def\bias{\hbox{bias}}

\def\squarebox#1{\hbox to #1{\hfill\vbox to #1{\vfill}}}

\def\bpi{{\boldsymbol \pi}}

\def\0{{\bf 0}}

\def\mR{\mathcal R}

\def\var{\hbox{var}}
\def\cov{\hbox{cov}}

\def\wh{\widehat}
\def\wt{\widetilde}

\def\log{\hbox{log}}
\def\bias{\hbox{bias}}

\begin{document}
\thispagestyle{empty}
\baselineskip=18pt
\title{Learning non-monotone optimal individualized treatment
regimes}
\author[a]{Trinetri Ghosh}
\author[b]{Yanyuan Ma}
\author[c]{Wensheng Zhu}
\author[d]{Yuanjia Wang}
\affil[a]{Department of Biostatistics and Medical Informatics, University of Wisconsin-Madison}
\affil[b]{Department of Statistics,  Pennsylvania State University}
\affil[c]{
	School of Mathematics and Statistics, Northeast Normal University}
\affil[d]{Department of Biostatistics, Columbia University}
        \maketitle

\begin{abstract}
We propose a new modeling and estimation approach
to select the optimal treatment regime 
from different options through constructing a
robust estimating equation. The method is protected against
misspecification of the propensity score model, the outcome regression
model for the non-treated group, or the potential non-monotonic treatment
difference model. Our method also allows residual errors to depend on covariates. A single index structure is incorporated
  to facilitate the nonparametric 
estimation of the treatment difference. We then 
identify the optimal treatment through  maximizing the
value function. Theoretical properties of the treatment
assignment strategy are established.  We
illustrate the performance and effectiveness of our proposed
estimators through extensive simulation studies and a real dataset 
on the effect of maternal smoking on baby birth weight.
\end{abstract}


\section{Introduction}

Due to the between-person heterogeneity, it is common to observe
different responses to the same treatment in different individuals. 
Various factors can contribute to the heterogeneity from one individual
to another, such as genetic risk factors,
age and individual-specific environmental exposures.
Thus, when different treatment options are available, it is desirable 
to select the best treatment regime that is specific to a
particular individual, which is one of the goals of precision medicine.  
Precision medicine aims to determine a strategic
assignment of treatments to patients according to their
characteristics and medical history. This goal can be achieved by
using an individualized treatment 
rule (ITR), i.e., a deterministic function of subject-specific factors that are responsible
for patients' heterogeneous responses to treatment.  The optimal ITR maximizes the
expected clinical outcome of interest under the ITR.  
Furthermore, an optimal dynamic treatment regime usually consists of a
set of sequential decision rules applied at a set of decision points.
There has been significant
research developments on estimating the optimal treatment regimes
in recent years, as will be elaborated subsequently.
In this work, we will focus on estimating the individualized
treatment regimes at a single decision time point.

Two popular model-based methods
to derive the optimal dynamic treatment regimes are Q-learning 
and A-learning, standing for quality-learning and
advantage-learning, respectively. Q-learning \citep{watkins1989,watkins_dayan_1992,nahumetal2012,zhaoetal2009,zhaoetal2011,murphy2005ge,qian_murphy_2011,song2015penalized,goldberg2012Q,chakraborty2010} 
is built on a postulated regression outcome model for the
outcome of interest and is implemented through a backward induction
fitting procedure. This approach was initially proposed by
\cite{watkins1989}, and later a detailed proof of convergence
was provided by \cite{watkins_dayan_1992}. The performance of the 
optimal treatment decision rule obtained by 
Q-learning depends on the correctly-specified outcome model. 
On the other hand, A-learning
\citep{murphy2003,blattetal2004,robins2004optimal,orellana2010,liang2018d}
maximizes 
estimating equations to estimate the contrast functions,
using the estimated probability of observed treatment 
assignment given patient information at each decision
point (i.e., treatment propensity scores). The performance of the 
optimal treatment decision rule obtained by 
A-learning thus relies on the suitable treatment assignment model.

Another approach, known as model free or policy (value) search
method
\citep{zhang2012r,zhang2012estimating,zhao2012estimating,jiang2017estimation,jiang2017doubly},
directly derives and maximizes a consistent estimator for the
value function over a prespecified class of treatment regimes
indexed by a finite dimensional parameter or over a class of
nonparametric treatment regimes. For example, \cite{zhang2012r} 
formulated the inverse propensity score weighted (IPW)
estimator and the
double-robust augmented IPW estimator for the value function
with a single decision time point. Later, \cite{zhangetal2013}
extended this idea to more than one decision
point. \cite{zhang2012estimating} and
\cite{zhao2012estimating} recast the original problem of
finding the optimal treatment regime as a weighted
classification problem. The former obtains the optimal
treatment regime by minimizing the expected weighted
misclassification error, whereas the later used
outcome-weighted support vector machine. Other relevant work
includes \cite{robins2004optimal}, \cite{foster2011subgroup},
\cite{zhao2013effectively}, \cite{matsouaka2014evaluate},
\cite{song2017}, \cite{bai2017optimal}, \cite{fanetal2017},
\cite{shi2018maximin}, and \cite{huang2020robust}.          

In this paper, we propose a new modeling and estimation method to
determine the optimal treatment regimes at a decision time point,
combining the advantages of Q-learning, A-learning and the
model-free approach.  In addition, our model has the advantage that 
it only assumes the treatment
difference as a smooth function of an index of the covariates, without requiring
the smooth function to be monotonic. This is
practically important.  For example, for a patient
with heart disease, low blood pressure and high blood pressure both
can increase the risk of heart attack, hence resulting in a possible
non-monotonic treatment difference model. Another example is the
relationship between BMI and health risks, where both underweight and
obese individuals have increased risks of a range of health
measures. Our model also 
has the flexibility of allowing the model error to be dependent on
the covariates. In practice, this is also an important
feature. Further, we consider a multi-robust estimating
equation as protection against misspecified propensity score function,
treatment difference model, and outcome 
regression model for the non-treated group. Benefiting from the smoothness of the
treatment difference function, our treatment regime identification rate is
$O_p(n^{-2/5})$, faster than the existing rate of $O_p(n^{-1/3})$ 
\citep{fanetal2017}, where treatment difference function is assumed to be 
monotonic.

The remainder of the paper is organized as follows. In Section
\ref{sec:method}, we introduce 
the estimation procedure and the algorithm for our proposed
method. Section \ref{sec:theo} provides the asymptotic properties of
the proposed estimators for $\bb$ and the treatment difference function,
$Q(\cdot)$. In Section \ref{sec:simu}, we summarize the finite-sample 
performance of the estimators for different designs, including well-specified
and misspecified models. We implemented our method on baby birth weight dataset,  where the research interest is to investigate whether the maternal smoking during pregnancy has any effect on  
birth weight in Section \ref{sec:real}. Section \ref{sec:discussion} concludes the paper. 

\section{Model and Estimation}\label{sec:method}

We consider the following treatment difference model
\be\label{eq:model}
Y_{i1}-Y_{i0}=Q(\bb\trans\X_i)+\epsilon_i, 
\ee
where $Y_{i1}$ is the potential outcome for individual $i$ if
treatment is received, $Y_{i0}$ is the potential outcome for
individual $i$ if no treatment is received, $\X_i\in \mR^{d_\bb}$ is the
set of covariates, the treatment difference function $Q(\cdot)$	is an
unknown smooth function and $E(\epsilon\mid\X)=0$, where $\epsilon$ is the model error. Here
$\bb\in\mR^{d_\bb}$ is a vector of  unknown parameters and $d_{\bb}$ is the dimension of $\bb$. 
Let $A_i$ be the treatment indicator. Our estimation is performed under the following two assumptions commonly assumed in the literature. 
\begin{Assump} 
	(Stable unit treatment value assumption) $Y_i = Y_{i1}A_i + Y_{i0}(1-A_i)$.
\end{Assump}
\begin{Assump} 
	(No-unmeasured-confounders assumption) $A_i \perp (Y_{i1},Y_{i0})\mid \X_i$.
\end{Assump}
For identifiability of $\bb$, we require $\bb$ to have the form
$\bb=(1,\bb_L\trans)\trans$, where the lower sub-vector is an arbitrary
vector of
length $d_{\bb}-1$. If $Y_{i1}$ and $Y_{i0}$ had been
  both available, we could have estimated $\bb$ through 
simultaneously solving
\bse  
&&\sumi
\left\{Y_{i1}-Y_{i0}-\wt Q(\bb\trans\X_i) \right\} \left\{\X_{Li}-E(\X_{Li}\mid\bb\trans\X_i) \right\}=\0,\\
&&\sumi K_h \left(\bb\trans\X_i-\bb\trans\X_j \right)
(Y_{i1}-Y_{i0}-c_j)
=\0 \mbox{ for }j=1, \dots, n.
\ese
Here $\X_L$ represents the sub-vector of $\X$ formed by its lower
$d_{\bb}-1$ components and $c_j=\wt Q(\bb\trans\X_j)$, 
$K_h(\cdot)=K(\cdot/h)/h$, where $K(\cdot)$ is a kernel function and
$h$ is a bandwidth. However, since we only observe $Y_i$, we may consider an inverse
probability weighting based estimator \citep{rrzhao1994} and modify the
above equations to 
\bse
&&\sumi
\left\{\frac{A_iY_i}{\pi(\X_i)}-\frac{(1-A_i)Y_i}{1-\pi(\X_i)}
-\wt Q(\bb\trans\X_i)\right\} \left\{\X_{Li}-E(\X_{Li}\mid\bb\trans\X_i) \right\}=\0,\\
&&\sumi K_h\left(\bb\trans\X_i-\bb\trans\X_j \right)
\left\{\frac{A_iY_i}{\pi(\X_i)}-\frac{(1-A_i)Y_i}{1-\pi(\X_i)}-c_j\right\}
=\0 \mbox{ for }j=1, \dots, n,
\ese
where $\pi(\X_i)$ a known propensity score model. To gain protection
against a misspecified $\pi(\X_i)$, we adopt models 
$\mu(\X_i, \ba)=E(Y_{i0}\mid\X_i)$ and $\pi(\X_i,\bg)=P(A_i\mid\X_i)$,
and obtain the estimate of $\bb$ by modifying the above
  equations to a double-robust augmented version
\citep{rrzhao1994}
\be
\sumi
\left[\frac{\{A_i-\pi(\X_i,\wh\bg)\}\{Y_i-\mu(\X_i,\wh\ba)\}}{\pi(\X_i,\wh\bg)\{1-\pi(\X_i,\wh\bg)\}}
+\left\{1-\frac{A_i}{\pi(\X_i,\wh\bg)}\right\}\wt Q(\bb\trans\X_i)
\right]&&  \label{eq:beta} \\
\times\left\{\X_{Li}-E(\X_{Li}\mid\bb\trans\X_i) \right\}&=&\0, \n
\ee
and
\bse
\sumi K_h\left(\bb\trans\X_i-\bb\trans\X_j \right)
\left[\frac{\{A_i-\pi(\X_i,\wh\bg)\}\{Y_i-\mu(\X_i,\wh\ba)\}}{\pi(\X_i,\wh\bg)\{1-\pi(\X_i,\wh\bg)\}}
-\frac{A_i}{\pi(\X_i,\wh\bg)} c_j
\right]
=\0 
\ese
for $j=1, \dots, n$. 
Here, the above relation can be equivalently written as
\be\label{eq:q}
&&\wt Q (\bb\trans\X_j,\bb,\wh\ba,\wh\bg) \\
&=& \frac{\sumi K_h (\bb\trans\X_i - \bb\trans\X_j) \{A_i - \pi(\X_i,\wh\bg)\}\{ Y_i - \mu(\X_i,\wh\ba) \} / [ \pi(\X_i,\wh\bg)\{ 1- \pi(\X_i,\wh\bg)\}  ] }
{\sumi K_h (\bb\trans\X_i - \bb\trans\X_j) A_i / \pi(\X_i,\wh\bg)}.\n
\ee
The $\bb$ estimator based in (\ref{eq:q}) is double
robust with respect to $\pi(\X,\bg)$ and $\mu(\X,\ba)$. 
Following the literature, we consider parametric models
  $\pi(\X,\bg)$ and $\mu(\X,\ba)$ for simplicity.
\begin{Pro}\label{pro:robust}
Under the model in (\ref{eq:model}), 
as long as one of the two models $\pi(\x,\bg)$ and $\mu(\x,\ba)$ is 
correct, then estimator for $\bb$ is
consistent. In addition, for the estimation of $\bb$, we can further
use a working model for the $Q(\cdot)$ function, which may 
be different from the true treatment difference function if both $\pi(\x,\bg)$ and $\mu(\x,\ba)$ are correctly specified. 
\end{Pro}

Note that in Proposition \ref{pro:robust}, we do not 
require known values of $\bg$ or $\ba$. Instead, $\bg$ and $\ba$ are 
unknown parameters. As long as one of the models in
$\pi(\x,\bg)$ and $\mu(\x,\ba)$ is correctly specified, the conclusion
of Proposition \ref{pro:robust} holds.
Note also that $\wh\bg$ can be obtained based on data $(\X_i,A_i), i=1,
\dots, n$ via, for
example, maximum likelihood estimator (MLE). Likewise, 
$\mu(\X,\ba)=E(Y_i\mid\X_i, A_i=0)$ and hence
$\wh\ba$ can be obtained based on data $(\X_i,
Y_i)$ for the $i$ where $A_i=0$, via, for example,  solving
generalized estimating equations (GEEs). 
Once we obtain $\wh Q(\cdot,\wh\bb,\wh\ba,\wh\bg)$ and $\wh\bb$, we
directly identify the optimal 
treatment regime by assigning treatment 1 if and only if $\wh
Q(\wh\bb\trans\X)>0$. This strategy automatically maximizes the expected value
function, $V\{Q(\cdot),\bb\}\equiv 
E[Y_{i1}I\{Q(\bb\trans\X_i)>0\}+Y_{i0}I\{Q(\bb\trans\X_i)\le0\}]$.

In solving (\ref{eq:beta}) to obtain $\wh\bb$, the
choice of bandwidth $h$ is flexible and can be any positive number as
long as 
$n^{-1/2}<<h<<n^{-1/4}$. However, once we obtain  $\wh\bb$, we estimate
$Q(\cdot)$ using an optimal bandwidth of order $n^{-1/5}$, which can be obtained
from cross-validation. We now describe the algorithm of the
estimation procedure in detail.

\textbf{Algorithm:}
\begin{enumerate}
	\item[Step 1.] Obtain the estimate of $\bg$, $\wh\bg$ by MLE based on
	the data $(\X_i,A_i)$, $i= 1, \dots, n$. 
	\item[Step 2.] Extract the observations with $A_i=0$. Denote the sub-dataset corresponding to $A_i=0$ as
	$(\X_i,Y_i^0), i=1, \dots, n_0$. Using the subset of the observations
	to compute the estimator of $\ba$, $\wh\ba$ by solving the GEEs
	\be\label{eq:gee_alpha}
	\sum_{i=1}^{n_0} \W(\X_i,\ba) \{ Y_i^0 - \mu(\X_i,\ba) \} =\0,
	\ee
	where $\W(\X_i,\ba)$ is an arbitrary $d_{\ba} \times 1$ matrix of functions of covariates $\X_i$,
	the parameter $\ba \in \mR^{d_\ba}$ and $d_{\ba}$ is the dimension of $\ba$.
	
	\item[Step 3.] Plug  $\wh\bg$ and $\wh\ba$ in (\ref{eq:q}) and obtain
	$\wt Q (\bb\trans\X_j,\bb,\wh\ba,\wh\bg)$.
	
	\item[Step 4.] Plug $\wh\bg$, $\wh\ba$ and $\wt
	Q(\bb\trans\X_i,\bb,\wh\ba,\wh\bg)$'s in
	(\ref{eq:beta}) and solve  (\ref{eq:beta}) to
	get $\wh\bb_L$. 
	\item[Step 5.] Select a bandwidth $h_{\rm opt}$.
	
	\item[Step 6.] Obtain 
	$
	\wh Q(\cdot,\wh\bb,\wh\ba,\wh\bg)
	$
	by (\ref{eq:q}), while plugging in $\wh\bg$,
	$\wh\ba$, $\wh\bb$ and $h_{\rm opt}$.
	
\end{enumerate}
In step 5, to estimate $Q(\cdot)$, we need a suitable bandwidth. We
adopt the leave-one-out cross-validation method to select
the bandwidth. Specifically, we estimate $Q(\cdot)$ by 
\bse
&&\wt Q_{-j} (\wh\bb\trans\X_j,\wh\bb,\wh\ba,\wh\bg) \\
&=& \frac{\sum_{i=1,i\neq j}^n K_h (\wh\bb\trans\X_i -
	\wh\bb\trans\X_j) \{A_i - \pi(\X_i,\wh\bg)\}\{ Y_i -
	\mu(\X_i,\wh\ba) \} / [ \pi(\X_i,\wh\bg)\{ 1- \pi(\X_i,\wh\bg)\}  ]
} 
{\sum_{i=1,i\neq j}^n K_h (\wh\bb\trans\X_i - \wh\bb\trans\X_j) A_i / \pi(\X_i,\wh\bg)},
\ese
where $\wt Q_{-j} (\cdot)$ denotes the estimator
with the $j\th$ observation left out.
Then we calculate the leave-one-out cross-validated prediction MSE as 
\bse
\hbox{CV}(h) 
= \frac{1}{n}\sumi \left[\frac{\{A_i-\pi(\X_i,\wh\bg)\}\{Y_i-\mu(\X_i,\wh\ba)\}}{\pi(\X_i,\wh\bg)\{1-\pi(\X_i,\wh\bg)\}}
-\frac{A_i}{\pi(\X_i,\wh\bg)}
\wt Q_{-i}(\wh\bb\trans\X_i,\wh\bb,\wh\ba,\wh\bg) \right]^2
\ese
and we choose $h$ to be the minimizer of CV$(h)$. 

After obtaining $\wh\bb$ and $\wh Q(\cdot,\wh\bb,\wh\ba,\wh\bg)$, we can estimate the maximum expected value function as
\be\label{eq:value}
&&\wh V\{\wh Q(\cdot),\wh\bb,\wh\ba,\wh\bg\}\n\\
&=&
n^{-1} \sumi 
\frac{ [ A_i I\{\wh Q(\wh\bb\trans\X_i,\wh\bb,\wh\ba,\wh\bg) > 0 \} 
	+ (1-A_i) I\{\wh Q(\wh\bb\trans\X_i,\wh\bb,\wh\ba,\wh\bg) \leq 0 \}  ]Y_i }
{\pi(\X_i,\wh\bg) I\{\wh Q(\wh\bb\trans\X_i,\wh\bb,\wh\ba,\wh\bg) > 0 \} 
	+ \{1-\pi(\X_i,\wh\bg)\} I\{\wh
	Q(\wh\bb\trans\X_i,\wh\bb,\wh\ba,\wh\bg) \leq 0 \}}\n\\
&&+n^{-1} \sumi 
\{\pi(\X_i,\wh\bg)-A_i\}\left[
\frac{\mu(\X_i,\wh\ba)+\wh
	Q(\wh\bb\trans\X_i,\wh\bb,\wh\ba,\wh\bg)}{\pi(\X_i,\wh\bg)}I\{\wh
Q(\wh\bb\trans\X_i,\wh\bb,\wh\ba,\wh\bg) > 0 \} \right.\n\\
&&\left.-
\frac{\mu(\X_i,\wh\ba)}{1-\pi(\X_i,\wh\bg)}I\{\wh
Q(\wh\bb\trans\X_i,\wh\bb,\wh\ba,\wh\bg) \le 0 \} \right].\n\\
&=&
n^{-1} \sumi \left(
\frac{ [ A_i 
	+ (1-2A_i) I\{\wh Q(\wh\bb\trans\X_i,\wh\bb,\wh\ba,\wh\bg) \leq 0 \}  ]Y_i }
{\pi(\X_i,\wh\bg) 
	+ \{1-2\pi(\X_i,\wh\bg)\} I\{\wh
	Q(\wh\bb\trans\X_i,\wh\bb,\wh\ba,\wh\bg) \leq 0 \}}+\{\pi(\X_i,\wh\bg)-A_i\}\right.\\
&&\left.\times
\frac{
	[\mu(\X_i,\wh\ba)+\wh Q(\wh\bb\trans\X_i,\wh\bb,\wh\ba,\wh\bg)
	-\{2 \mu(\X_i,\wh\ba)+
	\wh Q(\wh\bb\trans\X_i,\wh\bb,\wh\ba,\wh\bg)\}
	I\{\wh
	Q(\wh\bb\trans\X_i,\wh\bb,\wh\ba,\wh\bg)\le0 \}]}{
	\pi(\X_i,\wh\bg) 
	+ \{1-2\pi(\X_i,\wh\bg)\} I\{\wh
	Q(\wh\bb\trans\X_i,\wh\bb,\wh\ba,\wh\bg) \leq 0 \}}\right), \n
\ee 
which is a consistent estimator of the true value function 
$V\{Q(\cdot),\bb\}\equiv 
E[Y_{i1}I\{Q(\bb\trans\X_i)>0\}+Y_{i0}I\{Q(\bb\trans\X_i)\le0\}]$.

\section{Theoretical Properties}\label{sec:theo}

We now study the theoretical properties of the proposed
estimators. For notational simplicity, define
\bse
\W(\bg) &\equiv&  E\left(\frac{\partial^2 \log[
\pi(\X,\bg)^{A}\{1-\pi(\X,\bg)\}^{1-A}]
}{\partial \bg \partial \bg\trans}  \right),\\
\bphi_\bg(\X_i,A_i,\bg)
&\equiv&\W(\bg)^{-1}
\frac{\partial}{\partial \bg} \log \left[ \pi(\x_i,\bg)^{A_i} \{ 1-
\pi(\x_i,\bg) \}^{1-A_i} \right],\\
\bphi_\ba(\X_i,A_i,Y_i,\ba)
&\equiv&
\left[ E\left\{ \W(\X,\ba)\D(\X,\ba)\right\}\right]^{-1}
\W(\X_i,\ba) (1-A_i)\left\{Y_i - \mu(\X_i,\ba)\right\},
\ese
where $\W(\X_i,\ba)$ is an arbitrary weight matrix and $\D(\X,\ba)=\partial \mu(\X,\ba)/\partial \ba\trans$.
Through out the paper, $\a^{\otimes2} = \a\a\trans.$

\begin{Pro}\label{pro:gamma}
Write the conditional distribution of $A$ given $\X$  as $\pi(\x,\bg)^{a}\{1-\pi(\x,\bg)\}^{1-a}$. 
Regardless if  $\pi(\X,\bg)$ 
is the true propensity score model or not, there exists $\bg_0$
so that the MLE, $\wh\bg$ satisfies
\be\label{eq:expandgamma}
\sqrt{n}(\wh\bg-\bg_0)= \frac{1}{\sqrt{n}} \sumi 
\bphi_\bg(\X_i,A_i,\bg)
+ o_p(1),
\ee
hence
$\sqrt{n}(\wh\bg - \bg_0) \to N\{\0,\W(\bg_0)^{-1}\B(\bg_0)\W(\bg_0)^{-1}\}$ in
distribution when $n\to\infty$, where 
\bse
\B(\bg_0) \equiv E\left\{
\bphi_\bg(\X_i,A_i,\bg_0)^{\otimes2}
\right\}.
\ese
Specifically, when the
$\pi(\x,\bg)$ model is correct, $\bg_0$ is the true parameter value
which yields $\pi_0(\x)=\pi(\x,\bg_0)$,
and the covariance matrix simplifies to inverse of Fisher's
Information matrix $\I(\bg_0)$, which is 
$
\I(\bg_0) = -\W(\bg_0)^{-1}.
$
When the
$\pi(\x,\bg)$ model is incorrect, $\bg_0$ is the parameter vector
which minimizes the Kullback-Leibler distance 
$E\left\{\log\left(
[\pi_0(\X)^{A}\{1-\pi_0(\X)\}^{1-A}]/ \right.\right.$
$\left.\left.[\pi(\X,\bg)^{A}\{1-\pi(\X,\bg)\}^{1-A}] 
\right) \right\}$.
\end{Pro}

\begin{Pro}\label{pro:alpha}
Whether $\mu(\X,\ba)$ is the true model or not, let $\wh\ba$ be
estimator which solves the  estimating equation $\sum_{i=1}^{n_0}
\W(\X_i,\ba)\{ Y_i^{0}-\mu(\X_i,\ba)\}=\0$.
Then, 
\be\label{eq:expandalpha}
\sqrt{n_0}(\wh\ba-\ba_0)=\frac{1}{\sqrt{n_0}} \sum_{i=1}^{n_0} 
\bphi_\ba(\X_i,A_i,Y_i,\ba_0) + o_p(1)
=
\frac{1}{\sqrt{n_0}} \sumi
\bphi_\ba(\X_i,A_i,Y_i,\ba_0) + o_p(1)
\ee
hence
$\sqrt{n_0}(\wh\ba-\ba_0)\to N(\0,\V_\ba)$ in distribution when
$n\to\infty$, where
\bse
\V_\ba = [E\{\W(\X,\ba_0)\D(\X,\ba_0)\}]^{-1}E\{\W(\X,\ba_0)v(\X)\W(\X,\ba_0)\trans\}([E\{\W(\X,\ba_0)\D(\X,\ba_0)\}]^{-1})\trans,
\ese
and $v(\X)$ is
the conditional variance of $Y$ given $\X$. When the model
$\mu(\X,\ba)$ is correctly specified, $\ba_0$ satisfies
$\mu_0(\X)=\mu(\X,\ba_0)$, where $\mu_0(\x)$ is the true mean outcome
under $A=0$, while when  the model
$\mu(\X,\ba)$ is incorrectly specified, $\ba_0$ satisfies
$E[
\W(\X_i,\ba_0) \{ Y_i^0 - \mu(\X_i,\ba_0) \}] =0
$.
\end{Pro}

The results in Proposition \ref{pro:gamma} and
\ref{pro:alpha} are direct consequence from \cite{white1982} and
\cite{yi_reid2010}, hence we omit the detailed proofs. 
The asymptotic properties for the
estimators $\wh\bb$, $\wh Q(\cdot,\wh\bb,\wh\ba,\wh\bg)$, the root of 
$\wh Q(\cdot,\wh\bb,\wh\ba,\wh\bg)$, and  $\wh V \{\wh Q(\cdot),\wh\bb,\wh\ba,\wh\bg\}$ 
are developed under the 
following conditions.

\textbf{Regularity Conditions:}
\begin{enumerate}[label=\textnormal{(C\arabic*)}]

\item\label{con:C1}The true parameter value $\bb_0$ belongs to a compact set $\Omega$.

\item\label{con:C2} The univariate kernel $K(\cdot)$ is symmetric, has compact support and is Lipschitz continuous
on its support. It satisfies
\bse
\int K(u)du =1, \,\,\,\, \int u K(u)du = 0, \,\,\,\, 0\neq \int u^2 K(u)du < \infty.
\ese

\item\label{con:C3} The probability density function of $\bb\trans\X$, which is denoted by $f(\bb\trans\x)$, is bounded away from $0$ and $\infty$.

\item\label{con:C4} $E(\X\mid\bb\trans\x)f(\bb\trans\x)$ and
$Q(\bb\trans\x)$ are twice differentiable and their
second derivatives, as well as $f(\bb\trans\x)$ are locally
Lipschitz-continuous and bounded.

\item\label{con:C5} The bandwidth $h=O(n^{-\kappa})$ for $1/8 <
\kappa < 1/2$. 

\item\label{con:C6} The
treatment assignment probability satisfies $c<\pi_0(x)<1-c$, where
$c$ is a small positive constant.

\item\label{con:C7} The true treatment responses, $\mu_0(\x)$ for
  non-treated group and $\mu_1(\x)$ for treated group are bounded by a
  constant $C$.

\item\label{con:C8} The true treatment effect function
$Q(\bb_0\trans\x)$ has roots $r_1, \dots, r_K$, $K<\infty$. In
addition,  $Q'(r_k)\ne0$ for all $k=1, \dots, K$.

\end{enumerate}

Conditions \ref{con:C1}--\ref{con:C5} are standard
  conditions and they ensure sufficient convergence rate of the
nonparametric estimators. Condition \ref{con:C6} is also routinely
assumed to exclude weights near 0 and 1. Condition \ref{con:C7} is
very mild and is usually satisfied in practice.
Condition \ref{con:C8} allows roots for the function $Q(\cdot)$ and is
also very mild.

\begin{Lem}\label{lem:order}
Under Conditions \ref{con:C2}, \ref{con:C3} and
\ref{con:C4}, at any $\bb\in\Omega$ and for any function $H(\X_j,A_j,Y_j)$
such that $E\left\{ H(\X_j,A_j,Y_j) \mid \bb\trans\X_j\right\}$ is
twice  differentiable, we have
\bse
&&E\left\{ H(\X_j,A_j,Y_j) K_h(\bb\trans\X_j -\bb\trans\x)\right\} - E\left\{ H(\X_j,A_j,Y_j) \mid \bb\trans\x \right\}  f(\bb\trans\x) \\
&=& \frac{\partial^2}{(\partial \bb\trans\x)^2} [E\left\{ H(\X_j,A_j,Y_j) \mid \bb\trans\x \right\}
f(\bb\trans\x)]\frac{h^2}{2} \int z^2 K(z) dz +
o(h^2),\\ 
&&\var\{n^{-1} \sumj H(\X_j,A_j,Y_j)K_h(\bb\trans\X_j - \bb\trans\x)\} \\
&=& 
(nh)^{-1}E\{H^2(\X_j,A_j,Y_j) \mid \bb\trans\x\}f(\bb\trans\x)
\int K^2(z) dz+O(n^{-1}).
\ese
\end{Lem}

\begin{Lem}\label{lem:rateQ}
Assume
the regularity Conditions \ref{con:C1}-\ref{con:C5} hold. Then 
at any $\bb\in\Omega$
the kernel estimator  
$\wt Q(\bb\trans\x,\bb,\ba_0,\bg_0)$ satisfies
\bse
\wt Q(\bb\trans\x,\bb,\ba_0,\bg_0)  - Q(\bb\trans\x) = O_p\{h^2 + (nh)^{-1/2}\}.
\ese
\end{Lem}

Note that the convergence rate of $\wt Q(\bb\trans\x,\bb,\ba_0,\bg_0)$ is
slower than root-$n$ while 
$\wh\bg, \wh\ba$ have root-$n$ convergence rate. Thus,
estimating $\wt Q(\bb\trans\x,\bb,\wh\ba,\wh\bg)$, which is
based on $\wh\ba$ and  $\wh\bg$ instead of based on 
$\ba_0$ and  $\bg_0$, does not change the results in Lemma
\ref{lem:rateQ}.

To facilitate the statements in the following Theorems \ref{th:beta}
and \ref{th:value}, we first make a few definitions. 
Let
\bse
&&\bpsi_\bb(\X_i,Y_i,A_i,\bb,\ba,\bg)\\
&\equiv&
\left[ \frac{\{A_i - \pi(\x_i,\bg)\}\{Y_i - \mu(\x_i,\ba)\}}{\pi(\x_i,\bg)\{1-\pi(\x_i,\bg)\}} 
+ \left\{1 - \frac{A_i}{\pi(\x_i,\bg)}\right\} Q(\bb\trans\x_i)  \right] 
\left\{\x_{Li} - E(\X_{Li} \mid \bb\trans\x_i) \right\},\\
&&\bphi_\bb(\X_i,Y_i,A_i,\bb,\ba,\bg)\\
&\equiv&
\bpsi_\bb(\X_i,Y_i,A_i,\bb,\ba,\bg)
+
\frac{
E\left[\left\{\X_{Li} - E(\X_{Li} \mid \bb_0\trans\X_i) \right\}
\left\{1 - \pi_0(\X_i)/\pi(\X_i,\bg_0)\right\}\mid\bb_0\trans\X_i\right]}{E\{\pi_0(\X_i) /
\pi(\X_i,\bg_0)\mid\bb\trans\X_i\}}
\\
&&\times\left[\frac{
\{A_i - \pi(\X_i,\bg_0)\}\{ Y_i - \mu(\X_i,\ba_0) \} }{  \pi(\X_i,\bg_0)\{ 1- \pi(\X_i,\bg_0)\}  }
- \frac{Q(\bb_0\trans\X_i) A_i}{ \pi(\X_i,\bg_0)}\right]
+
Q(\bb_0\trans\X_i)\left\{E(\X_{Li} \mid \bb_0\trans\X_i) -\X_{Li}
\right\}.
\ese
Define also
\bse
\B_\bg&\equiv&
E\left[
\frac{\left\{\X_{L} -  E(\X_{L}  \mid
\bb_0\trans\X) \right\}\bpi_\bg\trans(\X,\bg_0)\{\mu(\X,\ba_0)-\mu_0(\X) \}}{\pi(\X,\bg_0) 
\{1-\pi(\X,\bg_0)\}} \right],
\ese
\bse
\B_\ba&\equiv&
E\left(
\left\{\X_{L} -E(\X_{L} \mid
\bb_0\trans\X) \right\}
\left\{1-\frac{\pi_0(\X)}{\pi(\X,\bg_0)}\right\}\left[
\frac{\bmu_\ba\trans(\X,\ba_0)}{1-\pi(\X,\bg_0)} \right.\right.\\
&&\left.\left.-\frac{E(\{\pi_0(\X) - \pi(\X,\bg)\} \bmu_\ba\trans(\X,\ba) / [\pi(\X,\bg)\{
1- \pi(\X,\bg)\}]\mid\bb\trans\X)}
{E\{\pi_0(\X) / \pi(\X,\bg)\mid\bb\trans\X\}} \right] \right),\\
\B&\equiv&
E\left[ 
\frac{-\pi_0(\X)\left\{\X_L - E(\X_L \mid \bb_0\trans\X)
\right\}
}{\pi(\X,\bg_0) f(\bb\trans\x)
E\{\pi_0(\X) / \pi(\X,\bg)\mid\bb\trans\X\}}\right.\\
&&\left.\times\frac{\partial}{\partial\bb\trans\x}\left\{E\left(\X_L\trans
\left[\frac{\pi_0(\X)}{\pi(\X,\bg) }-
E\left\{\frac{\pi_0(\X)}{ \pi(\X,\bg)}\mid\bb\trans\x\right\} 
\right]\mid\bb\trans\x\right) Q(\bb\trans\x) f(\bb\trans\x)\right\}
\right]\\
&&+E\left\{ 
\frac{Q(\bb_0\trans\X)}{f(\bb\trans\X)}\frac{\partial\left( E[\{\X_L-E(\X_L\mid\bb\trans\x)\}^{\otimes2}\mid\bb\trans\X] f(\bb\trans\X)\right)
}{\partial(\bb\trans\X)}\right\}.
\ese
\begin{Th}\label{th:beta}
Assume $\wh \bb_L$  solves (\ref{eq:beta}). 
Then, under the regularity
conditions \ref{con:C1}-\ref{con:C5},  
$\wh\bb_L$ satisfies $\sqrt{n}(\wh\bb_L-\bb_{L0})\rightarrow 
N\{\0,\B^{-1} \V_1(\B^{-1})\trans \}$
in distribution as $n\rightarrow\infty$, where
\bse
\V_1&\equiv&  E\left[\left\{\bphi_\bb(\X_i,Y_i,A_i,\bb_0,\ba_0,\bg_0)
+ \B_{\bg}\bphi_\bg(\X_i,A_i,\bg_0)
+\B_{\ba} \bphi_\ba(\X_i,A_i,Y_i,\ba_0)
\right\}^{\otimes2} \right].
\ese
\end{Th}
The first term in the variance expression $\V_1$ captures the
variability of estimating different functions, the second term
captures the variability in estimating $\bb$ due to the estimation of
$\bg$ and the third term captures the same induced by $\wh\ba$. 

\begin{Lem}\label{lem:Q}
Assume
the regularity Conditions \ref{con:C1}-\ref{con:C5} hold. Then 
the kernel estimator obtained from Step 6,
$\wh Q(\wh\bb\trans\x,\wh\bb,\wh\ba,\wh\bg)$ satisfies
\bse
&&\bias\{\wh Q(\wh \bb\trans\x,\wh\bb,\wh\ba,\wh\bg)\}  
=h_{\rm opt}^2
\left\{\frac{Q'(\bb_0\trans\x)
d[E\{\pi_0(\X_j)/\pi(\X_j,\bg_0)\mid \bb_0\trans\x\}f(\bb_0\trans\x)]/d(\bb_0\trans\x)
}{
E\{\pi_0(\X_j)/\pi(\X_j,\bg_0)\mid \bb_0\trans\x\}
f(\bb_0\trans\x)} 
+ \frac{Q''(\bb_0\trans\x)}{2}\right\}\\
&&\times\int z^2 K(z) dz
+ o(h_{\rm opt}^2 + n^{-1/2}h_{\rm opt}^{-1/2}), 
\ese
and
\bse
&&\var\{\wh Q(\wh\bb\trans\x,\wh\bb,\wh\ba,\wh\bg)\} \\
&=&\frac{1}{nh_{\rm opt}} \bigg( 
E\left[ \frac{\pi_0(\X_j)}{\pi^2(\X_j,\bg_0)} \{
Y_{1j}-\mu(\X_j,\ba_0)\}^2  \mid \bb_0\trans\x\right]\\
&&+ E\left[\frac{1-\pi_0(\X_j)}{\{1-\pi(\X_j,\bg_0)\}^2}  \{ Y_{0j} -
\mu(\X_j,\ba_0) \}^2   \mid \bb_0\trans\x\right]
\\
&& - Q^2(\bb_0\trans\x) 
E\left\{ \frac{\pi_0(\X_j)}{\pi^2(\X_j,\bg_0) }\mid \bb_0\trans\x \right\} 
-  2 Q(\bb_0\trans\x) 
E\left[\frac{\pi_0(\X_j)}{\pi^2(\X_j,\bg_0)}
\left\{\mu_0(\X_j) - \mu(\X_j,\ba_0) \right\} \mid \bb_0\trans\x \right]
\bigg)  \\
&& \times \frac{1}{f(\bb_0\trans\x)}\left[E\left\{ \frac{\pi_0(\X_j)}{\pi(\X_j,\bg_0)} \mid \bb_0\trans\x \right\} \right]^{-2}\int K^2(z) dz+ O(n^{-1}).
\ese
Here, for a generic function $r(\cdot)$, $r'(\cdot)$ and $r''(\cdot)$
are respectively its first and  second derivatives.
\end{Lem}

\begin{Th}\label{th:root}
Let $Q(z_0)=0$ and $\wh Q(\wh z,\wh\bb,\wh\ba,\wh\bg)=0$. Then as
$n\to\infty$, under the regularity Conditions 
\ref{con:C1}-\ref{con:C5},
$\wh z\to z_0$ at the rate $n^{-2/5}$. Specifically,  
the leading
term of the bias of $\wh z$ is
\bse
-h_{\rm opt}^2
\left\{\frac{
d[E\{\pi_0(\X)/\pi(\X,\bg_0)\mid \bb_0\trans\X=z_0\}f(z_0)]/d(z_0)
}{
E\{\pi_0(\X)/\pi(\X,\bg_0)\mid \bb_0\trans\X=z_0\}
f(z_0)} 
+ \frac{Q''(z_0)}{2 Q'(z_0)}\right\}\int z^2 K(z) dz,
\ese
and the leading term of the variance  of $\wh z$ is
\bse
&&\frac{1}{nh_{\rm opt}} \bigg( 
E\left[ \frac{\pi_0(\X)}{\pi^2(\X,\bg_0)} \{
Y_{1}-\mu(\X,\ba_0)\}^2  \mid \bb_0\trans\X=z_0\right]\\
&&+ E\left[\frac{1-\pi_0(\X)}{\{1-\pi(\X,\bg_0)\}^2}  \{ Y_{0} -
\mu(\X,\ba_0) \}^2   \mid \bb_0\trans\X=z_0\right]
\bigg)  \\
&& \times \frac{1}{f(z_0) Q'(z_0)^2}\left[E\left\{ \frac{\pi_0(\X)}{\pi(\X,\bg_0)} \mid \bb_0\trans\X=z_0\right\} \right]^{-2}\int K^2(z) dz.
\ese
\end{Th}
From Theorem \ref{th:root}, we can observe that our treatment region
identification rate is $O_p(n^{-2/5})$, better than the classical rate
$O_p(n^{-1/3})$ \citep{fanetal2017}. This is due to the smoothness
assumption made in Condition \ref{con:C4}. 
To further state Theorem \ref{th:value}, we make a few more
definitions. Let
\bse
\U_\bb&\equiv&
E\left(\frac{[1-E\{\pi_0(\X)/\pi(\X,\bg_0)\mid\bb_0\trans\X\}]
I\{Q(\bb_0\trans\X)>0\}}{f(\bb_0\trans\X)
E\{\pi_0(\X) / \pi(\X,\bg_0)\mid\bb_0\trans\X\}}\right.\n\\
&&\left.\times \frac{\partial}{\partial\bb_0\trans\X}
\left[
\cov\left\{\X_L,\frac{\pi_0(\X)}{\pi(\X,\bg_0) }\mid\bb_0\trans\X\right\}
Q(\bb_0\trans\X)
f(\bb_0\trans\X)\right]\right),\n\\
\U_\ba&\equiv&
E\left(
[1-E\{\pi_0(\X)/\pi(\X,\bg_0)\mid\bb_0\trans\X\}]
I\{Q(\bb_0\trans\X)>0\}
\right.\n\\
&&\left.\times\frac{E(
\{\pi(\X,\bg_0)-\pi_0(\X)\} \bmu_\ba(\X,\ba_0) / [ \pi(\X,\bg_0)\{ 1- \pi(\X,\bg_0)\}  ] \mid\bb_0\trans\X)}
{E\{\pi_0(\X) / \pi(\X,\bg_0)\mid\bb_0\trans\X\}}\right.\\
&&\left.+\frac{\{\pi(\X,\bg_0)-\pi_0(\X)\}\bmu_\ba(\X,\ba_0)[1-2I\{Q(\bb_0\trans\X)\le0\}]}
{\pi(\X,\bg_0)+\{1-2\pi(\X,\bg_0)\}I\{Q(\bb_0\trans\X)\le0\}}\right),\n\\
\U_\bg&\equiv&
E\left(
[1-E\{\pi_0(\X)/\pi(\X,\bg_0)\mid\bb_0\trans\X\}]
I\{Q(\bb_0\trans\X)>0\}\right.\n\\
&&\left.\times\frac{E(\bpi_\bg(\X,\bg_0)\{
\mu(\X,\ba_0)-\mu_0(\X) \} / [ \pi(\X,\bg_0)\{ 1- \pi(\X,\bg_0)\}  ]\mid\bb_0\trans\X) }
{E\{\pi_0(\X) / \pi(\X,\bg_0)\mid\bb_0\trans\X\}}\right)
\n\\
&&+E\left(\left[
\frac{
I\{
Q(\bb_0\trans\X) \leq 0 \}\{1-\pi_0(\X)\}}{
\{1-\pi(\X,\bg_0)\}^2}
-\frac{I\{
Q(\bb_0\trans\X)>0 \}\pi_0(\X)}{
\pi(\X,\bg_0)^2}\right]\right.\n\\
&&\left.\times \{\mu_0(\X)-\mu(\X,\ba_0)\}\bpi_\bg(\X,\bg_0)
\right).
\ese
Further,  define
\bse
&&\wt v_Q\{\X_i,A_i,Y_i,\bb,\ba,\bg,Q(\cdot)\}\n\\
&\equiv&
\left\{
E\left(\frac{J_a'\{Q(\bb_0\trans\X)\}
\{\pi(\X,\bg_0)-1\}\pi(\X,\bg_0)Q(\bb_0\trans\X)}{[\pi(\X,\bg_0)+\{1-2\pi(\X,\bg_0)\} J_a\{
Q(\bb_0\trans\X)\}]^2}\mid\bb_0\trans\X_i\right)\right.\n\\
&&\left.+
E\left(\frac{
I\{Q(\bb_0\trans\X)>0\}
\{\pi(\X,\bg_0)-\pi_0(\X)\}}{\pi(\X,\bg_0)}\mid\bb_0\trans\X_i\right)\right\}\n\\
&&\left(
\frac{ \{A_i - \pi(\X_i,\bg_0)\}\{ Y_i - \mu(\X_i,\ba_0) \} / \{ 1- \pi(\X_i,\bg_0)\}  
-A_i Q(\bb_0\trans\X_i)
}
{\pi(\X_i,\bg_0) E\{\pi_0(\X_i) /
\pi(\X_i,\bg_0)\mid\bb_0\trans\X_i\}}
\right),
\ese
\bse
&& v_Q\{\X_i,A_i,Y_i,\bb,\ba,\bg,Q(\cdot)\}\n\\
&\equiv&
\left\{
E\left(\frac{
I\{Q(\bb_0\trans\X)>0\}
\{\pi(\X,\bg_0)-\pi_0(\X)\}}{\pi(\X,\bg_0)}\mid\bb_0\trans\X_i\right)\right\}\n\\
&&\left(
\frac{ \{A_i - \pi(\X_i,\bg_0)\}\{ Y_i - \mu(\X_i,\ba_0) \} / \{ 1- \pi(\X_i,\bg_0)\}  
-A_i Q(\bb_0\trans\X_i)
}
{\pi(\X_i,\bg_0) E\{\pi_0(\X_i) /
\pi(\X_i,\bg_0)\mid\bb_0\trans\X_i\}}
\right),
\ese
and
\bse
&&v_0(\X_i,A_i,Y_i)\\
&\equiv&\left(
\frac{ [ A_i 
+ (1-2A_i) I\{Q(\bb_0\trans\X_i) \le0 \}  ]Y_i }
{\pi(\X_i,\bg_0) 
+ \{1-2\pi(\X_i,\bg_0)\} I\{
Q(\bb_0\trans\X_i) \le0 \}}
+\{\pi(\X_i,\bg_0)-A_i\}\right.\n\\
&&\left.\times
\frac{
[\mu(\X_i,\ba_0)+Q(\bb_0\trans\X_i)
-\{2 \mu(\X_i,\ba_0)+
Q(\bb_0\trans\X_i)\}
I\{Q(\bb_0\trans\X_i)\le0\}]}{
\pi(\X_i,\bg_0) 
+ \{1-2\pi(\X_i,\bg_0)\} I\{
Q(\bb_0\trans\X_i) \le0 \}}
-V\{Q(\cdot),\bb_0\}
\right).
\ese
\begin{Th}\label{th:value}
Under the regularity Conditions 
\ref{con:C1}-\ref{con:C8},
the optimal value function estimator given in (\ref{eq:value}) satisfies
\bse
n^{1/2}[\wh V\{\wh Q(\cdot),\wh\bb,\wh\ba,\wh\bg\}
-V\{Q(\cdot),\bb_0\}]\to N(0,\sigma^2)
\ese
in distribution when $n\to\infty$, where
\bse
&&\sigma^2\\
&=&E
\left[-\U_\bb\trans
\B^{-1}\left\{\bphi_\bb(\X_i,Y_i,A_i,\bb_0,\ba_0,\bg_0)
+ \B_{\bg}\bphi_\bg(\X_i,A_i,\bg_0)
+\B_{\ba} \bphi_\ba(\X_i,A_i,Y_i,\ba_0)
\right\}\right.\\
&&\left.+\U_\ba\trans \bphi_\ba(\X_i,A_i,Y_i,\ba_0)
+\U_\bg\trans \bphi_\bg(\X_i,A_i,\bg_0)
+ v_Q\{\X_i,A_i,Y_i,\bb_0,\ba_0,\bg_0,Q(\cdot)\}
+v_0(\X_i,A_i,Y_i)
\right]^2.
\ese
\end{Th}

We can understand the first term in
  $\sigma^2$ as the variability in value function due to
  $\bb$. The second term is related to the variability
  induced by $\ba$. The third term captures the
  variability due to the $\bg$ estimation. The fourth term
  measures the variability in the value function induced by
  estimating the
  treatment effect function. Lastly, the fifth term captures the
  variability in the value function inherited from the
    variability of the covariates.

\section{Simulations}\label{sec:simu}
We conducted simulation studies to compare the performance of the estimators discussed in Section \ref{sec:method}. We considered different scenarios, depending misspecification of either $\pi(\X,\bg)$ or $\mu(\X,\ba)$ to show the robustness property of our estimators. We used sample size $n=500$ with 1000 replicates. 

\subsection{Simulation 1}\label{subsec:simu1}

Our first simulation follows similar designs as those 
in \cite{fanetal2017}, where the monotonicity of $Q(\cdot)$ is
required. We set $d_\bb=4$ and generated 
the covariate vector $\X_i$ from the multivariate  normal distribution
with zero mean and identity covariance matrix. We generated the 
treatment indicator $A_i$ from a
Bernoulli distribution with probability $\pi_0(\X_i)=0.5$.
The response variables are formed from $Y_i=\mu_0(\X_i)+A_i
Q_0(\bb_0\trans\X_i)+\epsilon_i$, where
$\epsilon_i$ is generated from a centered normal distribution with
variance 0.25. Here, 
$Q_0(\bb_0\trans\x)=2\bb_0\trans\x$ and $\mu_0(\x)=1
+ \ba_0\trans\x$, where $\ba_0=(1,-1,1,1)\trans$ and
$\bb_0=(1,1,-1,1)\trans$.

To illustrate the robustness 
of our method, we considered four 
cases in estimation. In Case I,
we used the constant treatment probability model 
and a linear model for $\mu(\x,\ba)$ in the implementation. 
Note that these two models are both correctly
specified. In Case II, we used a constant model for 
$\mu(\x,\ba)$, which is a misspecified model, 
while keeping the treatment probability $\pi$ unchanged. In Case III, 
we fixed $\pi$ at $0.4$ and used the same $\mu$ model
as in Case I. 
Thus, $\mu$ is correctly specified, whereas $\pi$
is misspecified. Lastly, in Case IV,  both models are misspecified by
using the same model for $\mu$ as in Case II and 
setting $\pi$ as in Case III.

We followed the algorithm described in Section
\ref{sec:method}, where we used the Epanechnikov kernel in the
nonparametric implementation, and used the bandwidth $c\sigma n^{-1/3}$ to estimate
$\bb$, where $\sigma^2$ is the estimated variance of $\bb\trans\x$ and
$c$ is a constant between 7 and 7.5 in step 3. 

\begin{figure}[htbp]
	\includegraphics[width=0.45\textwidth]{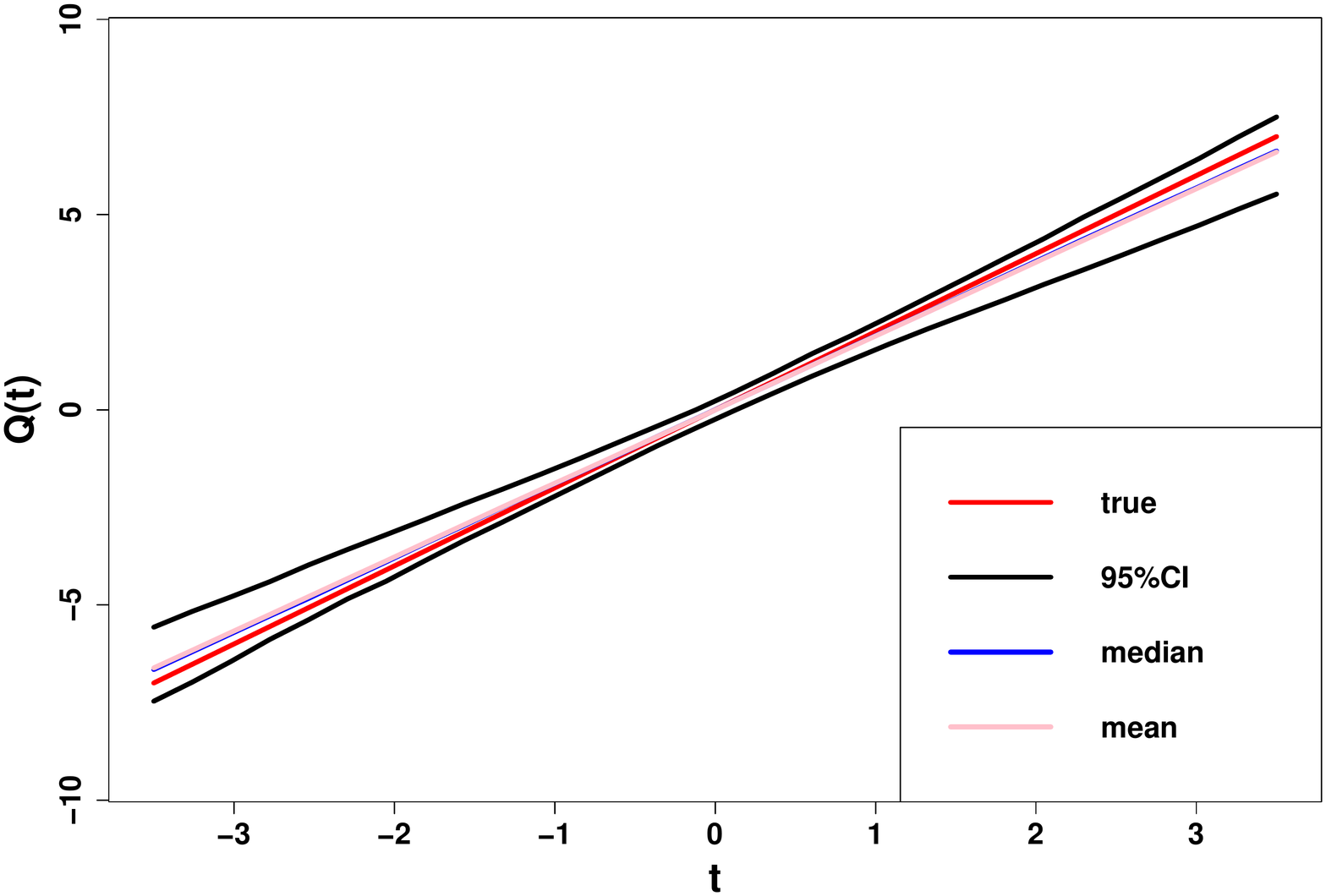}
	\hspace{0.6cm}
	\includegraphics[width=0.45\textwidth]{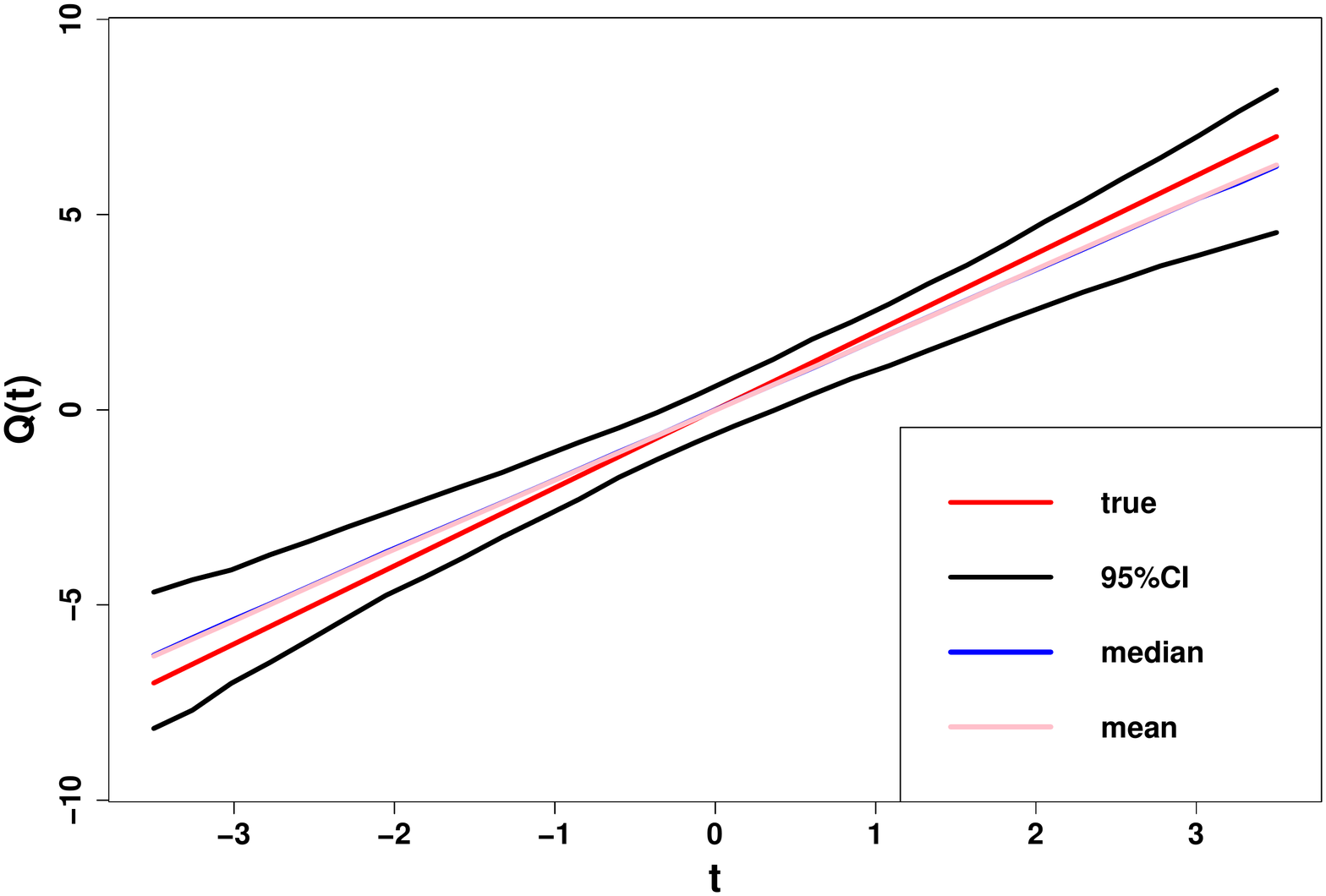}\\
	\includegraphics[width=0.45\textwidth]{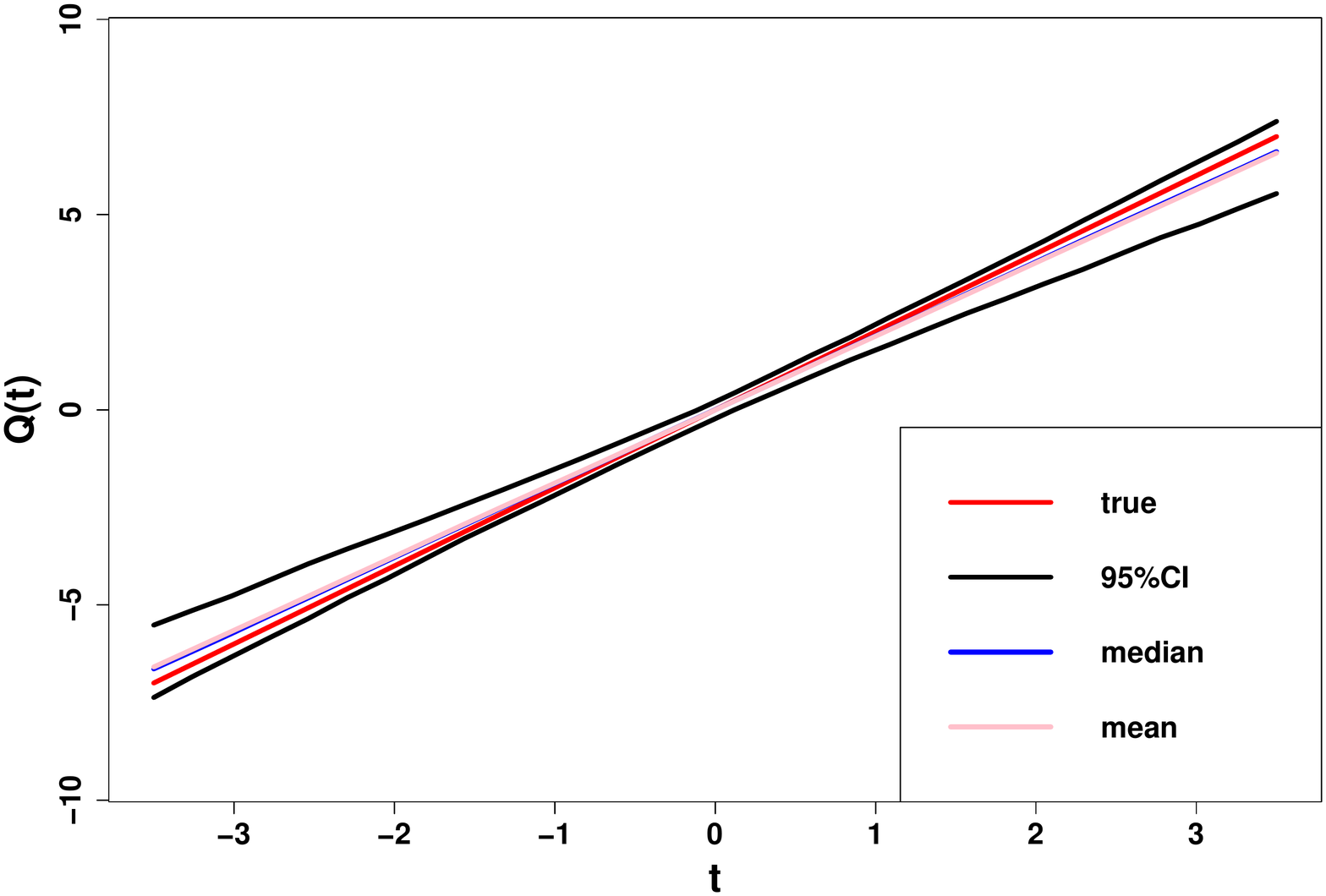}
	\hspace{0.6cm}
	\includegraphics[width=0.45\textwidth]{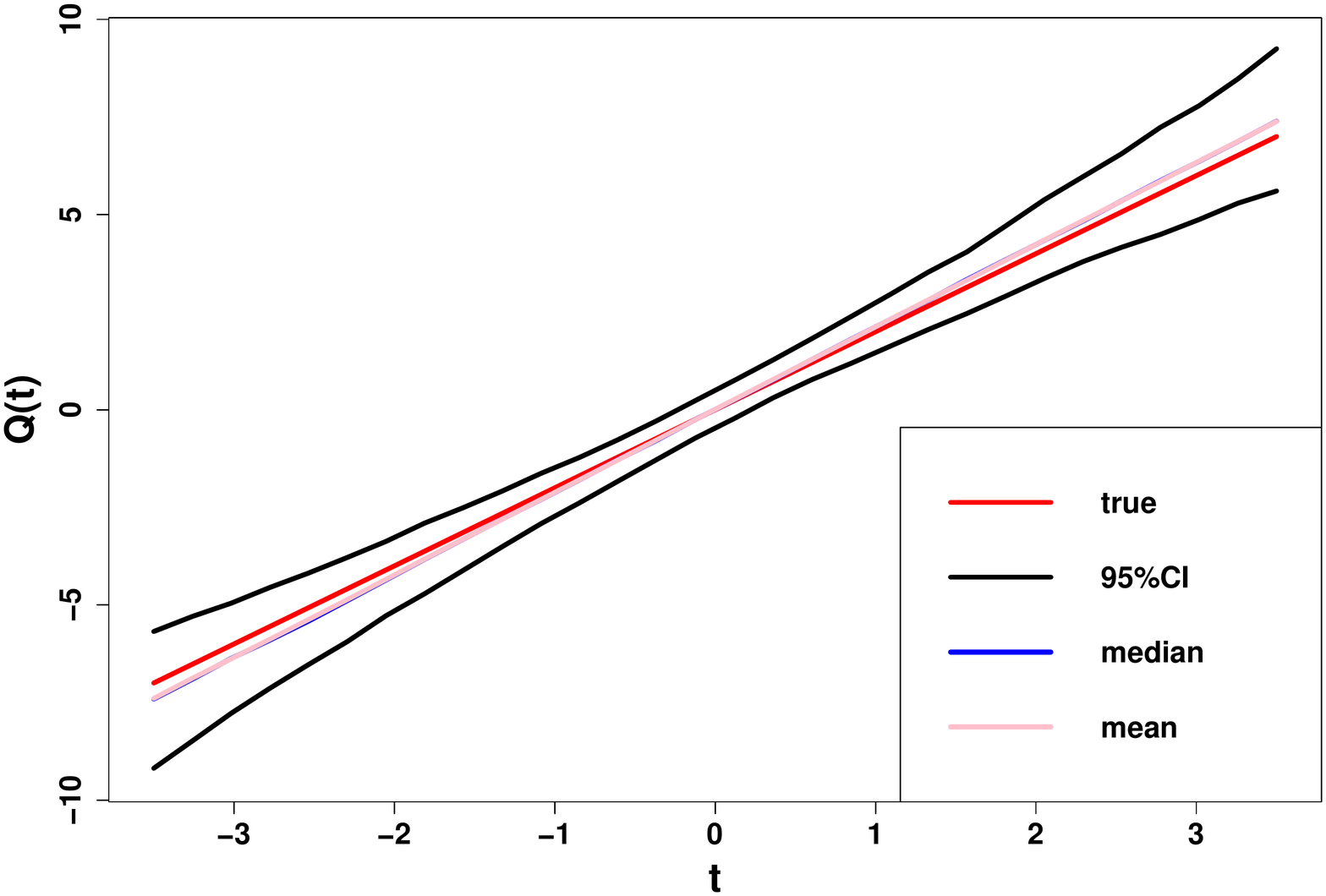}
	\caption{Simulation 1. Mean, median and  95\%
		confidence band of the estimators of $Q_0(t)=2t$, when
		(I) $\mu(\cdot)$ and $\pi(\cdot)$ are both correct
		(top-left), 
		(II) $\mu(\cdot)$ is misspecified and $\pi(\cdot)$ is correct
		(top-right),
		(III) $\pi(\cdot)$ is misspecified and $\mu(\cdot)$ is correct
		(bottom-left),
		(IV) both $\mu(\cdot)$ and $\pi(\cdot)$ are misspecified
		(bottom-right).}
	\label{fig:mono1}
\end{figure}

From the results summarized in Table \ref{table:mono1},
we  observe  
that in the first three cases, our estimation for $\bb$ yields small
bias. In contrast,
for Case IV, when both $\pi(\x)$ and $\mu(\x,\ba)$ are misspecified,
the estimation for $\bb$ is biased. In terms of inference,
the estimated standard deviations based on the asymptotic
properties match closely with the 
empirical variability of the estimators and the 95\% confidence
intervals have coverage close to the nominal level in the first three
cases.  
Interestingly, the value function estimator and the root of $Q(\cdot)$
perform well in all cases.
From Figure
\ref{fig:mono1}, we can also see that the $95\%$ confidence interval for
$Q(\cdot)$ includes the true function $Q_0(\cdot)$. 

\begin{table}[htbp]
	\centering
	\begin{tabular}{cccccccc} 
		\hline
		\multicolumn{8}{c}{ Results for $\bb$ and value function $V$}\\
		\hline
		\hline
		Case & parameters  & True  & Estimate & sd & $\hat{\rm sd}$ & cvg & MSE  \\ 
		\hline
		\multirow{4}{*}{I}& $\beta_2$      & 1     &    0.9960     &  0.0567  &      0.0563        &   95.1\%  &   0.0032   \\
		&  $\beta_3$      & -1    &    -0.9953      &  0.0563  &    0.0559          &   95.3\%  &   0.0032   \\
		&  $\beta_4$      & 1     &    0.9941      &  0.0559  &     0.0549         &  94.4\%   &   0.0032   \\ 
		& V  & 2.5958 &    2.5955      &  0.1453  &    0.1458           &  97.2\%   &   0.0211   \\
		\hline
		
		\multirow{4}{*}{II}& $\beta_2$      & 1     &    1.0256      &  0.1598  &      0.2033        &   93.4\%  &   0.0262   \\
		&  $\beta_3$      & -1    &    -1.0252      &  0.1561  &    0.1982          &   93.7\%  &   0.0250   \\
		&  $\beta_4$      & 1     &    1.0068      &  0.1361  &     0.1913         &  94.6\%   &   0.0186   \\ 
		& V   & 2.5958 &    2.6083      &  0.1926  &    0.1702           &  96.2\%   &   0.0373   \\
		\hline
		
		\multirow{4}{*}{III}& $\beta_2$      & 1     &    1.0049      &  0.0468  &      0.0470        &   95.3\%  &   0.0022   \\
		&  $\beta_3$      & -1    &    -1.0044      &  0.0470  &    0.0473          &   95.2\%  &   0.0022   \\
		&  $\beta_4$      & 1     &    1.0032      &  0.0482  &     0.0464         &  94.7\%   &   0.0023   \\ 
		& V & 2.5958 &    2.6176      &  0.1476  &    0.1471           &  96.9\%   &   0.0223   \\
		\hline
		
		\multirow{4}{*}{IV}& $\beta_2$      & 1     &    0.7494      &  0.1106  &      0.0940        &   41.3\%  &   0.0751   \\
		&  $\beta_3$      & -1    &    -0.7485      &  0.1073  &    0.0919          &   42.1\%  &   0.0748   \\
		& $\beta_4$      & 1     &    1.0243      &  0.0901  &     0.0939         &  95.8\%   &   0.0087   \\ 
		& V & 2.5958 &    2.6401      &  0.1665  &    0.1655           &  96.4\%   &   0.0297   \\
		\hline
	\end{tabular}
	\begin{tabular}{ c  c  c  c  c c  c  c c  }
		\hline
		\multicolumn{9}{c}{Results for the root of $Q_0(t)=2t$}\\
		\hline
		\hline
		Case & true & mean & bias & $\hat{\bias}$ & {\rm sd} & $\hat{\rm sd}$ & cvg & MSE \\\hline 
		I & 0 & 0.0023 & 0.0023 & -0.0013 & 0.0641 & 0.0663 & 96.6\% &
		0.0041\\
		
		II & 0 & -0.0004 & -0.0004 & -0.0001 & 0.1703 & 0.1693 & 93.7\% &
		0.0290\\
		
		III & 0 & 0.0015 & 0.0015 & -0.0004 & 0.0589 & 0.0594 & 95.7\% &
		0.0035\\
		
		IV & 0 & -0.0080 & -0.0080 & 0.0044 & 0.1242 & 0.1225 & 93.2\% &
		0.0155\\\hline 
	\end{tabular}
	\caption{Simulation 1. $Q_0(\bb_0\trans\x)=2\bb_0\trans\x$ and
		$\mu_0(\x)=1+\ba_0\trans\x$. Case I: $\mu(\cdot)$ and
		$\pi(\cdot)$ are correctly-specified, Case II:
		$\mu(\cdot)$ is misspecified, Case III: $\pi(\cdot)$
		is misspecified, Case IV: both models are
		misspecified. For the different cases, we also
		compute the mean of the estimated sd based on
		asymptotics ($\hat{\rm sd}$), empirical coverage
		obtained with 95\% confidence intervals based on
		these estimated sd (cvg), and mean squared error
		(mse).}
	\label{table:mono1}
\end{table}

\subsection{Simulation 2}\label{subsec:simu2}
In our second simulation study, we considered heteroscedastic
error. In generating data, the true treatment difference function is
$Q_0(\bb_0\trans\x_i)=(\bb_0\trans\x_i)+\sin(\bb_0\trans\x_i)$, $\mu_0(\x)= 1 + \sin(\ba_{10}\trans\x) + 0.5 (\ba_{20}\trans\x)^2$, and the
errors satisfy
$\epsilon_i \sim \N(0,\log\{(\bb_0\trans\x_i)^2+1\})$. 
Here $\bb_0=(1,1,-1,1)\trans$, $\ba_{10}=(1,-1,1,1)\trans$ and
$\ba_{20}=(1,0,-1,0)\trans$. All other aspects of the simulation
setting are identical to those in Simulation 1. 

Despite the heteroscedasticity, similar to Simulation 1, we considered
four cases to demonstrate the robustness property of our estimator. In
Case I, we used 
correctly specified models for both $\pi$ and
$\mu(\x,\ba)$. In Case II, we misspecify the
$\mu(\x,\ba)$ model to a linear one.
In Case III, we misspecified 
$\pi$ to 0.4. Finally, in Case IV, we used the misspecified  models for both
$\mu(\x,\ba)$ and $\pi$.

\begin{figure}[htbp]
	\includegraphics[width=0.45\textwidth]{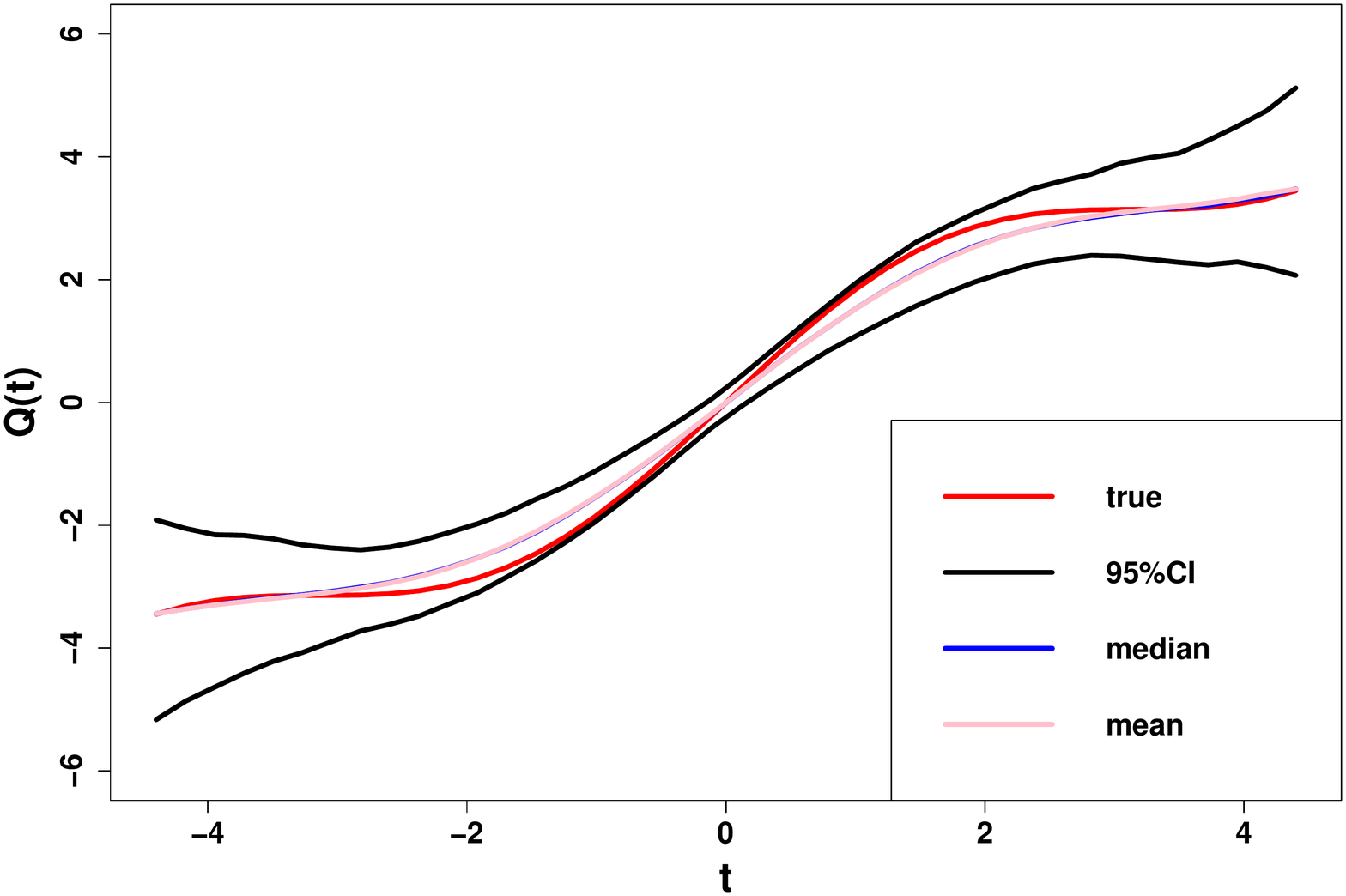}
	\hspace{0.6cm}
	\includegraphics[width=0.45\textwidth]{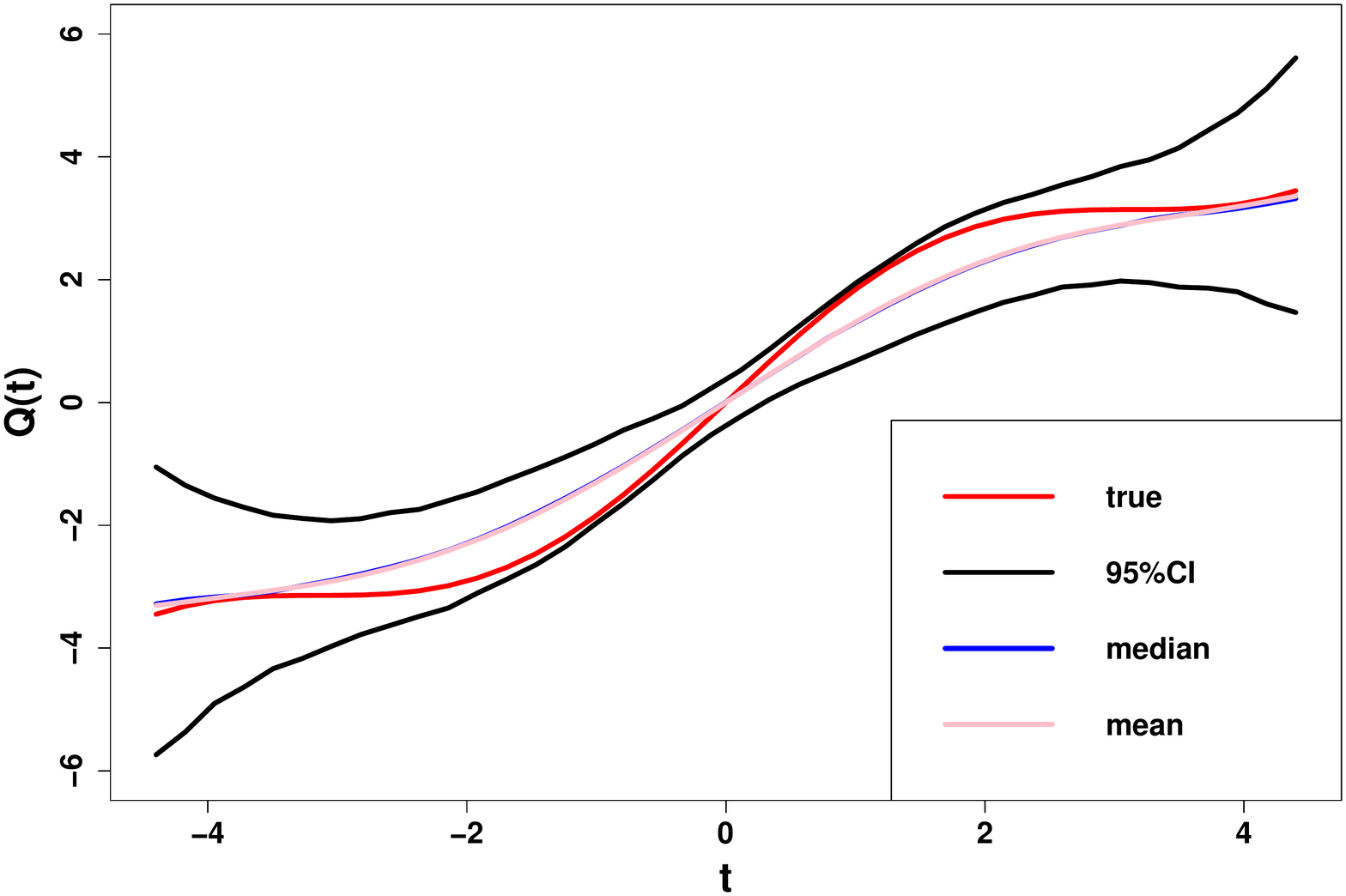}\\
	\includegraphics[width=0.45\textwidth]{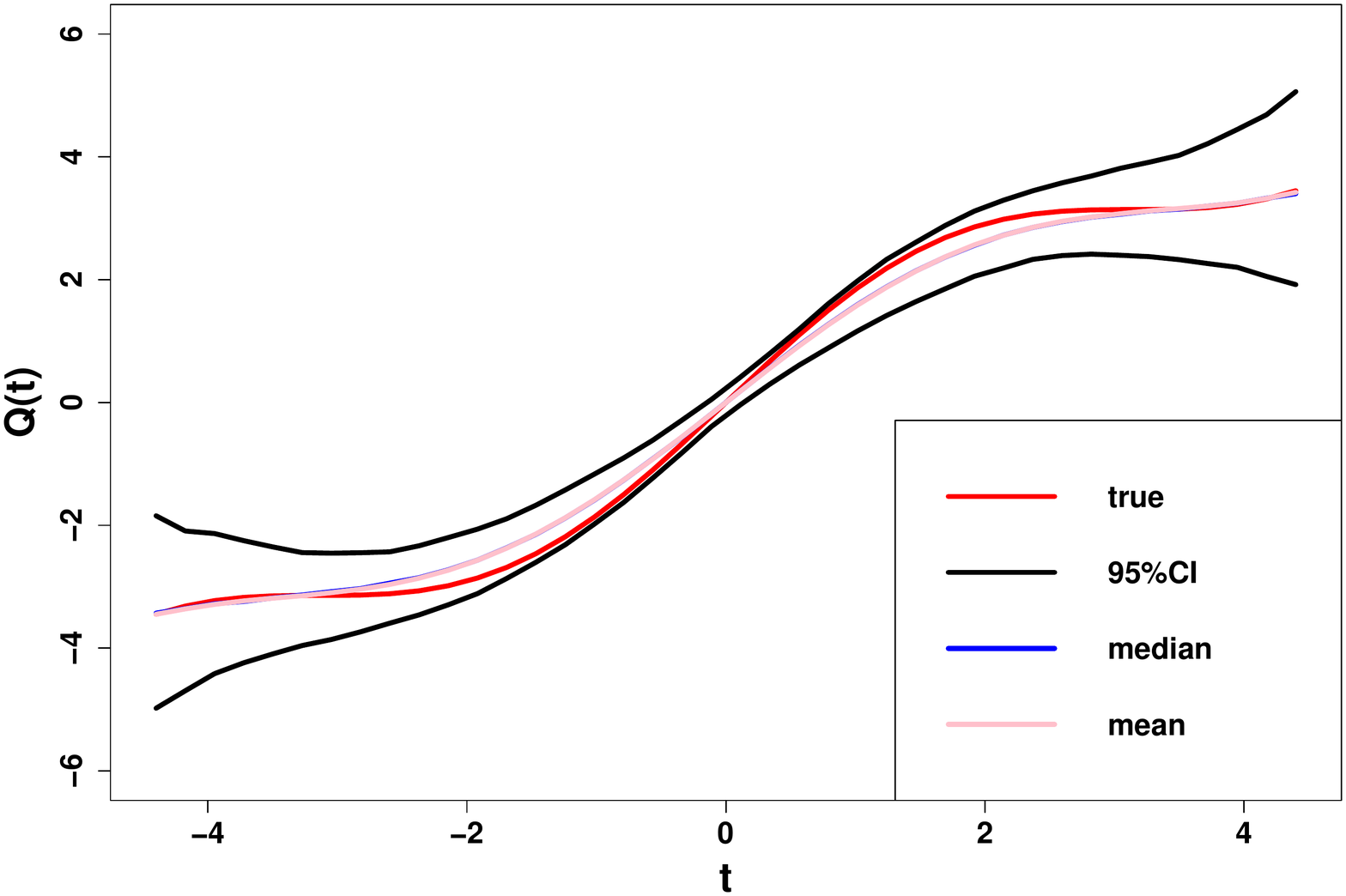}
	\hspace{0.6cm}
	\includegraphics[width=0.45\textwidth]{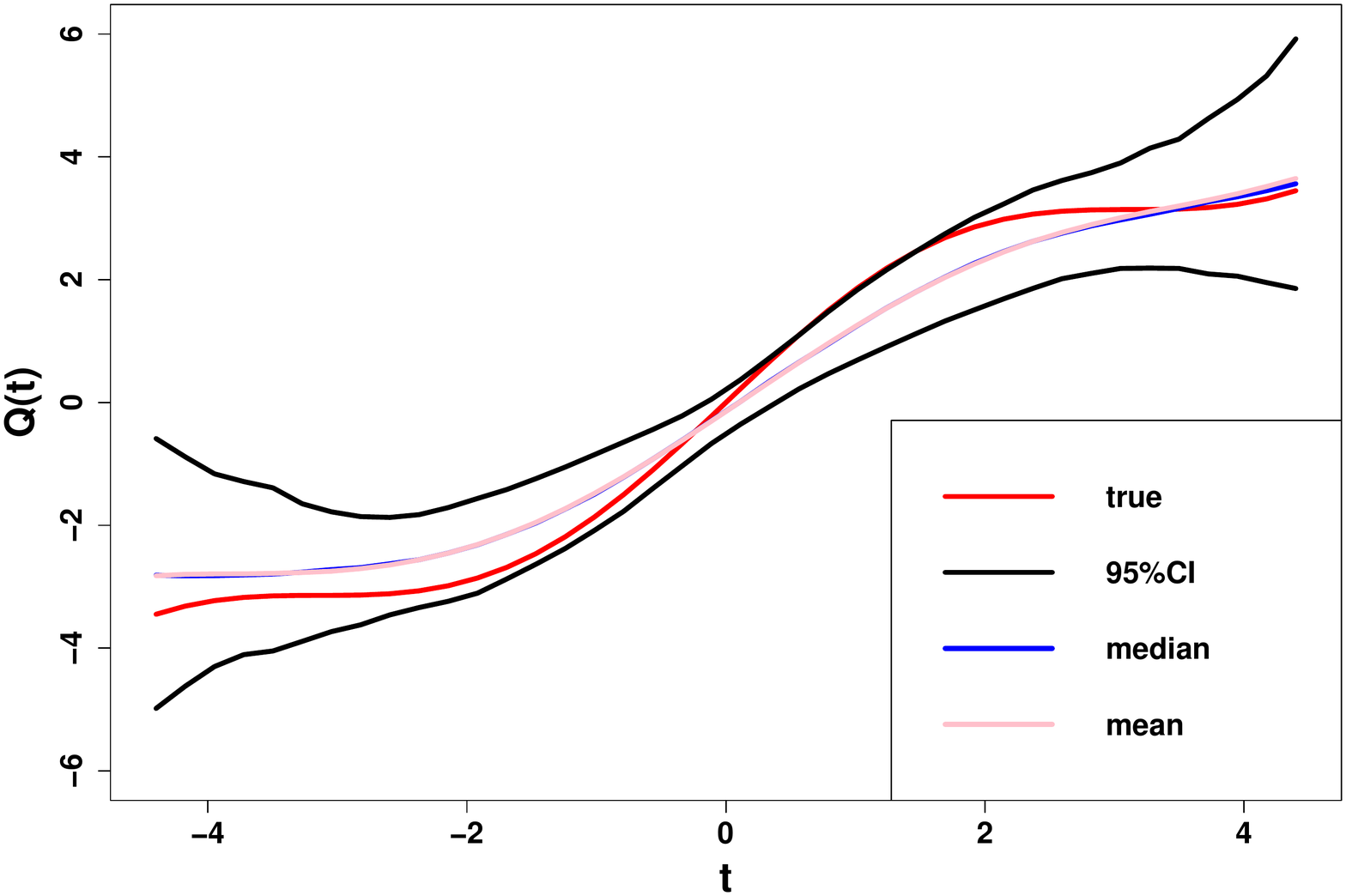}
	\caption{Simulation 2. Mean, median and  95\%
		confidence band of the estimators of $Q_0(t)=t+\sin(t)$, when
		(I) $\mu(\cdot)$ and $\pi(\cdot)$ are both correct
		(topleft), 
		(II) $\mu(\cdot)$ is misspecified and $\pi(\cdot)$ is correct
		(topright),
		(III) $\pi(\cdot)$ is misspecified and $\mu(\cdot)$ is correct
		(bottomleft),
		(IV) both $\mu(\cdot)$ and $\pi(\cdot)$ are misspecified
		(bottomright).}
	\label{fig:hetero}
\end{figure}

We used the same nonparametric estimation procedures as in Simulation 1 to
implement the algorithm in Section \ref{sec:method}. From the results
summarized in Table \ref{table:hetero}, we observe that despite of the
heteroscedastic error, in the first three cases, the estimation for
parameters $\bb$, the value function and $z$ yields very small bias.   
In addition, the estimated standard deviations are still  close to the
empirical version of the estimators, and the confidence intervals
have coverage close to the 
nominal levels. Surprisingly, the value function estimator performs
well even in Case IV. As a result, in Figure \ref{fig:hetero}, we can
see that the $95\%$ confidence interval  
for $Q(\cdot)$ includes the true function $Q_0(\cdot)$ in all four
cases. 

\begin{table}[htbp]
	\centering
	\begin{tabular}{cccccccc} 
		\hline
		\multicolumn{8}{c}{ Results for $\bb$ and value function $V$.}\\
		\hline
		\hline
		Case & parameters   & True  & Estimate & sd & $\hat{\rm sd}$ & cvg & MSE  \\ 
		\hline
		\multirow{4}{*}{I} & $\beta_2$      & 1     &    1.0128      &  0.1220  &      0.1553        &   93.1\%  &   0.0151   \\
		&$\beta_3$      & -1    &    -1.0076      &  0.1151  &    0.1533          &   93.0\%  &   0.0133   \\
		&$\beta_4$      & 1     &    1.0066      &  0.1187  &     0.1563       &  94.4\%   &   0.0141   \\ 
		&V   &3.0533 &    3.0506      & 0.1156  &  0.1212    &  96.7\%  &   0.0134  \\\hline
		
		\multirow{4}{*}{II}& $\beta_2$ & 1      & 1.0133   & 0.1454 & 0.1677 & 93.3\% & 0.0213 \\ 
		&    $\beta_3$ & -1     & -1.0085  & 0.1455 & 0.1652 & 93.8\% & 0.0212 \\ 
		&     $\beta_4$ & 1      & 1.0129   & 0.1489 & 0.1676 & 93.4\% & 0.0223 \\ 
		& V   & 3.0533 & 3.0327  & 0.1358 & 0.1408  & 95.6\%   & 0.0189 \\ \hline
		
		\multirow{4}{*}{III} &  $\beta_2$ & 1      & 1.0116   & 0.1070 & 0.1549 & 94.1\% & 0.0116 \\ 
		&   $\beta_3$ & -1     & -1.0129  & 0.1106 & 0.1489 & 92.9\% & 0.0124 \\ 
		&   $\beta_4$ & 1      & 1.0102   & 0.1111 & 0.1564 & 93.0\% &0.0125 \\ 
		& V & 3.0533 & 3.0696 & 0.1189 & 0.1220  & 96.6\%    & 0.0144 \\ \hline
		
		\multirow{4}{*}{IV} & $\beta_2$ & 1      & 1.0337   & 0.1532 & 0.1593 & 82.3\% & 0.0246 \\ 
		&       $\beta_3$ & -1     & -1.0315  & 0.1557 & 0.1773 & 84.0\% & 0.0252 \\ 
		&        $\beta_4$ & 1      & 1.0486  & 0.1612 & 0.1679 & 81.9\% & 0.0284 \\ 
		& V& 3.0533 & 3.0545 & 0.1343 & 0.1343  & 95.8\%    & 0.0180 \\ \hline
	\end{tabular}
	\begin{tabular}{ c  c  c  c  c  c  c  c  c }
		\hline
		\multicolumn{9}{c}{Results for the root of $Q_0(t)=t+\sin(t)$}\\
		\hline
		\hline
		Case &true & mean & bias & $\hat{\bias}$ & {\rm sd} & $\hat{\rm sd}$ & cvg & MSE \\\hline 
		I & 0 &  0.0015 & 0.0015 & 0.0015 & 0.0762 & 0.0767 & 95.6\% & 0.0058\\
		
		II & 0 & -0.0023 & -0.0023 & 0.0042 & 0.1498 & 0.1550 & 96.9\% & 0.0224 \\
		
		III & 0 & -0.0003 & -0.0003 & -0.0024 & 0.0680 & 0.0695 & 94.3\% & 0.0046 \\ 
		
		IV & 0 & 0.1027 & 0.1027 & -0.0229 & 0.1422 & 0.1314 & 82.9\% &  0.0308 \\ \hline
	\end{tabular}
	\caption{Simulation 2. $Q_0(\bb_0\trans\x)=(\bb_0\trans\x)+\sin(\bb_0\trans\x)$ and
		$\mu_0(\x)=1+\sin(\ba_{10}\trans\x)+0.5(\ba_{20}\trans\x)^2$. See also the caption of Table \ref{table:mono1}.}
	\label{table:hetero}
\end{table}

\subsection{Simulation 3}\label{subsec:simu3}

In Simulation 3, we examine the performance of  our estimators in the
presence of non-monotonic $Q_0(\cdot)$ function and heteroscedastic
error variance. We consider the same simulation setting as in
Simulation 2, except that we consider a non-monotonic function, 
$Q_0(\bb_0\trans\x_i)=(\bb_0\trans\x_i)^2 - 2$.  

\begin{figure}[htbp]
	\includegraphics[width=0.45\textwidth]{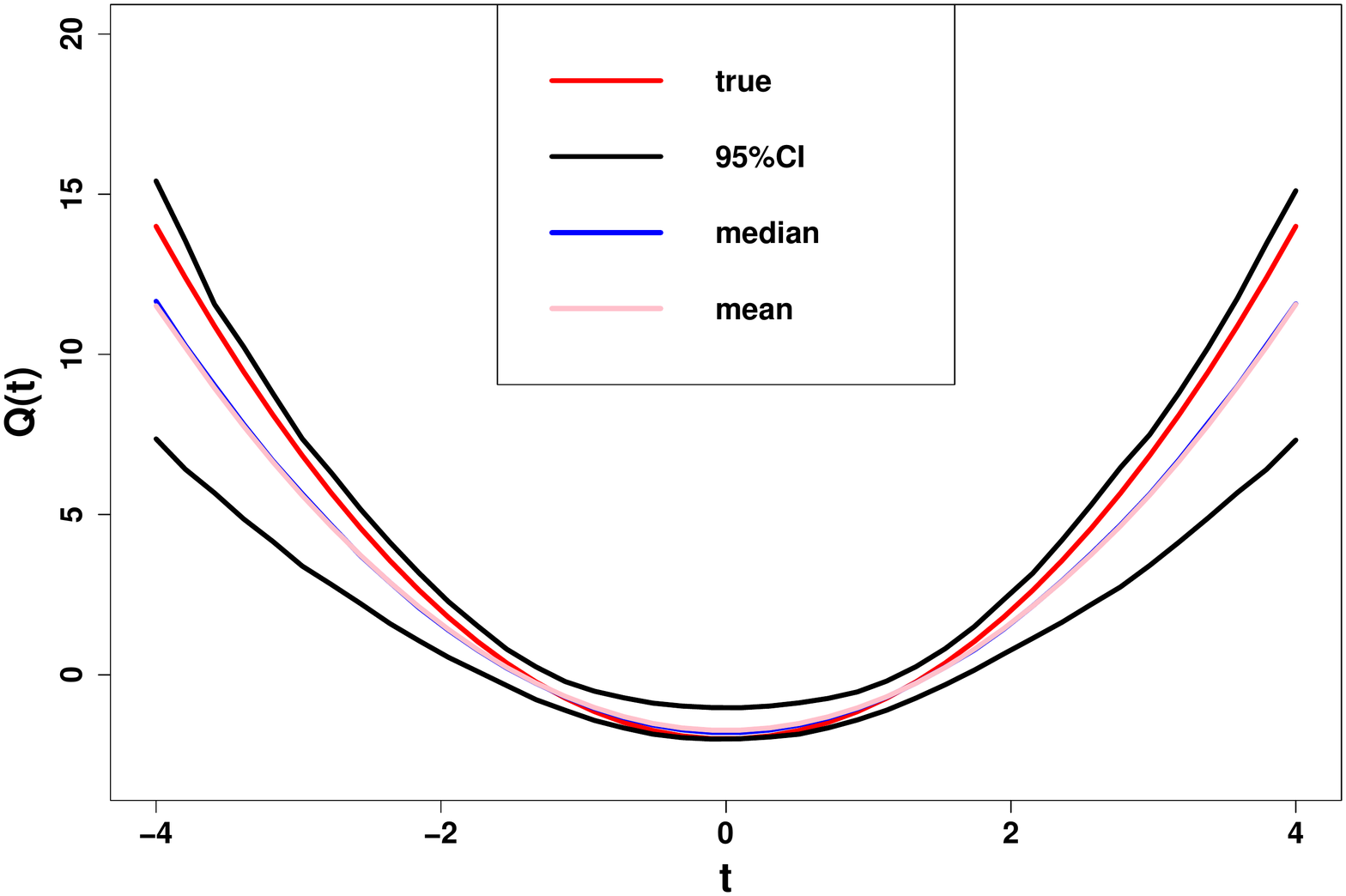}
	\hspace{0.6cm}
	\includegraphics[width=0.45\textwidth]{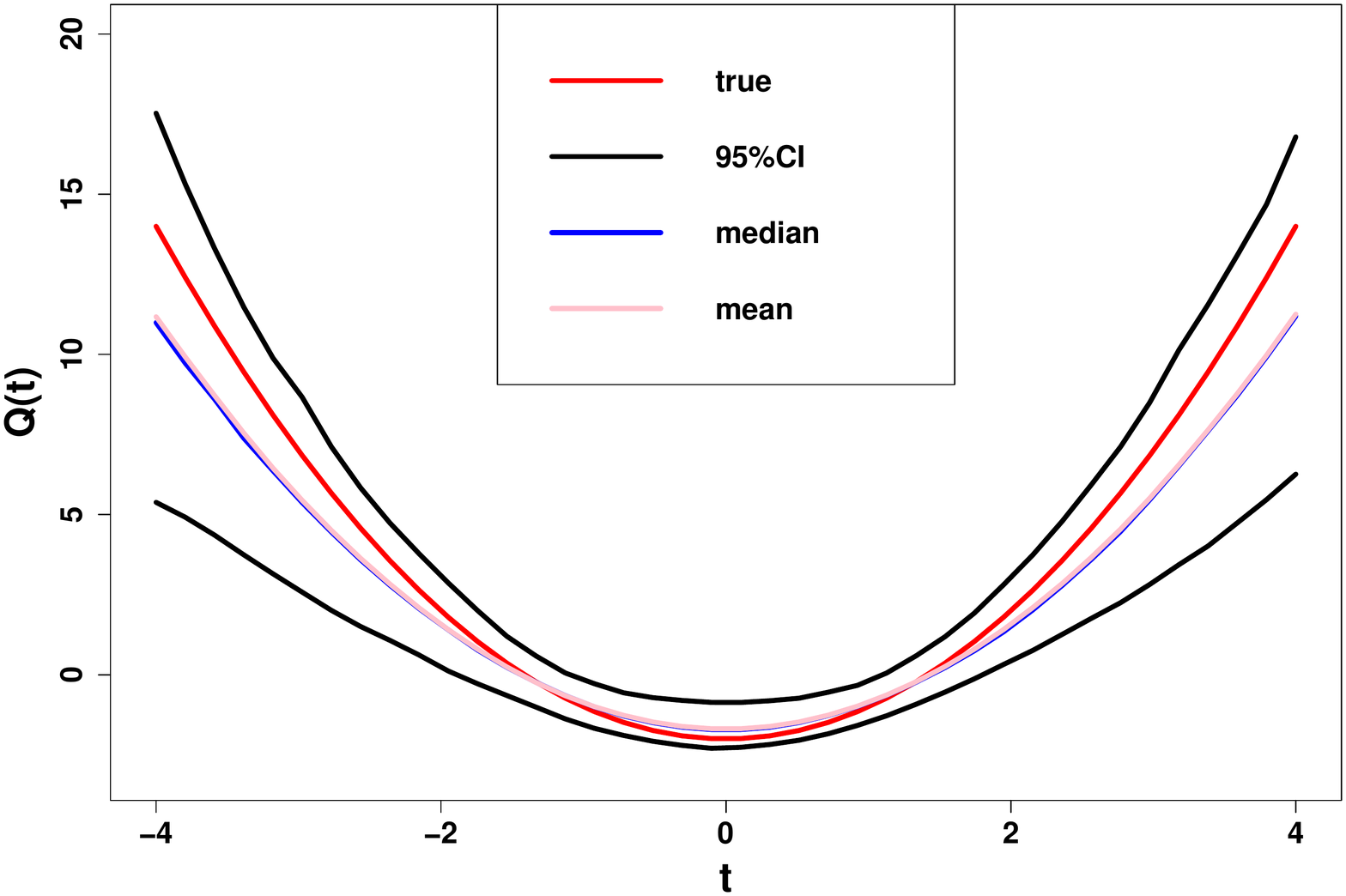}\\
	\includegraphics[width=0.45\textwidth]{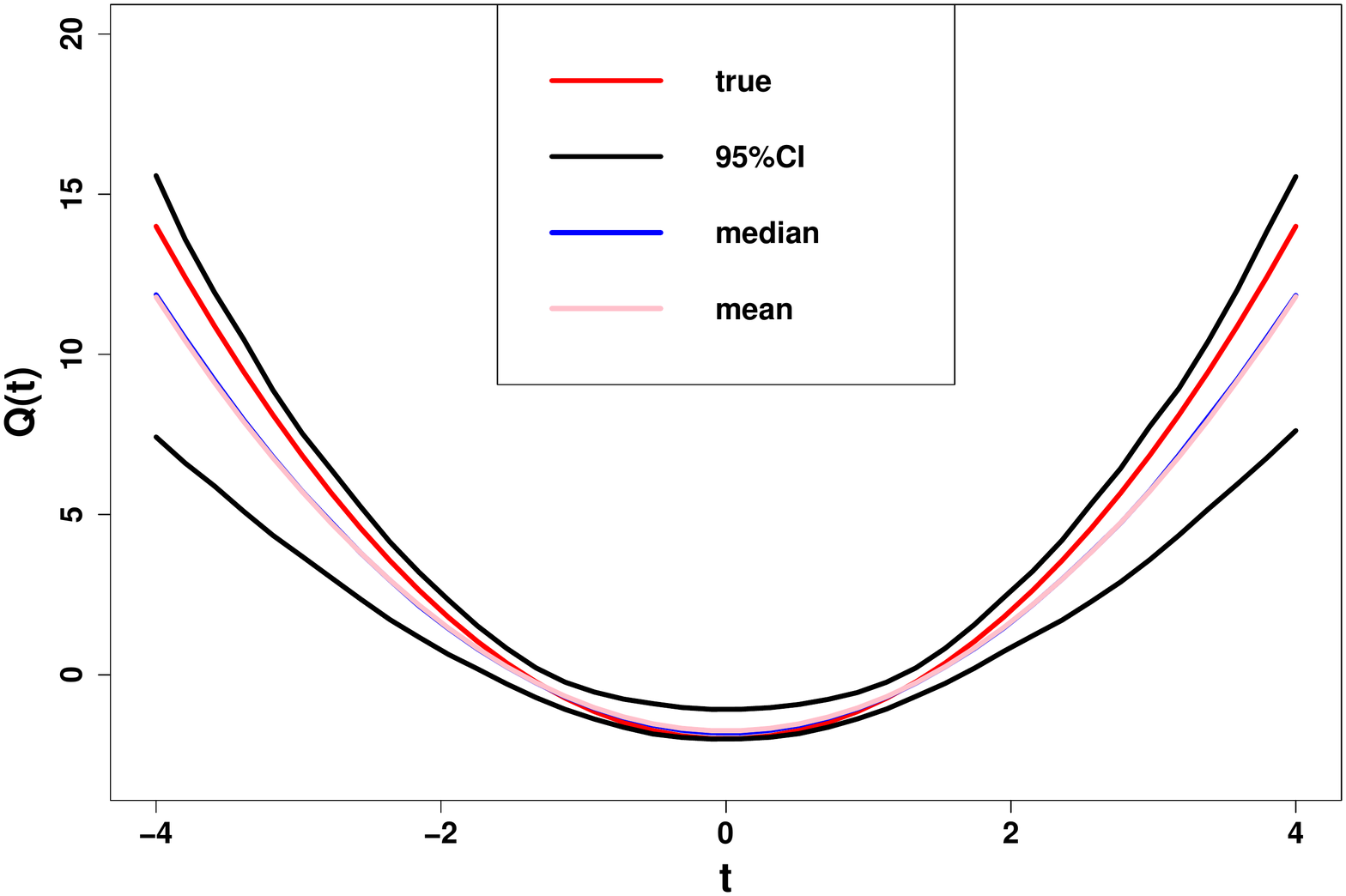}
	\hspace{0.6cm}
	\includegraphics[width=0.45\textwidth]{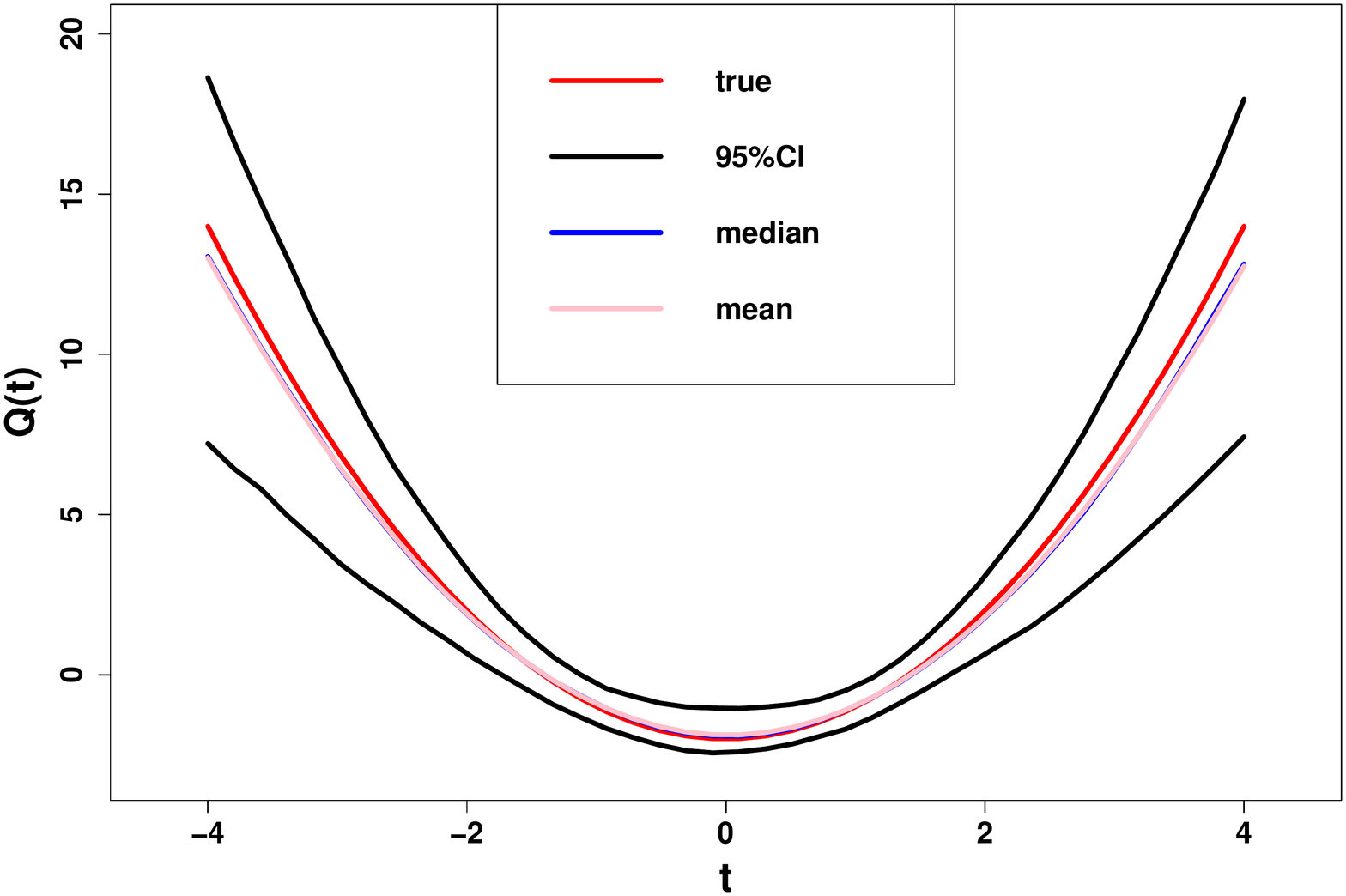}
	\caption{Simulation 3. Mean, median and  95\%
		confidence band of the estimators of $Q_0(t)=t^2-2$ with heteroscedastic error variance, when
		(I) $\mu(\cdot)$ and $\pi(\cdot)$ are both correct
		(topleft), 
		(II) $\mu(\cdot)$ is misspecified and $\pi(\cdot)$ is correct
		(topright),
		(III) $\pi(\cdot)$ is misspecified and $\mu(\cdot)$ is correct
		(bottomleft),
		(IV) both $\mu(\cdot)$ and $\pi(\cdot)$ are misspecified
		(bottomright).}
	\label{fig:norank}
\end{figure}

Similar to Simulation 2, we considered four cases and implemented our
estimation procedures to illustrate the robustness property of
estimators with the non-monotonic treatment difference function.
From the results summarized  in Table \ref{table:norank}, we 
observe that despite of both a non-monotonic $Q_0(\cdot)$
function and a heteroscedastic error variance, 
the estimation for parameters $\bb$, the value function and the two
roots of $Q_0(\cdot)$ yields very small bias in the first three cases. 
Also, the estimated
standard deviations are close to the empirical standard deviations and
the confidence intervals are close to nominal coverage levels. 
As expected, the estimation of $\bb$ in Case IV does not perform well, although
the performance of the value function and the two
roots of $Q(\cdot)$ show certain robustness even in Case IV. In Figure
\ref{fig:norank}, we note that the $95\%$ confidence interval
for $Q(\cdot)$ includes the true $Q_0(\cdot)$ function in all four
cases. 

\begin{table}[htbp]
	\centering
	\begin{tabular}{cccccccc} 
		\hline
		\multicolumn{8}{c}{ Results for $\bb$ and value function $V$.}\\
		\hline
		\hline
		Case &parameters & True  & Estimate & sd & $\hat{\rm sd}$ & cvg & MSE  \\ 
		\hline
		\multirow{4}{*}{I} &$\beta_2$      & 1     &    1.0364      &  0.0646  &      0.0911        &   93.5\%  &   0.0055   \\
		&  $\beta_3$      & -1    &    -1.0368      &  0.0653  &    0.0911         &   93.1\%  &   0.0056   \\
		& $\beta_4$      & 1     &    1.0330      &  0.0646  &     0.0893       &  92.9\%   &   0.0053   \\ 
		&V &4.7166 &    4.6415      & 0.2843  &  0.2970    &  94.9\%  &   0.0864  \\\hline
		
		\multirow{4}{*}{II} & $\beta_2$ & 1      & 1.0398   & 0.1124 & 0.1384 & 92.8\% & 0.0142 \\ 
		&                 $\beta_3$ & -1     & -1.0341  & 0.1062 & 0.1320 & 94.0\% & 0.0124 \\ 
		&                 $\beta_4$ & 1      & 1.0411   & 0.1166 & 0.1385 & 93.2\% & 0.0153 \\ 
		& V           & 4.7166 & 4.6160  & 0.3095 & 0.3053  & 94.4\%   & 0.1059 \\ \hline
		
		\multirow{4}{*}{III} & $\beta_2$ & 1      & 1.0295   & 0.0618 & 0.0833 & 92.6\% & 0.0047 \\ 
		&         $\beta_3$ & -1     & -1.0300  & 0.0640 & 0.0830 & 93.6\% & 0.0050 \\ 
		&         $\beta_4$ & 1      & 1.0262   & 0.0629 & 0.0826 & 94.2\% &0.0046 \\ 
		&V & 4.7166 & 4.7024 & 0.2967 & 0.3052  & 95.9\%    & 0.0882 \\ \hline
		
		\multirow{4}{*}{IV} & $\beta_2$ & 1      & 0.9549   & 0.0894 & 0.0990 & 86.4\% & 0.0100 \\ 
		&      $\beta_3$ & -1     & -1.0283  & 0.0865 & 0.1020 & 92.0\% & 0.0083 \\ 
		&      $\beta_4$ & 1      & 0.9563  & 0.0910 & 0.1007 & 89.4\% & 0.0102 \\ 
		& V      & 4.7166 & 4.7518 & 0.3122 & 0.3174  & 96.1\%    & 0.0987 \\ \hline
	\end{tabular}
	\begin{tabular}{c  c c  c  c  c  c  c  c }
		\hline
		\multicolumn{9}{c}{Results for the two roots of $Q_0(t)=t^2-2$}\\
		\hline
		\hline
		Case &true & mean & bias & $\hat{\bias}$ & {\rm sd} & $\hat{\rm sd}$ & cvg & MSE \\\hline 
		\multirow{2}{*}{I} & -1.4142 &  -1.4469 & -0.0327 & -0.0109 & 0.1498 & 0.1120 & 93.2\% & 0.0235\\
		& 1.4142 &  1.4480 & 0.0338 & 0.0047 & 0.1259 & 0.1108 & 93.7\% & 0.0170\\\hline 
		
		\multirow{2}{*}{II} & -1.4142 &  -1.4482 & -0.0340 & -0.0144 & 0.2096 & 0.1943 & 94.7\% & 0.0451\\
		& 1.4142 &  1.4408 & 0.0266 & 0.1191 & 0.1854 & 0.1830 & 95.2\% & 0.0351\\\hline 
		
		\multirow{2}{*}{III} & -1.4142 &  -1.4423 & -0.0281 & -0.0170 & 0.1104 & 0.1033 & 93.1\% & 0.0130\\
		& 1.4142 &  1.4436 & 0.0294 & 0.0113 & 0.1112 & 0.1019 & 93.3\% & 0.0132\\\hline 
		
		\multirow{2}{*}{IV} & -1.4142 &  -1.4087 & 0.0056 & -0.0180 & 0.1585 & 0.1745 & 97.4\% & 0.0252\\
		& 1.4142 &  1.4363 & 0.0221 & -0.0145 & 0.1497 & 0.1662 & 97.5\% & 0.0229\\\hline 
	\end{tabular}
	\caption{Simulation 3. $Q_0(\bb_0\trans\x)=(\bb_0\trans\x)^2-2$ and
		$\mu_0(\x)=1+\sin(\ba_{10}\trans\x)+0.5(\ba_{20}\trans\x)^2$, when error variance is heteroscedastic. See also the caption of Table \ref{table:mono1}.}
	\label{table:norank}
\end{table}

\subsection{Simulation 4}\label{subsec:simu4}

In the previous simulation settings, we only considered the situation
where the true propensity score is a constant.
We now consider non-constant propensity score case, which better
reflects the situation in observational studies. Specifically, we let 
$\pi(\X_i)=\exp(\bg_0\trans\X_i)/\{1+\exp(\bg_0\trans\X_i)\}$, where
$\bg_0=(0.1,0,-0.1,0)\trans$. All other data generation procedure is
indentical as in Simulation 1. 

Similar to Simulation 1, we considered four cases and implemented our
estimation procedures to illustrate the robustness property of
estimators. In case I, we use an expit function to estimate the
treatment probability model and a linear model for $\mu(\X,\ba)$ in
the implementation. In Case II, we use a constant model for
$\mu(\X,\ba)$, while keeping the treatment probability model
$\pi(\X,\bg)$ unchanged. Note that here only $\mu(\X,\ba)$ is
misspecified. In Case III, we consider a constant model for
$\pi(\X,\bg)$, which is a misspecified model and used the same linear
model for $\mu(\X,\ba)$ as in Case I. Lastly, in Case IV, both models
are misspecified by using the same model for $\mu(\X,\ba)$ as in Case
II and considering $\pi(\X,\bg)$ as in Case III.  
 
\begin{figure}[htbp]
	\begin{center}
		\includegraphics[width=0.36\textwidth]{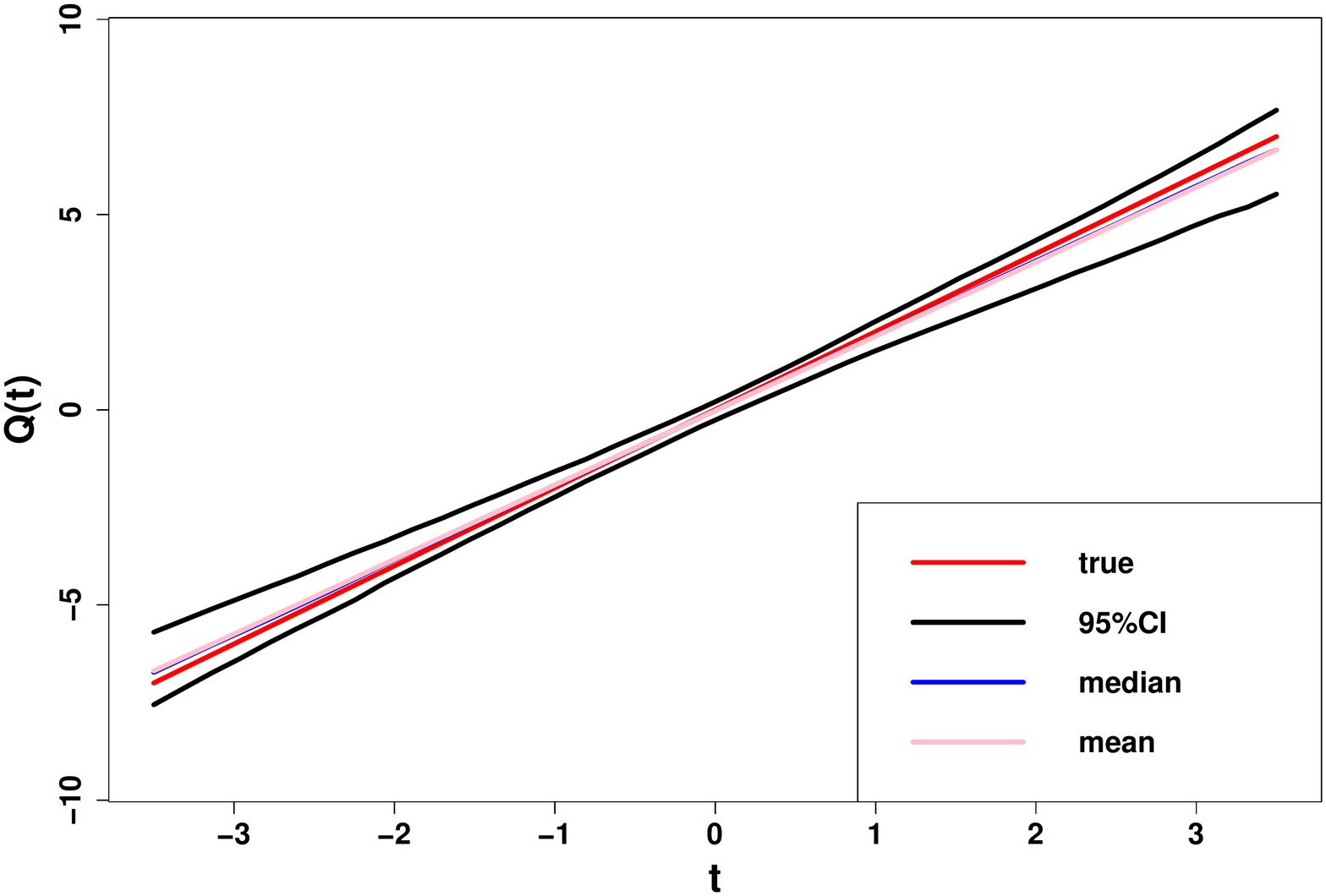}
		\hspace{0.2cm}
		\includegraphics[width=0.36\textwidth]{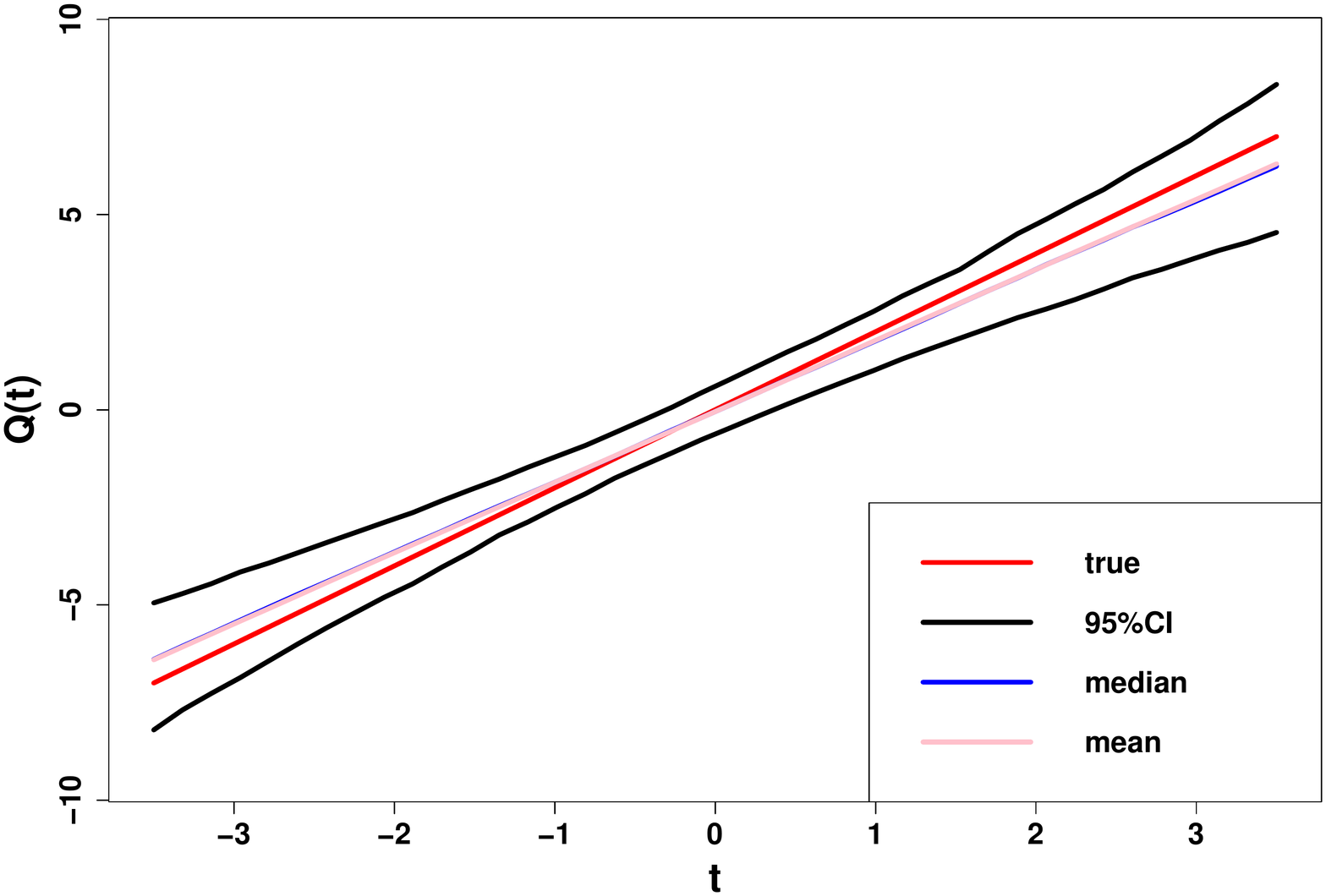}\\
		\includegraphics[width=0.36\textwidth]{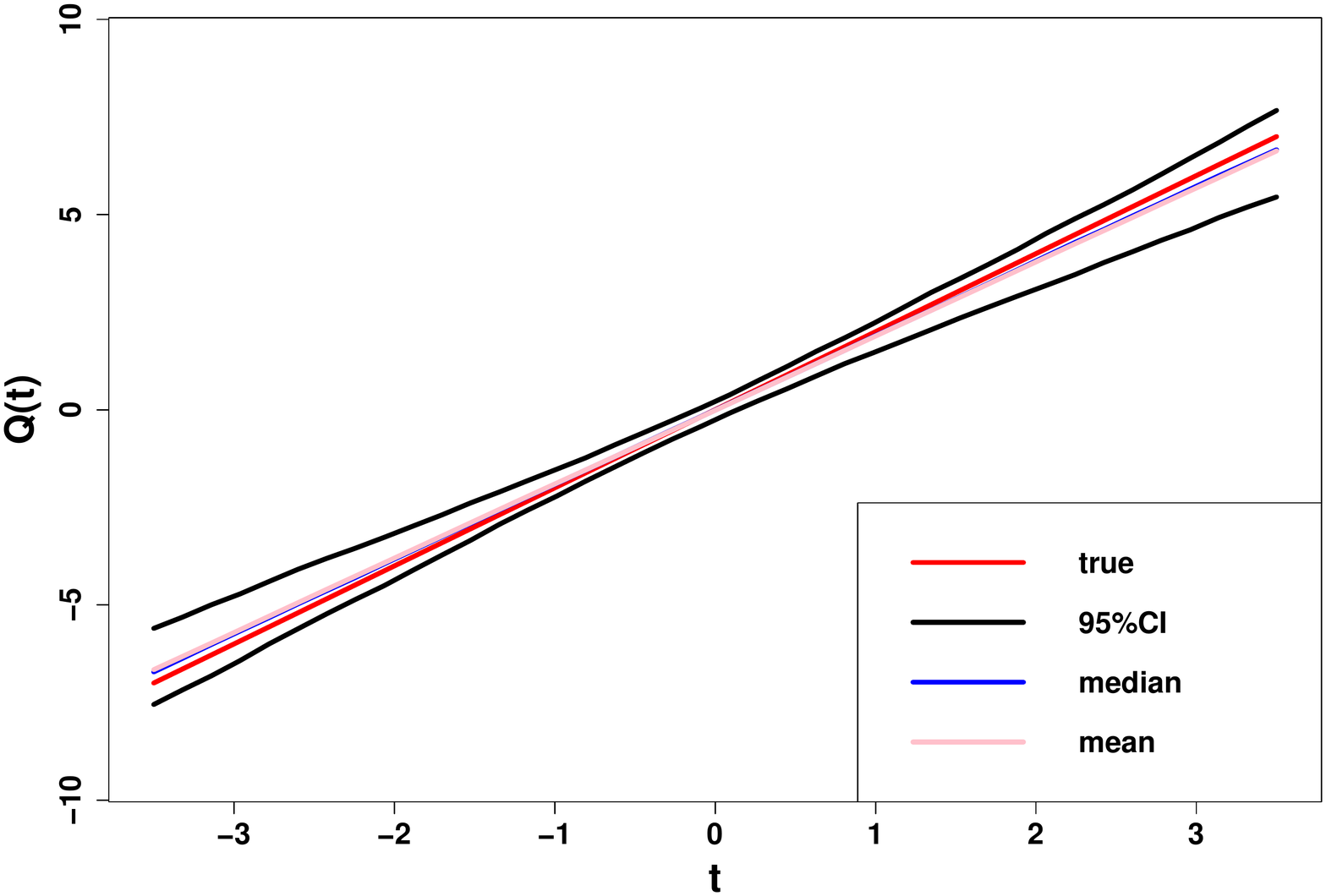}
		\hspace{0.2cm}
		\includegraphics[width=0.36\textwidth]{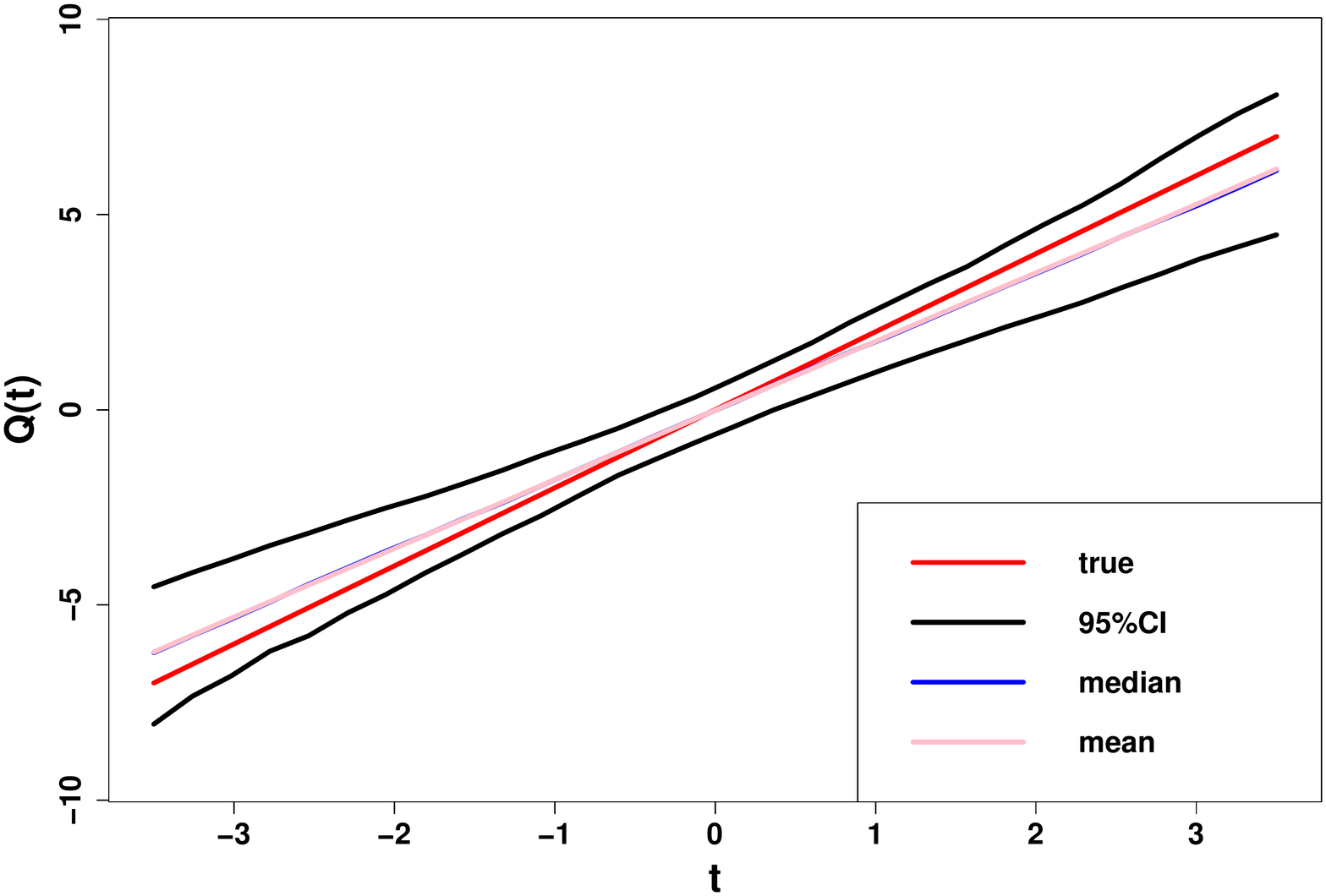}
		\caption{Simulation 4. Mean, median and  95\%
			confidence band of the estimators of $Q_0(t)=2t$ with non-constant propensity score model. See also the caption of Figure \ref{fig:mono1}.}
		\label{fig:pi}
	\end{center}
\end{figure}

We followed the algorithm described in Section \ref{sec:method} and
summarized the results in Table \ref{table:pi}. From the Table
\ref{table:pi}, we observe that the estimations for parameters $\bb$,
the value function $V$ and the root of the treatment difference
function yield small bias in the first three cases, as expected. In
terms of inference, the estimated standard deviations match closely
with empirical standard deviations and the confidence intervals have
coverage close to the nominal level. Interestingly, the estimation of
the value function and root of $Q(\cdot)$ perform well in all
cases. From Figure \ref{fig:pi}, we can also observe that the $95\%$
confidence interval for $Q(\cdot)$ contains the true function
$Q_0(\cdot)$ in all four cases.  

\begin{table}[htbp]
	\centering
	\begin{tabular}{ccccccccc} 
		\hline
		\multicolumn{8}{c}{ Results for $\bb$ and value function $V$.}\\
		\hline
		\hline
		Case &parameters  & True  & Estimate & sd & $\hat{\rm sd}$ & cvg & MSE  \\ 
		\hline
		\multirow{4}{*}{I} & $\beta_2$      & ~1     &    0.9840      &  0.0688  &      0.0795      &   95.3\%  &   0.0050   \\
		&  $\beta_3$      & -1    &    -0.9828    & 0.0696 &    0.0814    &   94.5\%  &   0.0051   \\
		&  $\beta_4$      & ~1     &    0.9891     &  0.0701  &   0.0808    &  94.9\%   &   0.0021   \\ 
		&V                 &2.5958 &    2.5733   &  0.9326  &  0.9961    &  95.4\%  &   0.0219  \\ \hline
		
		\multirow{4}{*}{II}&$\beta_2$      & 1     &    1.0150     &  0.1376  &      0.1751        &   92.9\%  &   0.0191   \\
		&$\beta_3$      & -1    &   -1.0024      & 0.1270  &    0.1697          &   92.1\%  &   0.0161   \\
		&$\beta_4$      & 1     &    0.9915      & 0.1230  &     0.1666         &  92.8\%   &   0.0152   \\ 
		&V& 2.5958 &    2.5908     &  0.1824  &    0.1656           &  95.0\%   &   0.0333  \\
		\hline
		
		\multirow{4}{*}{III} &  $\beta_2$ & ~1      & 0.9882  & 0.0732 &0.0890 & 95.5\% & 0.0055 \\ 
		& $\beta_3$ & -1     & -0.9869  & 0.0732 & 0.0873 & 96.2\% & 0.0055 \\ 
		&  $\beta_4$ & ~1      & 0.9879  &0.0752 & 0.0892 & 95.1\% & 0.0058 \\ 
		& V          & 2.5958 & 2.5866 & 0.1383& 0.1461  & 97.3\%    & 0.0192 \\ \hline
		
		\multirow{4}{*}{IV} & $\beta_2$ & ~1      & 1.0508   & 0.1698 & 0.1965 & 95.4\% & 0.0314 \\ 
		&  $\beta_3$ & -1     & -1.0565 & 0.1727 & 0.1982 & 95.1\% &0.0330 \\ 
		&  $\beta_4$ & ~1      & 1.0231  & 0.1445 & 0.1674 & 95.8\% & 0.0214 \\ 
		&V             & 2.5958 & 2.5940 & 0.1719 & 0.1688  & 95.4\%    & 0.0295 \\ \hline
		
	\end{tabular}
	\vspace{0.6cm}
	\begin{tabular}{ c  c  c  c  c  c  c  c  c }
		\hline
		\multicolumn{9}{c}{Results for the root of $Q_0(t)=2t$}\\
		\hline
		\hline
		Case &true & mean & bias & $\hat{\bias}$ & {\rm sd} & $\hat{\rm sd}$ & cvg & MSE \\\hline
		I & 0 & 0.0132 & 0.0132 & -0.0022 & 0.0646 & 0.0673 & 94.5\% & 0.0044\\ 
		
		II & 0 & 0.0257 & 0.0257 & -0.0036 &0.1626 & 0.1517 & 92.1\% &
		0.0271\\
		
		III & 0 & 0.0076 & 0.0076 & -0.0018 & 0.0642 & 0.0670 & 93.3\% & 0.0042\\ 
		
		IV & 0 & 0.0097 & 0.0097 &  -0.0139 & 0.1758 & 0.1748 & 95.0\% & 0.0310 \\ \hline
	\end{tabular}
	\caption{Simulation 4. $Q_0(\bb_0\trans\x)=2\bb_0\trans\x$, $\pi_0(\x)=\exp(\bg_0\trans\X)/\{1+\exp(\bg_0\trans\X)\}$ and
	$\mu_0(\x)=1+\ba_{0}\trans\x$. See also the caption of Table \ref{table:mono1}.}
	\label{table:pi}
\end{table}

\subsection{Simulation 5}\label{subsec:simu5}

In this simulation, we keep data generation identical to Simulation 2, except we consider non-constant propensity score model similar to Simulation 4. 

To show the robustness of our method, we consider four cases similar to Simulation 4. In Case I, we used correctly specified models for both $\pi(\x,\bg)$ and $\mu(\x,\ba)$. In Case II, we misspecify the $\mu(\x,\ba)$ model to a linear one. In Case III, we use a constant model for $\pi(\x,\bg)$. Finally, in Case IV, we used the misspecified models for both $\mu(\x,\ba)$ and $\pi(\x,\bg)$.

\begin{figure}[htbp]
	\begin{center}
		\includegraphics[width=0.36\textwidth]{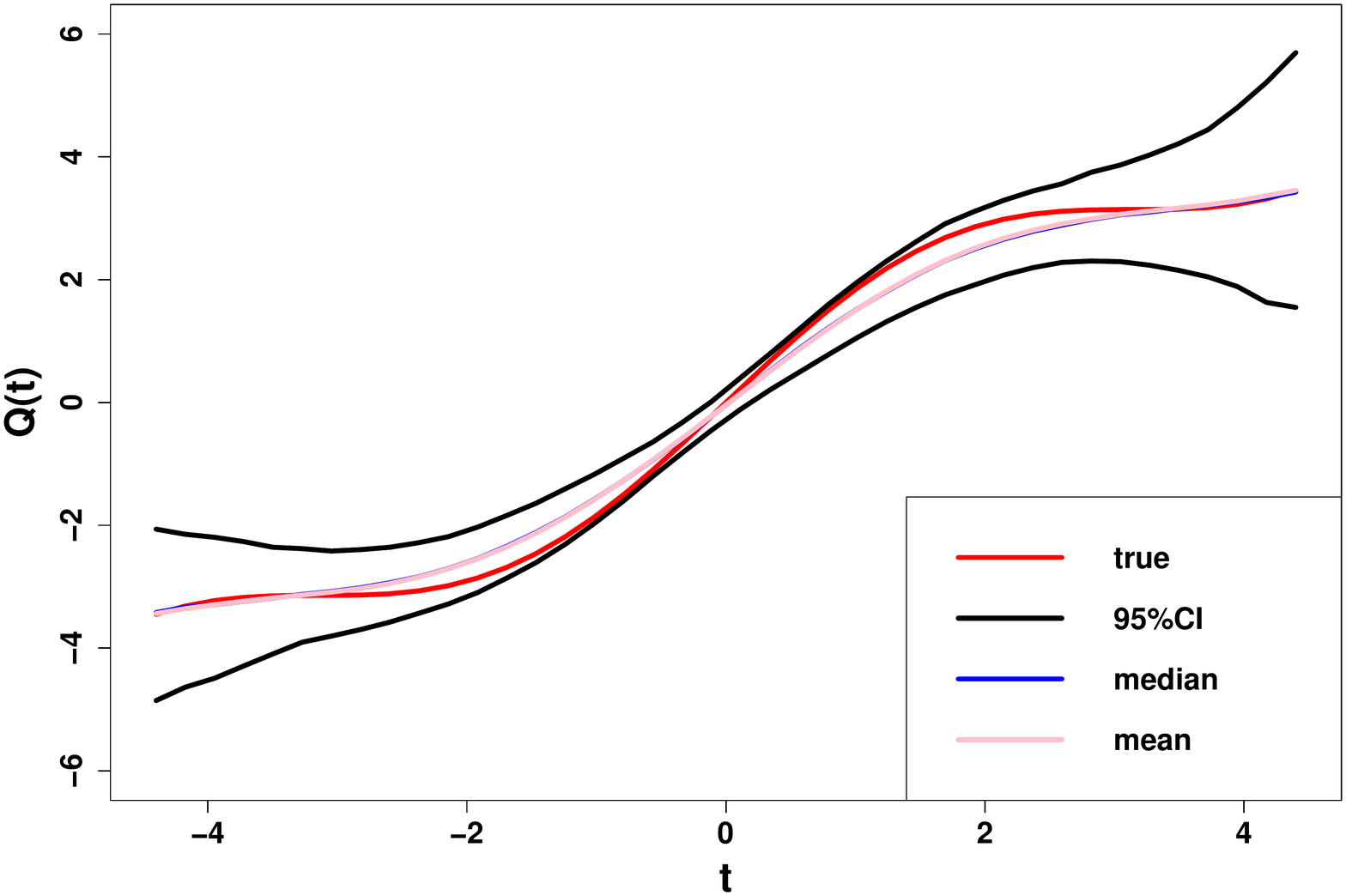}
		\hspace{0.2cm}
		\includegraphics[width=0.36\textwidth]{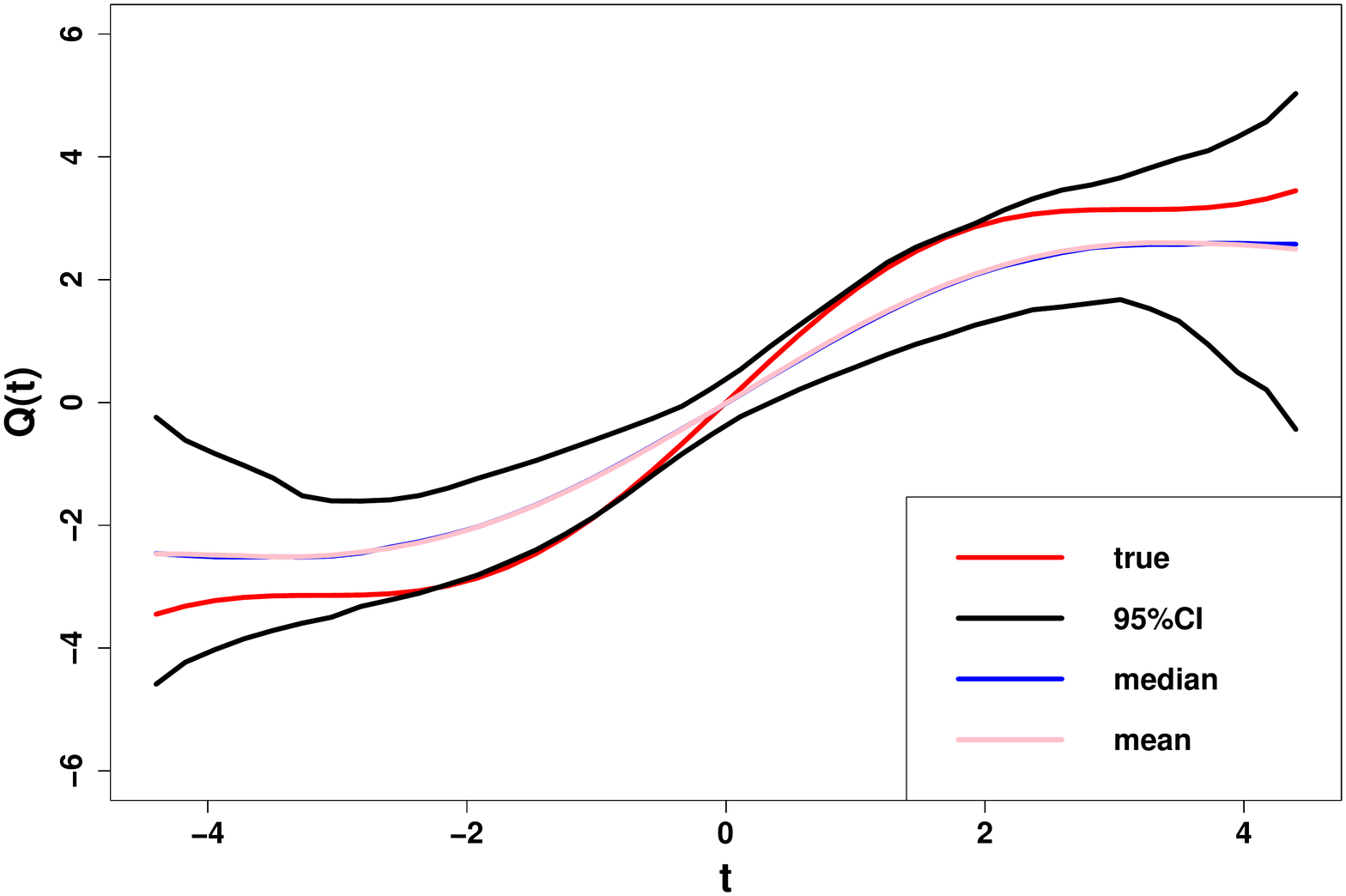}\\
		\includegraphics[width=0.36\textwidth]{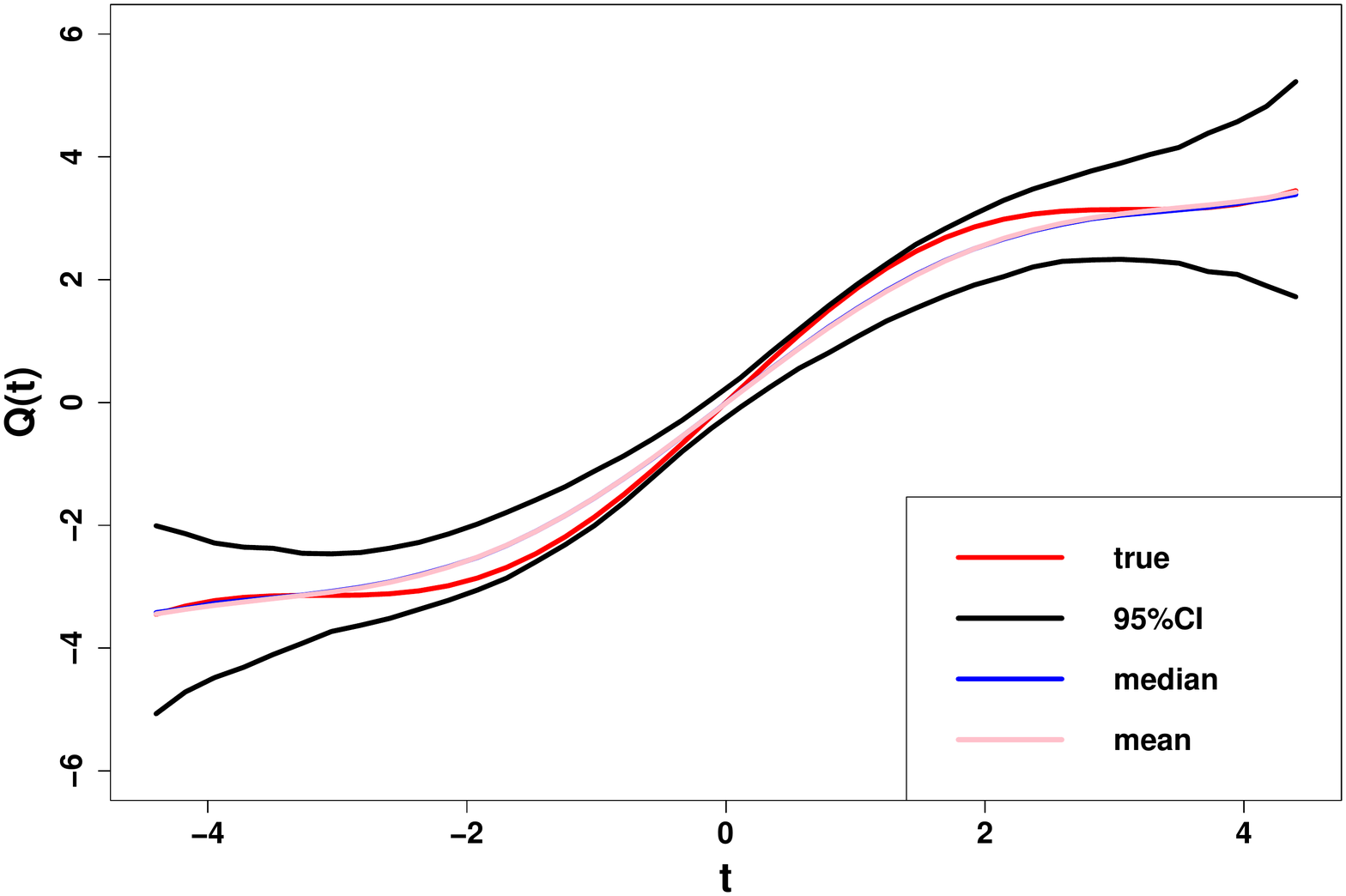}
		\hspace{0.2cm}
		\includegraphics[width=0.36\textwidth]{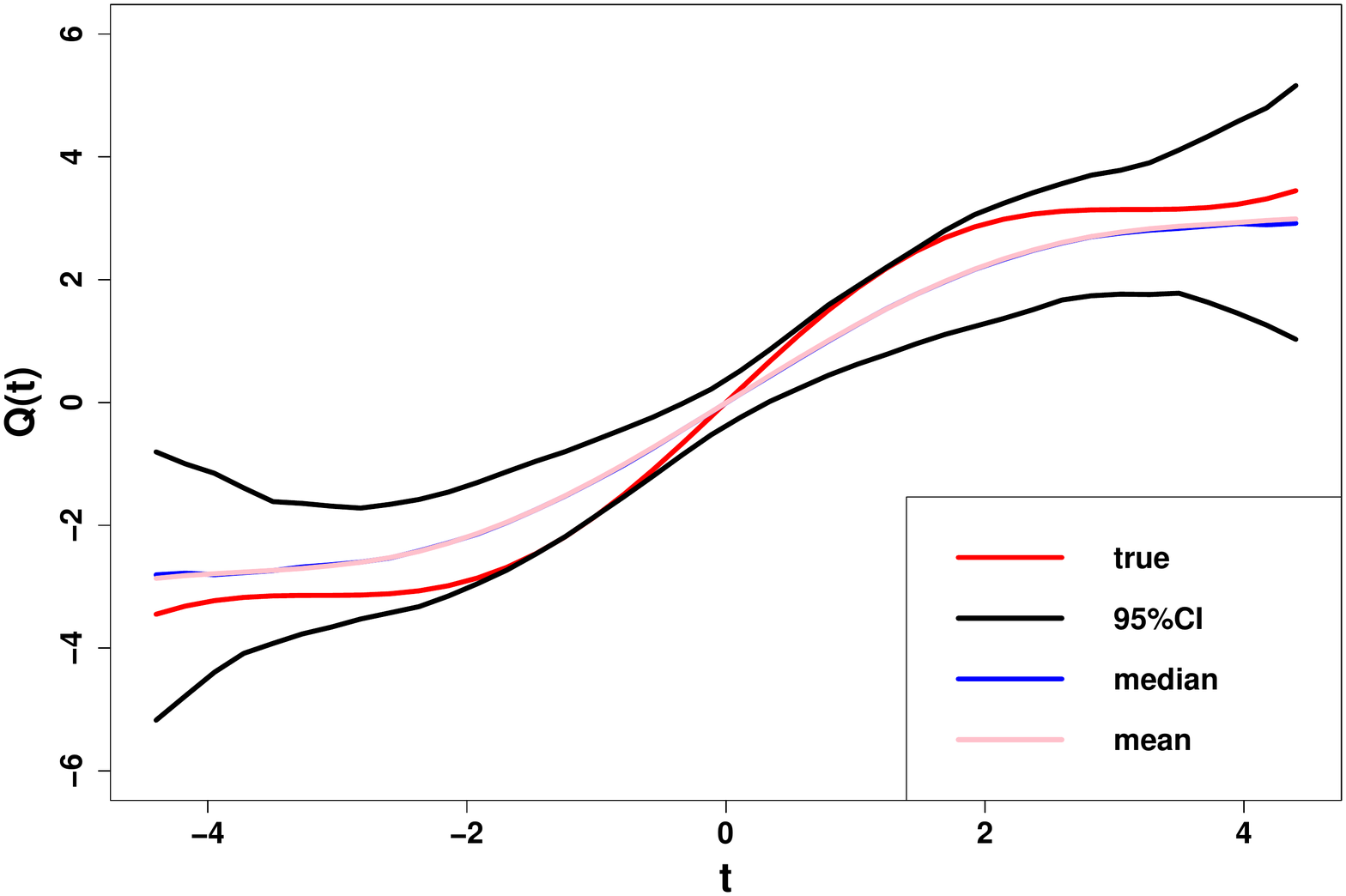}
		\caption{Simulation 5. Mean, median and  95\%
			confidence band of the estimators of $Q_0(t)=t+\sin(t)$ with non-constant propensity score model. See also the caption of Figure \ref{fig:mono1}.}
		\label{fig:pihetero}
	\end{center}
\end{figure}

We use the same nonparametric estimation procedures as in Simulation 4 to implement the algorithm in Section \ref{sec:method}. From the results summarized in Table \ref{table:pihetero}, we observe that despite the heteroscedastic error and non-constant propensity score model, in the first three cases, the estimations for parameters $\bb$, the value function, $V$ and the root of the treatment difference function yield small bias. In addition, the estimated standard deviations are still close to the empirical version of the estimators, and the confidence intervals have coverage close to the nominal levels. Interestingly, the estimator of the root of $Q(\cdot)$ performs well even in Case IV. In Figure \ref{fig:pihetero}, we note that the $95\%$ confidence interval for $Q(\cdot)$ includes the true $Q_0(\cdot)$ function in all four cases. 

\begin{table}[htbp]
	\centering
	\begin{tabular}{ccccccccc} 
		\hline
		\multicolumn{8}{c}{ Results for $\bb$ and value function $V$.}\\
		\hline
		\hline
		Case &parameters  & True  & Estimate & sd & $\hat{\rm sd}$ & cvg & MSE  \\ 
		\hline
		\multirow{4}{*}{I} & $\beta_2$& ~1& 1.0100 & 0.1197 & 0.1603 & 93.2\%  &  0.0144\\
		&  $\beta_3$      & -1    &   -1.0093 &  0.1177&  0.1625 &   93.5\%  &   0.0139\\
		&  $\beta_4$      & ~1     &   1.0120  & 0.1227 &  0.1640 &  93.3\%   &   0.0152 \\
		&V                 &3.0533 &   3.0345  & 0.1254 & 0.1287 &  95.7\%  &   0.0161  \\ \hline
		
		\multirow{4}{*}{II}&$\beta_2$ & 1 & 1.0418 &  0.1742 & 0.1896 &  93.9\%  & 0.0321\\
		&$\beta_3$      & -1    &  -0.9983 &0.1696&  0.2087 &   93.0\%  & 0.0287 \\
		&$\beta_4$      & 1     & 1.0467  &0.1749 & 0.1866 &  93.1\%   & 0.0327  \\ 
		&V& 3.0533&  2.9907& 0.1157 & 0.1359 &  95.5\%   &  0.0173 \\\hline
		
		\multirow{4}{*}{III} &  $\beta_2$ & 1      & 1.0220 & 0.1278& 0.1757& 92.4\% & 0.0168 \\ 
		& $\beta_3$ & -1     & -1.0170& 0.1243& 0.1755& 93.2\% & 0.0157 \\ 
		&  $\beta_4$ & 1      &1.0168&0.1276& 0.1765& 93.7\% &0.0165\\ 
		& V          &3.0533& 3.0440 & 0.1440& 0.1203& 95.0\%    &0.0208\\ \hline
		
		\multirow{4}{*}{IV} & $\beta_2$ & 1  &1.0512&0.2019&0.1727& 80.5\% &0.0434 \\ 
		&  $\beta_3$ & -1     &-1.0083&0.1976&0.1929 & 82.7\% &0.0391 \\ 
		&  $\beta_4$ & 1      &1.0575 &0.2094& 0.1837& 82.6\% & 0.0472 \\ 
		&V             & 3.0533& 2.9647& 0.1402 & 0.1406 & 90.6\%    & 0.0275 \\ \hline
	\end{tabular}
	\vspace{0.6cm}
	\begin{tabular}{ c  c  c  c  c  c  c  c  c }
		\hline
		\multicolumn{9}{c}{Results for the root of $Q_0(t)=t+\sin(t)$}\\
		\hline
		\hline
		Case &true & mean & bias & $\hat{\bias}$ & {\rm sd} & $\hat{\rm sd}$ & cvg & MSE \\\hline
		I & 0 & 0.0271 & 0.0271& 0.0147 & 0.0795& 0.0767& 93.8\% & 0.0070\\ 		
		II & 0 & 0.0105&  0.0105 & 0.0179&0.1564&0.1669 & 97.2\% & 0.0245\\		
		III & 0 & 0.0078 & 0.0078 &-0.0013& 0.0769&0.0823& 96.2\% & 0.0059\\ 		
		IV & 0 & 0.0089 & 0.0089 & 0.0186&0.1726& 0.1650 & 97.3\% & 0.0299 \\ \hline
	\end{tabular}
	\caption{Simulation 5. $Q_0(\bb_0\trans\x)=(\bb_0\trans\x)+\sin(\bb_0\trans\x)$, $\pi_0(\x)=\exp(\bg_0\trans\X)/\{1+\exp(\bg_0\trans\X)\}$ and
	$\mu_0(\x)=1+\sin(\ba_{10}\trans\x)+0.5(\ba_{20}\trans\x)^2$. See also the caption of Table \ref{table:mono1}.}
	\label{table:pihetero}
\end{table}

\subsection{Simulation 6}\label{subsec:simu6}

Here, we consider a non-constant propensity score model similar to Simulation 4. Other data generation procedure is identical to Simulation 3. Thus, we consider a non-constant propensity score model with a non-monotonic treatment difference function and heteroscedastic error variance in this simulation.

\begin{figure}[htbp]
	\begin{center}
		\includegraphics[width=0.36\textwidth]{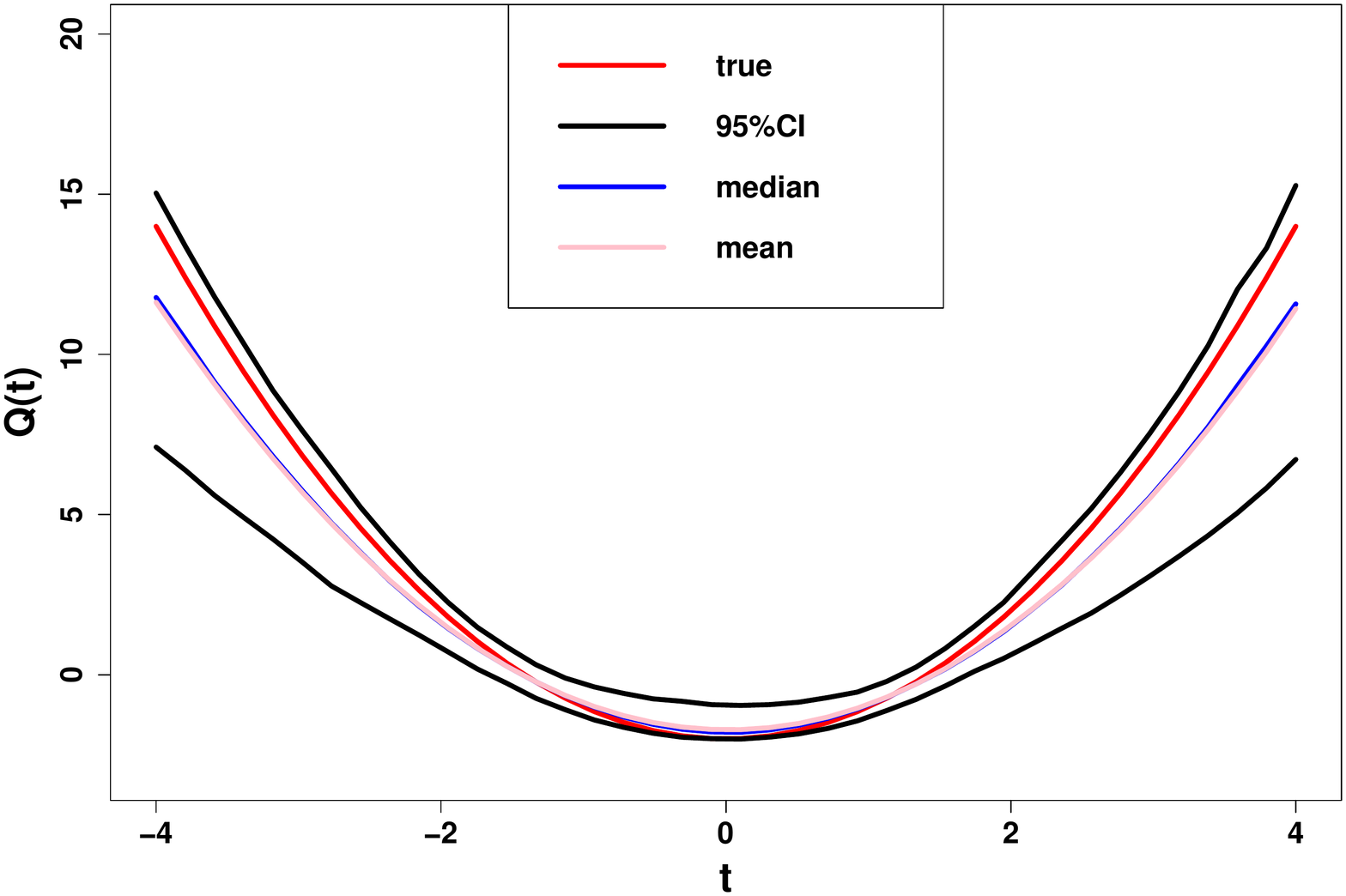}
		\hspace{0.2cm}
		\includegraphics[width=0.36\textwidth]{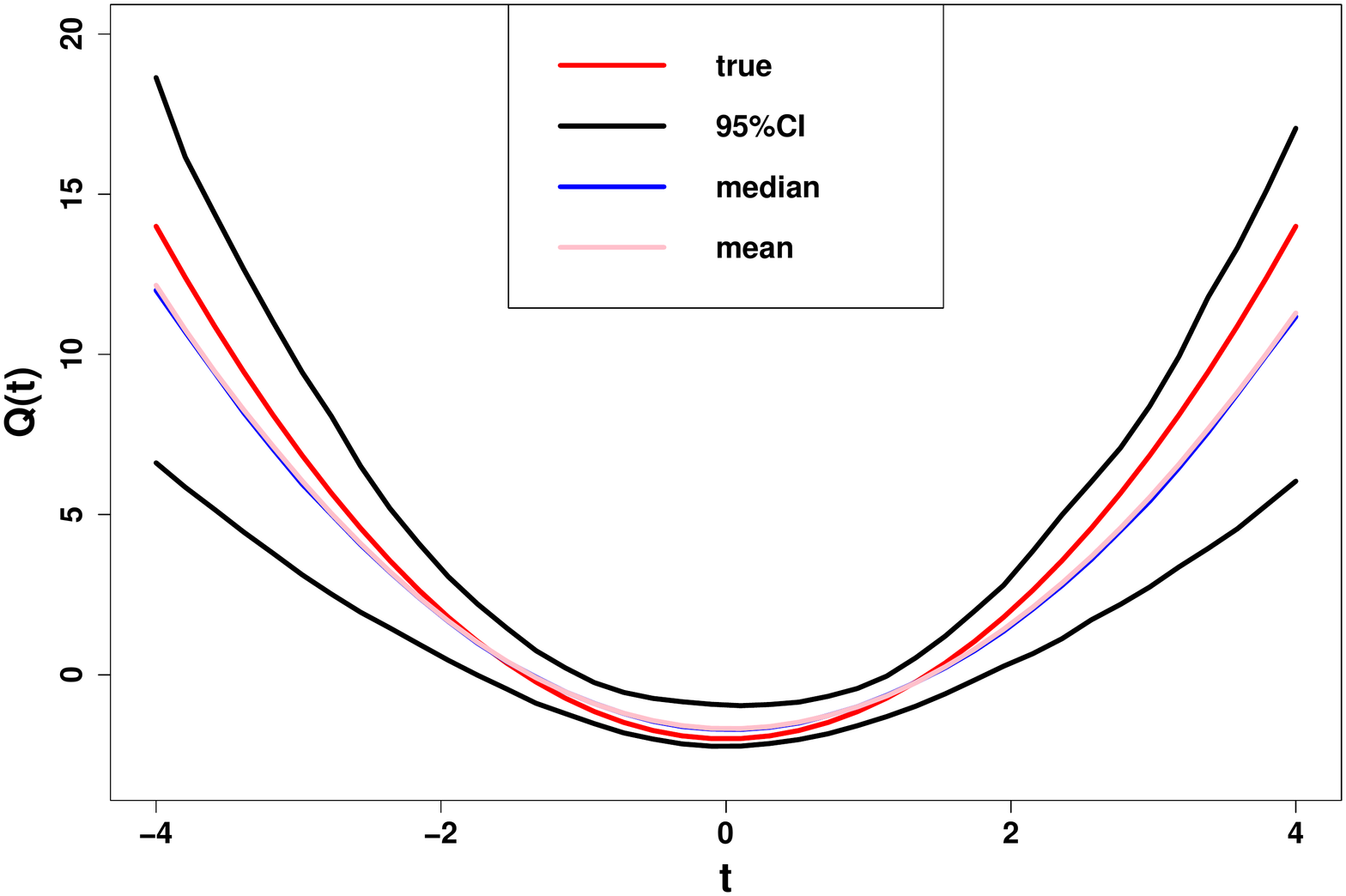}\\
		\includegraphics[width=0.36\textwidth]{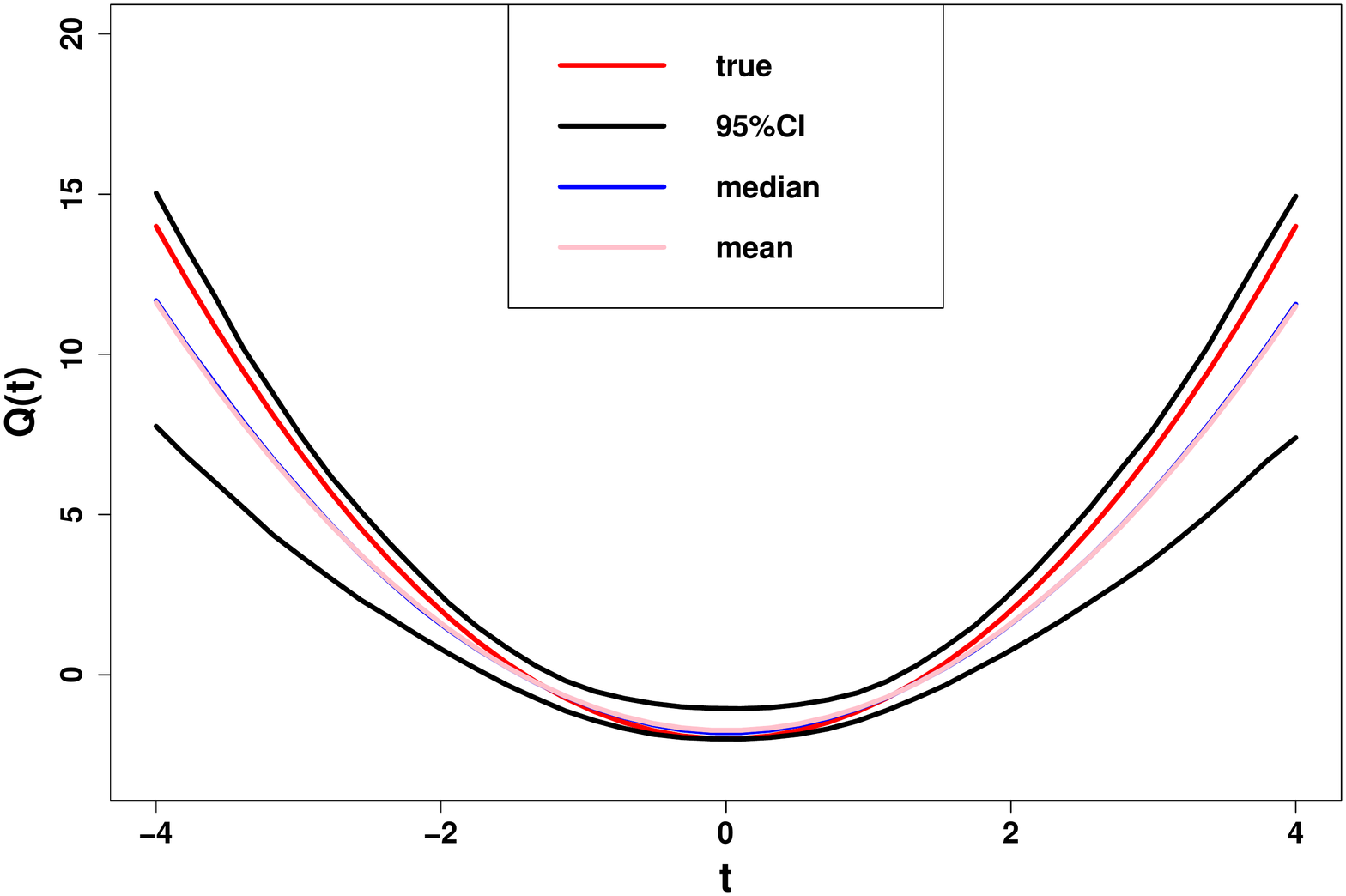}
		\hspace{0.2cm}
		\includegraphics[width=0.36\textwidth]{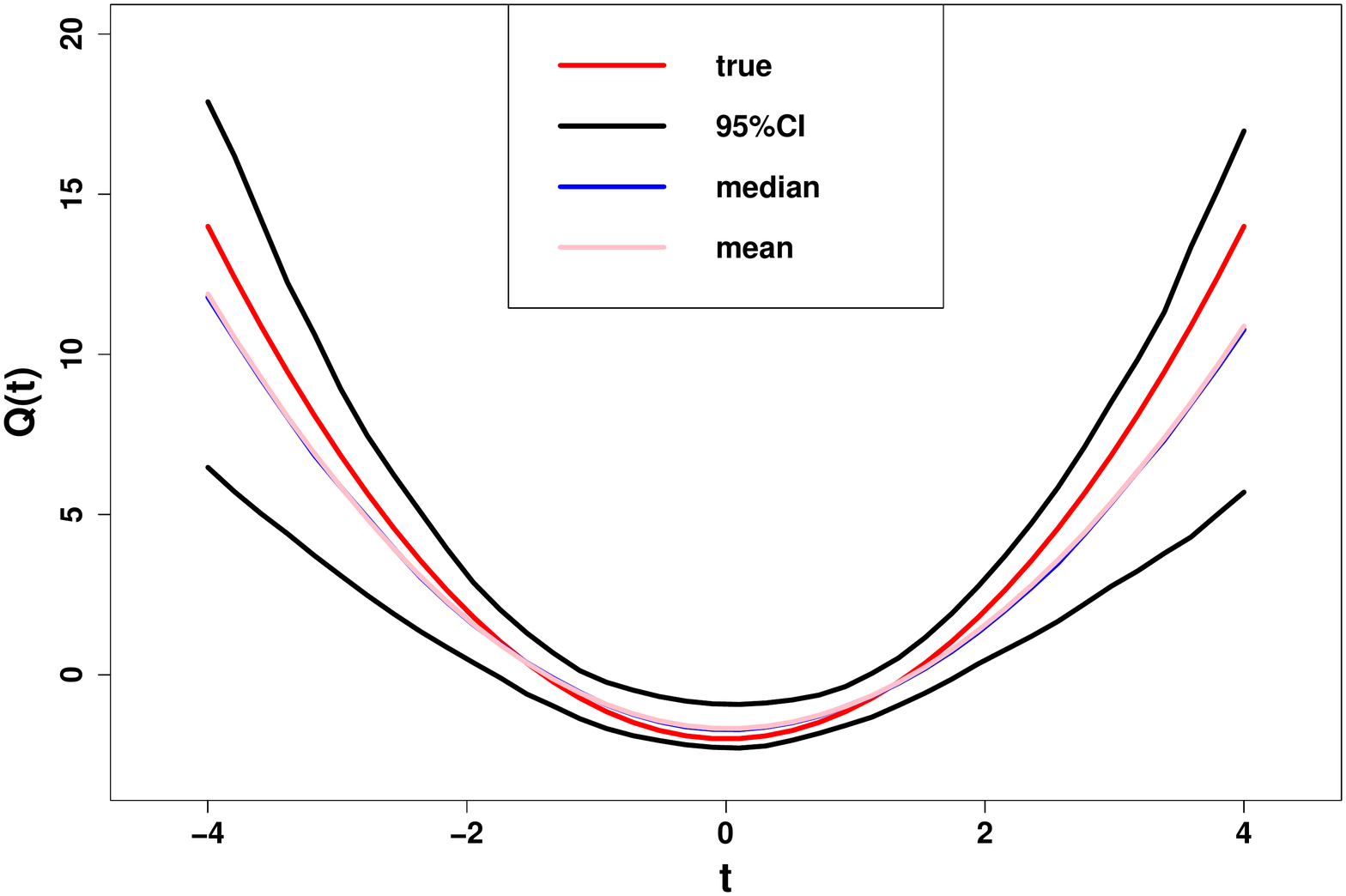}
		\caption{Simulation 6. Mean, median and  95\%
			confidence band of the estimators of $Q_0(t)=t^2-2$ with non-constant propensity score model and heteroscedastic error variance. See also the caption of Figure \ref{fig:mono1}.}
		\label{fig:pinorank}
	\end{center}
\end{figure}

Similar to Simulation 5, we consider four cases and implemented our estimation procedures to illustrate the robustness property of estimators. From the results summarized in Table \ref{table:pinorank}, we observe that the estimations for parameters $\bb$, the value function, $V$, and the root of the treatment difference function result in small bias in the first three cases, as expected. Also, the estimated standard deviations are close to the empirical standard deviations and the confidence intervals are close to nominal coverage levels. Interestingly, the estimation and inference of $\bb$, $V$, and the two roots perform well in Case IV. In Figure \ref{fig:pinorank}, we note that the $95\%$ confidence interval for $Q(\cdot)$ includes the true $Q_0(\cdot)$ function in all four cases.

For comparison, we implemented the methods in \cite{fanetal2017} for
all simulations. These results are summarized in Tables
\ref{table: CAL-DR1} to \ref{table:CAL-DR6} in the Appendix. As we can
see, when the monotonicity assumption is violated, 
\cite{fanetal2017} deteriorates and performs worse than our proposed method. 

\begin{table}[htbp]
	\centering
	\begin{tabular}{ccccccccc} 
		\hline
		\multicolumn{8}{c}{ Results for $\bb$ and value function $V$.}\\
		\hline
		\hline
		Case &parameters  & True  & Estimate & sd & $\hat{\rm sd}$ & cvg & MSE  \\ 
		\hline
		\multirow{4}{*}{I} & $\beta_2$      & 1     &    1.0316      &  0.0659  &      0.0870      &   92.7\%  &   0.0053   \\
		&  $\beta_3$      & -1    &   -1.0346  & 0.0644 &   0.0890   &   92.8\%  &   0.0053   \\
		&  $\beta_4$      & 1     &  1.0346   & 0.0635  &  0.0874   &  93.9\%   &   0.0052   \\ 
		&V                 &4.7166 &  4.6549  & 0.3035  &  0.3294   &  96.3\%  &   0.0959  \\ \hline
		
		\multirow{4}{*}{II}&$\beta_2$      & 1     &    1.0236     &  0.1117  &     0.1377      &   93.4\%  &   0.0130   \\
		&$\beta_3$      & -1    &   -1.0186     & 0.1111 &    0.1362    &   93.6\%  &   0.0127   \\
		&$\beta_4$      & 1     &    1.0173     & 0.1143  &    0.1349        &  92.9\%   &   0.0133  \\ 
		&V& 4.7166&    4.6134    &  0.2952  &    0.3269  &  96.3\%   &   0.0977 \\
		\hline
		
		\multirow{4}{*}{III} &  $\beta_2$ & 1      & 1.0361  & 0.0637 &0.0889 & 93.0\% & 0.0054 \\ 
		& $\beta_3$ & -1     & -1.0343  & 0.0637& 0.0895 & 95.0\% & 0.0052 \\ 
		&  $\beta_4$ & 1      & 1.0327  &0.0642 & 0.0877 & 94.0\% & 0.0052 \\ 
		& V          & 4.7166 & 4.6381 & 0.2930& 0.3020  & 95.8\%    & 0.0920 \\ \hline
		
		\multirow{4}{*}{IV} & $\beta_2$ & 1      & 1.0396   & 0.1219 & 0.1368 & 93.5\% & 0.0164 \\ 
		&  $\beta_3$ & -1     & -1.0245 & 0.1103 & 0.1321 & 94.0\% &0.0128 \\ 
		&  $\beta_4$ & 1      & 1.0343  & 0.1215 & 0.1371 & 92.3\% & 0.0159 \\ 
		&V             & 4.7166 & 4.6004 & 0.2905 & 0.3056  & 94.2\%    & 0.0979 \\ \hline
	\end{tabular}
	\vspace{0.6cm}
	\begin{tabular}{ c  c  c  c  c  c  c  c  c }
		\hline
		\multicolumn{9}{c}{Results for the two roots of $Q_0(t)=t^2-2$}\\
		\hline
		\hline
		Case &true & mean & bias & $\hat{\bias}$ & {\rm sd} & $\hat{\rm sd}$ & cvg & MSE \\ \hline
		\multirow{2}{*}{I} & -1.4142 &  -1.4268 & -0.0126 & -0.0199 & 0.1201 & 0.1085 & 95.3\% & 0.0146\\
		& 1.4142 & 1.4556 & 0.0414 & 0.0238 & 0.1115 & 0.1172 & 94.8\% & 0.0142\\ \hline
		
		\multirow{2}{*}{II} & -1.4142 &  -1.3810 & 0.0331 & 0.0368 &0.1811 & 0.1885 & 94.4\% & 0.0339\\
		& ~1.4142 & 1.4553 & 0.0411 & 0.0086 & 0.1792 & 0.1835 & 93.2\% & 0.0338\\ \hline
		
		\multirow{2}{*}{III} & -1.4142 &  -1.4410 & -0.0268 & -0.0159 & 0.1221 & 0.1169& 93.2\% & 0.0156\\
		& 1.4142 &  1.4554 & 0.0412 & 0.0137 & 0.1160 & 0.1206& 93.6\% & 0.0151\\ \hline
		
		\multirow{2}{*}{IV} & -1.4142 &  -1.3941 & 0.0201 & -0.0065 & 0.1905 & 0.1950 & 94.9\% & 0.0367\\
		& 1.4142 & 1.4508 & 0.0366 & -0.0080 & 0.1769 & 0.1885 & 95.9\% & 0.0326\\ \hline
	\end{tabular}
	\caption{Simulation 6. $Q_0(\bb_0\trans\x)=(\bb_0\trans\x)^2-2$, $\pi_0(\x)=\exp(\bg_0\trans\X)/\{1+\exp(\bg_0\trans\X)\}$ and
	$\mu_0(\x)=1+\sin(\ba_{10}\trans\x)+0.5(\ba_{20}\trans\x)^2$. See also the caption of Table \ref{table:mono1}.}
	\label{table:pinorank}
\end{table}


\section{Real Data Application}\label{sec:real}

We now apply our proposed method to a study of  the effect of smoking
during pregnancy on baby's birth weight. The primary outcomes is
birth weight (in grams) of singleton births in Pennsylvania, USA
\citep{Almond_2005}. This study aims to determine whether pregnant
women  should quit smoking to ensure healthy birth in terms of baby's
birth-weight. For illustration purpose, we consider a subset of 1394
unmarried mothers. The data set contains
the maternal smoking habit during pregnancy which is treated as
treatment $A_i$ ($1=$Non-smoking, $0=$Smoking). The covariates
observed are  mother's age (mage), an indicator variable for alcohol
consumption during pregnancy (alcohol), an indicator variable of
previous birth in which the infant died (deadkids), mother's education
(medu), father's education (fedu), number of prenatal care visits 
(nprenatal), months since last birth (monthslb), mother's race (mrace)
and an indicator variable of first born child (fbaby).  

To estimate the propensity score, mean outcome model for non-treated
group and treatment difference function, first we normalize all the
continuous covariates. We use the expit model $\pi(\X,\bg)$ to describe
the propensity score, and use MLE to estimate $\bg$. In addition, we
consider a linear model for the mean outcome model for the non-treatment
group $\mu(\X,\ba)$, and solved GEE to obtain $\wh\ba$. Lastly, we
estimate the treatment difference model 
$Q(\bb\trans\X)$ using the proposed method.  

In implementing the algorithm described in Section
\ref{sec:method}, we used the quartic kernel in the nonparametric
implementation to estimate $\bb$ with bandwidth $c\sigma n^{-1/3}$,
where $\sigma^2$ is the estimated variance of $\bb\trans\X$, and
$c=0.05$. 

\begin{figure}[htbp]
	\centering
	\includegraphics[width=0.75\textwidth]{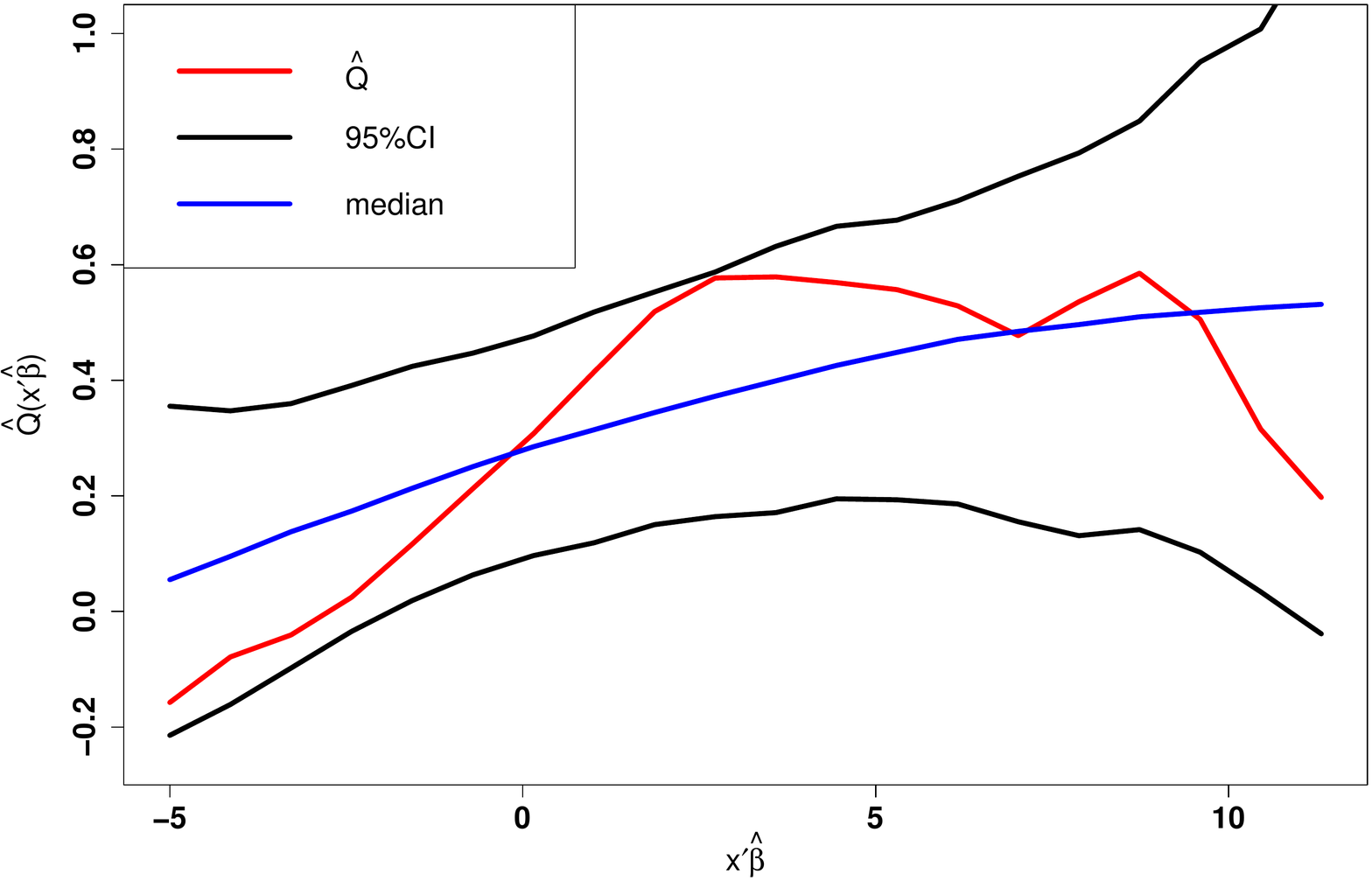}
	\caption{Data analysis. 
		The estimated treatment difference model, $\wh Q(\wh\bb\trans\x)$, its median, 95\% confidence bands based on low baby birth weight data set.}
	\label{fig:realQCI}
\end{figure}

For identifiability purpose, we fix the coefficient of the first
covariate (here, mage) to be $1$ and estimate the remaining eight
coefficients. The estimated parameters in $\bb$, their standard
errors, and the value function are summarized in Table
\ref{table:real}. From the 95\% confidence interval for $\bb$, we can
conclude that all covariates are significant, except the indicator
variable for the alcohol consumption. We provide the 
estimated treatment difference model, $\wh Q(\wh\bb\trans\X)$ in
Figure \ref{fig:realQCI}. It shows a higher baby birth weight for
those mothers who did not smoke during their pregnancy, once $\wh
Q(\wh\bb\trans\X)$ is greater than $0$. 
We further construct 95\% confidence band for the difference
function $Q(t)$ based on 500 bootstrap samples by resampling the
residuals. 
The results obtained
from CAL and CAL-DR method \cite{fanetal2017} are summarized in Table
\ref{table:CAL-DRreal}. It can be seen that no significant covariate
is detected by CAL or CAL-DR. Also the variability in
estimating the value function using CAL or CAL-DR is
higher than that from the proposed method. Further, we
observe that the $95\%$ confidence intervals computed by CAL and
CAL-DR methods include the estimated value function obtained by our
method. 

\begin{table}[htbp]
	\centering
	\caption{Birth-weight study analysis: Results for $\bb$ and value function $V$ with $95\%$ CI.}
	\begin{tabular}{lccr} 
		\hline
		parameters     & Estimate& $\wh{\rm sd}$ & Confidence interval  \\ 
		\hline
		$\beta_2$ (alcohol)  &  0.2965& 0.4842&(-0.6525,1.2455)\\    
		$\beta_3$  (deadkids)  &  0.3406&0.0233&(0.2950,0.3862)\\  
		$\beta_4$ (medu)  &  -0.1972&0.0073&(-0.2116,-0.1828)\\  
		$\beta_5$ (fedu)   & -0.0947&0.0005&(-0.0957,-0.0938)\\ 
		$\beta_6$ (nprenatal)  & 0.2822&0.0061&(0.2703,0.2941)\\  
		$\beta_7$ (monthslb)  & 0.0183&0.0002&(0.0178,0.0188)\\  
		$\beta_8$ (mrace)   & 3.0882&0.3753&(2.3527,3.8237)\\ 
		$\beta_9$ (fbaby)  & -1.8505&0.4267&(-2.6868,-1.0143)\\  
		Value function    &  3274.9  &25.439&(3225.1,3324.8)\\     
		\hline
	\end{tabular}
	\label{table:real}
\end{table}

\begin{table}[H]
	\centering
	\caption{Application to birth-weight study using CAL and CAL-DR methods.}
	\begin{tabular}{lccr} 
		\hline
		parameters  & Estimate& $\wh{\rm sd}$   & Confidence interval  \\ 
		\hline
		\multicolumn{4}{c}{ CAL estimates for $\bb$ and value function $V$ for the real data analysis.}\\
		\hline
		$\beta_2$ (alcohol)  &-0.0223&7.0207&(-13.783,13.738)\\      
		$\beta_3$  (deadkids)  &0.2088&5.9428&(-11.439,11.857)\\  
		$\beta_4$ (medu)  &-0.1998&0.6850&(-1.5424,1.1428)\\  
		$\beta_5$ (fedu)  &-0.0902&0.2929&(-0.6643,0.4838)\\  
		$\beta_6$ (nprenatal) &0.2966&0.5499&(-0.7812,1.3743)\\ 
		$\beta_7$ (monthslb)  &0.0192&0.1653&(-0.3048,0.3431)\\
		$\beta_8$ (mrace)  &3.1975&7.1149&(-10.747,17.142)\\ 
		$\beta_9$ (fbaby)  &-2.0682&3.8728&(-9.6587,5.5222)\\ 
		Value function    &3244.4&30.345&(3185.0,3303.9)\\     
		\hline
		\hline
		\multicolumn{4}{c}{ CAL-DR estimates for $\bb$ and value function $V$ for the real data analysis.}\\
		\hline
		$\beta_2$ (alcohol)   &-0.1111&1.7382&(-3.5179,3.2958)\\  
		$\beta_3$  (deadkids)  &0.2589&0.9554&(-1.6136,2.1314)\\ 
		$\beta_4$ (medu)  &-0.2058&0.3147&(-0.8226,0.4109)\\
		$\beta_5$ (fedu) &-0.0815&0.1545&(-0.3843,0.2213)\\ 
		$\beta_6$ (nprenatal) &0.2759&0.2782&(-0.2693,0.8212)\\  
		$\beta_7$ (monthslb)  &0.0183&0.0383&(-0.0568,0.0933)\\ 
		$\beta_8$ (mrace)  &3.2234&2.4674&(-1.6127,8.0594)\\ 
		$\beta_9$ (fbaby)  &-2.0459&1.7858&(-5.5460,1.4542)\\ 
		Value function     &3241.9&30.243&(3182.6,3301.2)\\      
		\hline
	\end{tabular}
	
	\label{table:CAL-DRreal}
\end{table}


\section{Discussion}\label{sec:discussion}

We proposed a new multi-robust estimation to estimate the optimal
treatment regimes for a single decision time point under weak
conditions, i.e., our treatment
difference model $Q(\cdot)$ does not need to be monotonic and we only require
$E(\epsilon \mid \X)=0$. Our method enjoys the protection against
misspecification of either the propensity score model or the
outcome regression model for the non-treated group or the
non-monotonic treatment difference model. We use nonparametric kernel-based
estimator to obtain the treatment difference model and show that the
treatment identification rate is $O_p(n^{-2/5})$. 
The extensive simulation studies illustrate superior  performance
under various scenarios. 

In this paper, we consider parametric models for the propensity score
function and the mean outcome model for the non-treated
group. Alternatively, one
can use the semiparametric or nonparametric method to obtain these two
functions. For example,  one can use the semiparametric estimation
procedure in \cite{mazhu2013} to estimate propensity score and the
mean outcome model for the non-treated group to obtain
$n^{1/2}$-consistent estimators for $\bg$ and $\ba$. The
treatment identification rate will  remain unchanged.  

Many extensions following this work can be interesting and  worth
pursuing. For example, we may consider multiple treatment decision
times while incorporating the usual backward
induction to obtain the optimal dynamic treatment regimes. 
We may also consider multiple treatment choices sharing the same
index. Further studies along these lines can be challenging and
fruitful.

\section*{Supplementary Material}
The online Supplementary Material contains proofs for Proposition \ref{pro:robust}, Lemma \ref{lem:order}--\ref{lem:Q}, and Theorems \ref{th:beta}--\ref{th:value}.

\baselineskip=14pt
\bibliographystyle{agsm}
\bibliography{notrank}

\end{document}


\thispagestyle{empty}
\baselineskip=18pt
\title{Supplementary Appendix for ``Learning non-monotone optimal individualized treatment regimes''}
\author[a]{Trinetri Ghosh}
\author[b]{Yanyuan Ma}
\author[c]{Wensheng Zhu}
\author[d]{Yuanjia Wang}
\affil[a]{Department of Biostatistics and Medical Informatics, University of Wisconsin-Madison}
\affil[b]{Department of Statistics,  Pennsylvania State University}
\affil[c]{
School of Mathematics and Statistics, Northeast Normal University}
\affil[d]{Department of Biostatistics, Columbia University}
        \maketitle

\setcounter{equation}{0}\renewcommand{\theequation}{A.\arabic{equation}}
\setcounter{subsection}{0}\renewcommand{\thesubsection}{A.\arabic{subsection}}
\setcounter{section}{0}\renewcommand{\thesection}{A.\arabic{section}}
\setcounter{table}{0}\renewcommand{\thetable}{A.\arabic{table}}
\setcounter{figure}{0}\renewcommand{\thefigure}{A.\arabic{figure}}

\subsection{Proof of Proposition \ref{pro:robust}}
 It is easy to show that when all the models, $\pi(\X,\bg)$, 
$\mu(\X,\ba)$ and $Q(\bb\trans\X)$ are correct, the expectation  
of the estimating equation is indeed $\0$. So, we need to show that if 
any one of these aforementioned 
models is misspecified, still we get the expectation to be 
$\0$. 

First, we show the robustness property against the misspecification of 
the propensity model. 
Let the propensity score function model 
be misspecified, i.e.  $E(A_i \mid \X_i) \neq \pi(\X_i,\bg)$. Suppose the 
true propensity function is $\pi_0(\X_i)$. Then, 
\bse 
&&E\bigg( \bigg[ \frac{\{A_i - \pi(\X_i,\bg)\}\{Y_i-\mu(\X_i,\ba)\}}{\pi(\X_i,\bg)\{1-\pi(\X_i,\bg)\}} 
+ \bigg\{1 - \frac{A_i}{\pi(\X_i,\bg)} \bigg\}Q(\bb\trans\X_i)\bigg] \{\X_{Li}-E(\X_{Li}\mid\bb\trans\X_i)\} \bigg) \\
&=& E\bigg( E \bigg[ \frac{\{A_i - \pi(\X_i,\bg)\}\{Y_i-\mu(\X_i,\ba)\}}{\pi(\X_i,\bg)\{1-\pi(\X_i,\bg)\}} 
+ \bigg\{1 - \frac{A_i}{\pi(\X_i,\bg)} \bigg\}Q(\bb\trans\X_i) \mid \X_i \bigg] 
\{\X_{Li}-E(\X_{Li}\mid\bb\trans\X_i)\}\bigg) \\
&=& E\bigg\{ E \bigg( \frac{\{A_i - \pi(\X_i,\bg)\}[A_i\{Y_{i1}-\mu(\X_i,\ba)\}
+(1-A_i)\{Y_{i0}-\mu(\X_i,\ba)\}]}{\pi(\X_i,\bg)\{1-\pi(\X_i,\bg)\}} \\
&&+ \bigg\{1 - \frac{A_i}{\pi(\X_i,\bg)} \bigg\}Q(\bb\trans\X_i) \mid \X_i \bigg) \{\X_{Li}-E(\X_{Li}\mid\bb\trans\X_i)\} \bigg\} \\
&=& E\bigg\{ \bigg( \frac{E[\{A_i-\pi(\X_i,\bg)\}A_i \mid \X_i] Q(\bb\trans\X_i) }{\pi(\X_i,\bg)\{1-\pi(\X_i,\bg)\}} 
+ \bigg\{1 - \frac{\pi_0(\X_i)}{\pi(\X_i,\bg)} \bigg\}Q(\bb\trans\X_i) \bigg) 
\{\X_{Li}-E(\X_{Li}\mid\bb\trans\X_i)\} \bigg\} \\
&=& E\bigg( \bigg[ \frac{\pi_0(\X_i)}{\pi(\X_i,\bg)} Q(\bb\trans\X_i) 
+ \bigg\{1 - \frac{\pi_0(\X_i)}{\pi(\X_i,\bg)} \bigg\}Q(\bb\trans\X_i) \bigg] 
\{\X_{Li}-E(\X_{Li}\mid\bb\trans\X_i)\} \bigg) \\
&=& E[Q(\bb\trans\X_i)\{\X_{Li} - E(\X_{Li} \mid \bb\trans\X_i)\}] \\
&=& E(E[Q(\bb\trans\X_i)\{\X_{Li} - E(\X_{Li} \mid \bb\trans\X_i)\} \mid \bb\trans\X_i])\\
&=& E[Q(\bb\trans\X_i)\{E(\X_{Li} \mid \bb\trans\X_i) - E(\X_{Li} \mid \bb\trans\X_i)\}] = 0. 
\ese 
The third equality holds because $E\{Y_{i1} - \mu(\X_i,\ba) \mid \X_i \} = Q(\bb\trans\X_i)$ and 
$E\{Y_{i0} - \mu(\X_i,\ba) \mid \X_i \} = 0$.

Next, we show the robustness property against the misspecification of 
the outcome model. 
Let the model $\mu(\X_i,\ba)\ne E(Y_{i0}\mid\X_i)$, i.e. it is misspecified. Assume the true 
outcome is $E(Y_{i0} \mid \X_i) =\mu_0(\X_i)$. Then 
\bse 
&&E\bigg( \bigg[ \frac{\{A_i - \pi(\X_i,\bg)\}\{Y_i-\mu(\X_i,\ba)\}}{\pi(\X_i,\bg)\{1-\pi(\X_i,\bg)\}} 
+ \bigg\{1 - \frac{A_i}{\pi(\X_i,\bg)} \bigg\}Q(\bb\trans\X_i)\bigg] \{\X_{Li}-E(\X_{Li}\mid\bb\trans\X_i)\} \bigg) \\
&=& E\bigg( E \bigg[ \frac{\{A_i - \pi(\X_i,\bg)\}\{Y_i-\mu(\X_i,\ba)\}}{\pi(\X_i,\bg)\{1-\pi(\X_i,\bg)\}} 
+ \bigg\{1 - \frac{A_i}{\pi(\X_i,\bg)} \bigg\}Q(\bb\trans\X_i) \mid \X_i \bigg] 
\{\X_{Li}-E(\X_{Li}\mid\bb\trans\X_i)\} \bigg) \\
&=& E\bigg\{ E \bigg( \frac{\{A_i - \pi(\X_i,\bg)\}[A_i\{Y_{i1}-\mu(\X_i,\ba)\}
+(1-A_i)\{Y_{i0}-\mu(\X_i,\ba)\}]}{\pi(\X_i,\bg)\{1-\pi(\X_i,\bg)\}} \\
&&+ \bigg\{1 - \frac{A_i}{\pi(\X_i,\bg)} \bigg\}Q(\bb\trans\X_i) \mid \X_i \bigg) \{\X_{Li}-E(\X_{Li}\mid\bb\trans\X_i)\} \bigg\} \\
&=& E\bigg\{ \bigg( \frac{E[\{A_i - \pi(\X_i,\bg)\}A_i \mid \X_i] E\{Y_{i1}-\mu(\X_i,\ba) \mid \X_i\}}{\pi(\X_i,\bg)\{1-\pi(\X_i,\bg)\}} \\
&&+\frac{E[\{A_i - \pi(\X_i,\bg)\}(1-A_i)\mid \X_i] E\{Y_{i0}-\mu(\X_i,\ba)\mid \X_i\}]}{\pi(\X_i,\bg)\{1-\pi(\X_i,\bg)\}} \bigg) 
  \{\X_{Li}-E(\X_{Li}\mid\bb\trans\X_i)\} \bigg\} \\
&=& E[Q(\bb\trans\X_i)\{\X_{Li} - E(\X_{Li} \mid \bb\trans\X_i)\}] \\
&=& E(E[Q(\bb\trans\X_i)\{\X_{Li} - E(\X_{Li} \mid \bb\trans\X_i)\} \mid \bb\trans\X_i])\\
&=& E[Q(\bb\trans\X_i)\{E(\X_{Li} \mid \bb\trans\X_i) - E(\X_{Li} \mid \bb\trans\X_i)\}] = 0. 
\ese 
The third equality is due to $E(A_i \mid \X_i)=\pi(\X_i,\bg)$. The fourth equality holds because, 
$E[\{A_i - \pi(\X_i,\bg)\}A_i \mid \X_i]=-E[\{A_i - \pi(\X_i,\bg)\}(1-A_i) \mid \X_i]= \pi(\X_i,\bg)\{1- \pi(\X_i,\bg)\}$, 
$E\{Y_{i1}-\mu(\X_i,\ba) \mid \X_i\}=Q(\bb\trans\X_i)+\mu_0(\X_i)-\mu(\X_i,\ba)$ and 
$E\{Y_{i0}-\mu(\X_i,\ba) \mid \X_i\}=\mu_0(\X_i)-\mu(\X_i,\ba)$.

Lastly, we consider the situation when $Q(\cdot)$ is 
misspecified while both $\pi(\X_i,\bg)$ and $\mu(\X_i,\ba)$ models are 
correct. Suppose the true model is $Q_0(\cdot)$. Then 
\bse 
&&E\bigg( \bigg[ \frac{\{A_i - \pi(\X_i,\bg)\}\{Y_i-\mu(\X_i,\ba)\}}{\pi(\X_i,\bg)\{1-\pi(\X_i,\bg)\}} 
+ \bigg\{1 - \frac{A_i}{\pi(\X_i,\bg)} \bigg\}Q(\bb\trans\X_i)\bigg] \{\X_{Li}-E(\X_{Li}\mid\bb\trans\X_i)\} \bigg) \\
&=& E\bigg( E \bigg[ \frac{\{A_i - \pi(\X_i,\bg)\}\{Y_i-\mu(\X_i,\ba)\}}{\pi(\X_i,\bg)\{1-\pi(\X_i,\bg)\}} 
+ \bigg\{1 - \frac{A_i}{\pi(\X_i,\bg)} \bigg\}Q(\bb\trans\X_i) \mid \X_i \bigg] 
\{\X_{Li}-E(\X_{Li}\mid\bb\trans\X_i)\} \bigg) \\
&=& E\bigg\{ E \bigg( \frac{\{A_i - \pi(\X_i,\bg)\}[A_i\{Y_{i1}-\mu(\X_i,\ba)\}
+(1-A_i)\{Y_{i0}-\mu(\X_i,\ba)\}]}{\pi(\X_i,\bg)\{1-\pi(\X_i,\bg)\}} \\
&&+ \bigg\{1 - \frac{A_i}{\pi(\X_i,\bg)} \bigg\}Q(\bb\trans\X_i) \mid \X_i \bigg) \{\X_{Li}-E(\X_{Li}\mid\bb\trans\X_i)\} \bigg\} \\
&=& E\bigg(  \frac{E[\{A_i-\pi(\X_i,\bg)\}A_i \mid \X_i] Q_0(\bb\trans\X_i) }{\pi(\X_i,\bg)\{1-\pi(\X_i,\bg)\}} 
 \{\X_{Li}-E(\X_{Li}\mid\bb\trans\X_i)\} \bigg) \\
&=& E[Q_0(\bb\trans\X_i)\{\X_{Li} - E(\X_{Li} \mid \bb\trans\X_i)\}] \\
&=& E(E[Q_0(\bb\trans\X_i)\{\X_{Li} - E(\X_{Li} \mid \bb\trans\X_i)\} \mid \bb\trans\X_i])\\
&=& E[Q_0(\bb\trans\X_i)\{E(\X_{Li} \mid \bb\trans\X_i) - E(\X_{Li} \mid \bb\trans\X_i)\}] = 0. 
\ese 
The third equality holds because $E\{Y_{i1} - \mu(\X_i,\ba) \mid \X_i \} = Q_0(\bb\trans\X_i)$, 
$E\{Y_{i0} - \mu(\X_i,\ba) \mid \X_i \} = 0$ and $E(A_i \mid 
\X_i)=\pi(\X_i,\bg)$. 
\qed

\subsection{Proof of Lemma \ref{lem:order}}
 Let $z=(\bb\trans\x_j - \bb\trans\x)/h$, then 
$\bb\trans\x_j = \bb\trans\x + zh$. Let the probability density 
function of $\bb\trans\X_j$ be $f(\bb\trans\X_j)$. We have 
\bse 
&&E\left\{ H(\X_j,A_j,Y_j) K_h(\bb\trans\X_j -\bb\trans\x)\right\} \\
&=& E\left[E\left\{ H(\X_j,A_j,Y_j) \mid \bb\trans\X_j\right\} K_h(\bb\trans\X_j -\bb\trans\x) \right] \\
&=& \int E\left\{ H(\X_j,A_j,Y_j) \mid \bb\trans\x_j \right\} 
h^{-1} K\{(\bb\trans\x_j -\bb\trans\x)/h\} f(\bb\trans\x_j) d(\bb\trans\x_j) \\
&=& \int E\left\{ H(\X_j,A_j,Y_j) \mid \bb\trans\x + zh \right\} 
K(z) f(\bb\trans\x + zh) dz\\
&=& E\left\{ H(\X_j,A_j,Y_j) \mid \bb\trans\x \right\}
f(\bb\trans\x) \\
&&+ \frac{\partial^2}{(\partial \bb\trans\x)^m} [E\left\{ H(\X_j,A_j,Y_j) \mid \bb\trans\x \right\}
f(\bb\trans\x)]|_{\bb\trans\x} \frac{h^2}{2} \int z^2 K(z) dz +
o(h^2)\\
&=& E\left\{ H(\X_j,A_j,Y_j) \mid \bb\trans\x \right\}
f(\bb\trans\x) \\
&&+ \frac{\partial^2}{(\partial \bb\trans\x)^m} [E\left\{ H(\X_j,A_j,Y_j) \mid \bb\trans\x \right\}
f(\bb\trans\x)]\frac{h^2}{2} \int z^2 K(z) dz +
o(h^2). 
\ese 
In addition, 
\bse 
&&\var\left\{ n^{-1} \sumj H(\X_j,A_j,Y_j) K_h(\bb\trans\X_j -\bb\trans\x) \right\} \\
&=&  n^{-2} \sumj \var\left\{H(\X_j,A_j,Y_j) K_h(\bb\trans\X_j -\bb\trans\x) \right\} \\
&=&  n^{-1} \var\left\{H(\X_j,A_j,Y_j) K_h(\bb\trans\X_j 
  -\bb\trans\x) \right\} \\
&=& n^{-1} E\left[\{ H(\X_j,A_j,Y_j) K_h(\bb\trans\X_j 
  -\bb\trans\x) \}^2\right] - n^{-1} \left[E\{ H(\X_j,A_j,Y_j) K_h(\bb\trans\X_j 
  -\bb\trans\x)\} \right]^2  \\
&=& n^{-1} \int E\left\{ H^2(\X_j,A_j,Y_j) \mid \bb\trans\x_j \right\} K_h^2(\bb\trans\x_j - \bb\trans\x) 
f(\bb\trans\x_i)d\mu(\bb\trans\x_i) \\
&& - n^{-1} \left[E\left\{ H(\X_j,A_j,Y_j) \mid \bb\trans\x \right\}
f(\bb\trans\x) +O(h^2)\right]^2\\
&=& n^{-1} \int E\left\{ H^2(\X_j,A_j,Y_j) \mid \bb\trans\x + zh \right\} h^{-1}K^2(z)f(\bb\trans\x + zh)dz- O(n^{-1})  \\
&=& (nh)^{-1}E\{H^2(\X_j,A_j,Y_j) \mid \bb\trans\x\}f(\bb\trans\x) 
\int K^2(z) dz+O(n^{-1}). 
\ese 
\qed

{\subsection{Proof of Lemma \ref{lem:rateQ}}
Let the probability density function of 
$\bb\trans\X_j$ be $f(\bb\trans\X_j)$. Denote 
\bse 
H_1(\X_j,A_j,Y_j,\bg_0,\ba_0) &=& \frac{\{A_j-\pi(\X_j,\bg_0)\}\{Y_j-\mu(\X_j,\ba_0)\}}{\pi(\X_j,\bg_0)\{1-\pi(\X_j,\bg_0)\}}, \\
H_2(\X_j,A_j,\bg_0) &=& \frac{A_j}{\pi(\X_j,\bg_0)}. 
\ese 
Then, 
\be\label{eq:expectH1}
&&E\left\{ H_1(\X_j,A_j,Y_j,\bg_0,\ba_0)\mid \bb\trans\X_j\right\} \n\\
&=&E\left(E\left[
      \frac{\{A_j-\pi(\X_j,\bg_0)\}\{Y_j-\mu(\X_j,\ba_0)\}}{\pi(\X_j,\bg_0)\{1-\pi(\X_j,\bg_0)\}}
      \mid\X_j \right]\mid \bb\trans\X_j\right) \n\\
&=&E\left\{E\left(
      \frac{\{A_j-\pi(\X_j,\bg_0)\}[A_j\{Y_{1j}-\mu(\X_j,\ba_0)\}+(1-A_j)\{Y_{0j}-\mu(\X_j,\ba_0)\}]}{\pi(\X_j,\bg_0) 
      \{1-\pi(\X_j,\bg_0)\}}
      \mid\X_j \right)\mid \bb\trans\X_j\right\} \n\\
&=&E\bigg[
      \frac{\{1-\pi(\X_j,\bg_0)\}\pi_0(\X_j)\{Q(\bb\trans\X_j)+\mu_0(\X_j)-\mu(\X_j,\ba_0)\}}
      {\pi(\X_j,\bg_0)\{1-\pi(\X_j,\bg_0)\}}  \n\\
&& -\frac{\pi(\X_j,\bg_0)\{1-\pi_0(\X_j)\}\{\mu_0(\X_j)-\mu(\X_j,\ba_0)\}}{\pi(\X_j,\bg_0)\{1-\pi(\X_j,\bg_0)\}}
      \mid \bb\trans\X_j\bigg] \n\\
&=& E\bigg[ \frac{\pi_0(\X_j)}{\pi(\X_j,\bg_0)}Q(\bb\trans\X_j) +
 \{\mu_0(\X_j)-\mu(\X_j,\ba_0)\}\left\{\frac{\pi_0(\X_j)}{\pi(\X_j,\bg_0)} - \frac{1-\pi_0(\X_j)}{1-\pi(\X_j,\bg_0)} \right\} 
 \mid \bb\trans\X_j\bigg] \n\\
&=& E\left\{ \frac{\pi_0(\X_j)}{\pi(\X_j,\bg_0)} \mid \bb\trans\X_j \right\} Q(\bb\trans\X_j) , 
\ee 
where the last equality follows because at least one of the two models, $\pi(\X_j,\bg_0)$ and $\mu(\X_j,\ba_0)$ 
is correctly specified. Further, 
\be\label{eq:expectH2}
E\left\{ H_2(\X_j,A_j,\bg_0)\mid \bb\trans\X_j \right\}
&=& E\left[E\left\{ \frac{A_j}{\pi(\X_j,\bg_0)} \mid\X_j \right\} \mid \bb\trans\X_j\right] \n\\
&=& E\left\{\frac{\pi_0(\X_j)}{\pi(\X_j,\bg_0)}\mid \bb\trans\X_j\right\}. 
\ee 
Thus,  Lemma \ref{lem:order} and (\ref{eq:expectH1}) lead to 
\bse 
&&E\left\{ H_1(\X_j,A_j,Y_j,\bg_0,\ba_0) K_h(\bb\trans\X_j -\bb\trans\x_i) \right\} \n\\
&=& E\left\{ H_1(\X_j,A_j,Y_j,\bg_0,\ba_0)\mid \bb\trans\x_i\right\} f(\bb\trans\x_i) 
 + O(h^2) \n\\
&=& E\left\{ \frac{\pi_0(\X_j)}{\pi(\X_j,\bg_0)} \mid \bb\trans\x_i \right\} Q(\bb\trans\x_i) f(\bb\trans\x_i) + O(h^2), 
\ese 
and Lemma \ref{lem:order} and (\ref{eq:expectH2}) lead to 
\bse 
&&E\left\{ H_2(\X_j,A_j,\bg_0) K_h(\bb\trans\X_j -\bb\trans\x_i)\right\} \n\\
&=& E\left\{ H_2(\X_j,A_j,\bg_0)\mid\bb\trans\x_i \right\} f(\bb\trans\x_i) + O(h^2) \n\\
&=& E\left\{ \frac{\pi_0(\X_j)}{\pi(\X_j,\bg_0)} \mid \bb\trans\x_i \right\}  f(\bb\trans\x_i) + O(h^2). 
\ese 
Applying Lemma \ref{lem:order}, we obtain 
\be 
&&n^{-1}\sumj H_1(\X_j,A_j,Y_j,\bg_0,\ba_0)K_h(\bb\trans\X_j - \bb\trans\x) - E\left\{ \frac{\pi_0(\X_j)}{\pi(\X_j,\bg_0)} \mid \bb\trans\x\right\} Q(\bb\trans\x) f(\bb\trans\x)\n\\
& =& O_p(h^2 + n^{-1/2}h^{-1/2}),\label{eq:prop_Q1}\\
&&n^{-1}\sumj H_2(\X_j,A_j,\bg_0)K_h(\bb\trans\X_j - \bb\trans\x) - 
E\left\{ \frac{\pi_0(\X_j)}{\pi(\X_j,\bg_0)} \mid \bb\trans\x \right\} f(\bb\trans\x)\n\\
&=& O_p(h^2 + n^{-1/2}h^{-1/2}).\label{eq:prop_Q2}
\ee 
Combining these results, we get 
\bse 
&&\wt Q(\bb\trans\x,\wh\bb,\wh\ba,\wh\bg) - Q(\bb\trans\x) \\
&=& \frac{n^{-1}\sumj H_1(\X_j,A_j,Y_j,\bg_0,\ba_0)K_h(\bb\trans\X_j - \bb\trans\x)}{
n^{-1} \sumj H_2(\X_j,A_j,\bg_0)K_h(\bb\trans\X_j - \bb\trans\x)}
- 
\frac{E\{\pi_0(\X_j)/ \pi(\X_j,\bg_0) \mid \bb\trans\x\}Q(\bb\trans\x)f(\bb\trans\x) }
{E\{\pi_0(\X_j)/ \pi(\X_j,\bg_0) \mid \bb\trans\x\}f(\bb\trans\x) } \\
&=&  \frac{n^{-1}\sumj H_1(\X_j,A_j,Y_j,\bg_0,\ba_0)K_h(\bb\trans\X_j - \bb\trans\x) - 
E\{\pi_0(\X_j)/ \pi(\X_j,\bg_0) \mid \bb\trans\x\}Q(\bb\trans\x)f(\bb\trans\x) }
{E\{\pi_0(\X_j)/ \pi(\X_j,\bg_0) \mid \bb\trans\x\}f(\bb\trans\x)} \\
&& - \frac{E\{\pi_0(\X_j)/ \pi(\X_j,\bg_0) \mid \bb\trans\x\}Q(\bb\trans\x)f(\bb\trans\x)}
{[E\{\pi_0(\X_j)/ \pi(\X_j,\bg_0) \mid \bb\trans\x\}f(\bb\trans\x)]^2} \bigg[
n^{-1} \sumj H_2(\X_j,A_j,\bg_0)K_h(\bb\trans\X_j - \bb\trans\x)   \\
&&  - E\{\pi_0(\X_j)/ \pi(\X_j,\bg_0) \mid \bb\trans\x\}f(\bb\trans\x) \bigg] 
+ O_p(h^4+n^{-1}h^{-1}) \\
&=& O_p(h^2 +n^{-1/2}h^{-1/2}) + O_p(h^4+n^{-1}h^{-1}) \\
&=& O_p(h^2 +n^{-1/2}h^{-1/2}), 
\ese 
where 
the third equality is due to (\ref{eq:prop_Q1}) and (\ref{eq:prop_Q2}). 
\qed

\subsection{Proof of Theorem \ref{th:beta}} 
Let the probability density function of 
$\bb\trans\X$ be $f(\bb\trans\x)$. 
Using Lemma \ref{lem:rateQ}, we have 
$\wt Q(\bb\trans\x,\bb,\wh\ba,\wh\bg)-Q(\bb\trans\x)=O_p\{h^2+(n_2h)^{-1/2}\}$. 
Further, by applying Lemma \ref{lem:order}
 with $H(\X_j,A_j,Y_j)=\X_{Lj}$, we get 
\be\label{eq:expectX1}
\frac{1}{n} \sumj \X_{Lj} K_h(\bb\trans\x_j -\bb\trans\x_i) - E(\X_{Li}\mid\bb\trans\x_i)f(\bb\trans\x_i) 
&=& O_p(h^2 + n^{-1/2}h^{-1/2}) 
\ee 
and with $H(\X_j,A_j,Y_j)=1$, we get 
\be\label{eq:expectX2}
\frac{1}{n} \sumj K_h(\bb\trans\x_j -\bb\trans\x_i) - f(\bb\trans\x_i) 
&=& O_p(h^2 + n^{-1/2}h^{-1/2}). 
\ee 
Combining these results, we get 
\be\label{eq:orderEx}
&&\wh E(\X_{Lj}\mid\bb\trans\x) - E(\X_{Lj} \mid\bb\trans\x) \n\\
&=& \frac{n^{-1}\sumj \X_{Lj} K_h(\bb\trans\X_j - \bb\trans\x)}{
n^{-1} \sumj K_h(\bb\trans\X_j - \bb\trans\x)}
- \frac{E(\X_{Lj} \mid \bb\trans\x)f(\bb\trans\x) }{f(\bb\trans\x) } \n\\
&=&  \frac{n^{-1}\sumj \X_{Lj} K_h(\bb\trans\X_j - \bb\trans\x) - 
E(\X_{Lj} \mid \bb\trans\x)f(\bb\trans\x) }
{f(\bb\trans\x)} \n\\
&& - \frac{E(\X_{Lj}\mid \bb\trans\x)f(\bb\trans\x)}
{\{f(\bb\trans\x)\}^2} \bigg\{
n^{-1} \sumj K_h(\bb\trans\X_j - \bb\trans\x) - f(\bb\trans\x) \bigg\}
+ O_p(h^4+n^{-1}h^{-1}) \n\\
&=& O_p(h^2 +n^{-1/2}h^{-1/2}) + O_p(h^4+n^{-1}h^{-1}) \n\\
&=& O_p(h^2 +n^{-1/2}h^{-1/2}). 
\ee 
The estimating equation (\ref{eq:beta}) satisfies 
\be\label{eq:expandmain} 
\0 &=& \frac{1}{\sqrt{n}} \sumi \left[ \frac{\{A_i - \pi(\x_i,\wh\bg)\}\{Y_i - \mu(\x_i,\wh\ba)\}}{\pi(\x_i,\wh\bg)\{1-\pi(\x_i,\wh\bg)\}} 
+ \left\{1 - \frac{A_i}{\pi(\x_i,\wh\bg)}\right\}\wt
Q(\wh\bb\trans\x_i,\wh\bb,\wh\ba,\wh\bg) \right]\n\\
&&\times \left\{\x_{Li} - \wh E(\X_{Li} \mid \wh\bb\trans\x_i) \right\} \n\\
&=& \frac{1}{\sqrt{n}} \sumi \left[ \frac{\{A_i - \pi(\x_i,\bg_0)\}\{Y_i - \mu(\x_i,\ba_0)\}}{\pi(\x_i,\bg_0)\{1-\pi(\x_i,\bg_0)\}} 
+ \left\{1 - \frac{A_i}{\pi(\x_i,\bg_0)}\right\}\wt
Q(\bb_0\trans\x_i,\bb_0,\ba_0,\bg_0) \right]\n\\
&&\times \left\{\x_{Li} - \wh E(\X_{Li} \mid \bb_0\trans\x_i) \right\}  \n\\
&& + \frac{1}{n} \sumi \frac{\partial}{\partial \bb\trans} \left.\left(\left[ \frac{\{A_i - \pi(\x_i,\bg)\}\{Y_i - \mu(\x_i,\ba)\}}{\pi(\x_i,\bg)\{1-\pi(\x_i,\bg)\}} 
+ \left\{1 - \frac{A_i}{\pi(\x_i,\bg)}\right\}\wt Q(\bb\trans\x_i,\bb,\ba,\bg) \right] \right.\right. \n\\
&&\left.\left.\times\left\{\x_{Li} - \wh E(\X_{Li} \mid \bb\trans\x_i) \right\} \right) \right |_{\bb=\wt\bb,\bg=\wt\bg,\ba=\wt\ba} \sqrt{n}(\wh\bb_L-\bb_{L0}) \n\\
&& + \frac{1}{n} \sumi \frac{\partial}{\partial \bg\trans} \left.\left(\left[ \frac{\{A_i - \pi(\x_i,\bg)\}\{Y_i - \mu(\x_i,\ba)\}}{\pi(\x_i,\bg)\{1-\pi(\x_i,\bg)\}} 
+ \left\{1 - \frac{A_i}{\pi(\x_i,\bg)}\right\}\wt Q(\bb\trans\x_i,\bb,\ba,\bg) \right] \right.\right. \n\\
&&\left.\left.\times\left\{\x_{Li} - \wh E(\X_{Li} \mid \bb\trans\x_i) \right\} \right) \right |_{\bb=\wt\bb,\bg=\wt\bg,\ba=\wt\ba} \sqrt{n}(\wh\bg-\bg_0) \n\\
&& + \frac{1}{n} \sumi \frac{\partial}{\partial \ba\trans} \left.\left(\left[ \frac{\{A_i - \pi(\x_i,\bg)\}\{Y_i - \mu(\x_i,\ba)\}}{\pi(\x_i,\bg)\{1-\pi(\x_i,\bg)\}} 
+ \left\{1 - \frac{A_i}{\pi(\x_i,\bg)}\right\}\wt Q(\bb\trans\x_i,\bb,\ba,\bg) \right] \right.\right. \n\\
&&\left.\left.\times\left\{\x_{Li} - \wh E(\X_{Li} \mid \bb\trans\x_i) \right\} \right) \right |_{\bb=\wt\bb,\bg=\wt\bg,\ba=\wt\ba} \sqrt{n}(\wh\ba-\ba_0) \n\\
&=&\frac{1}{\sqrt{n}} \sumi \left[ \frac{\{A_i - \pi(\x_i,\bg_0)\}\{Y_i - \mu(\x_i,\ba_0)\}}{\pi(\x_i,\bg_0)\{1-\pi(\x_i,\bg_0)\}} 
+ \left\{1 - \frac{A_i}{\pi(\x_i,\bg_0)}\right\}\wt Q(\bb_0\trans\x_i,\bb_0,\ba_0,\bg_0) \right] \n\\
&&\times\left\{\x_{Li} - \wh E(\X_{Li} \mid \bb_0\trans\x_i) \right\}\n\\
&& + \left[E\left\{ \frac{\partial}{\partial \bb_0\trans} \left(\left[ \frac{\{A_i - \pi(\X_i,\bg_0)\}\{Y_i - \mu(\X_i,\ba_0)\}}{\pi(\X_i,\bg_0) 
\{1-\pi(\X_i,\bg_0)\}} 
+ \left\{1 - \frac{A_i}{\pi(\X_i,\bg_0)}\right\}\wt Q(\bb_0\trans\X_i,\bb_0,\ba_0,\bg_0) \right] \right.\right. \right.\n\\
&&\left.\left.\left.\times\left\{\X_{Li} - \wh E(\X_{Li} \mid \bb_0\trans\X_i) \right\} \right) \right\}+o_p(1)\right]
 \sqrt{n}(\wh\bb_L-\bb_{L0})\n\\
&& + \left[E\left\{ \frac{\partial}{\partial \bg_0\trans} \left(\left[ \frac{\{A_i - \pi(\X_i,\bg_0)\}\{Y_i - \mu(\X_i,\ba_0)\}}{\pi(\X_i,\bg_0) 
\{1-\pi(\X_i,\bg_0)\}} 
+ \left\{1 - \frac{A_i}{\pi(\X_i,\bg_0)}\right\}\wt Q(\bb_0\trans\X_i,\bb_0,\ba_0,\bg_0) \right] \right.\right. \right.\n\\
&&\left.\left.\left.\times\left\{\X_{Li} - \wh E(\X_{Li}  \mid \bb_0\trans\X_i) \right\} \right) \right\}+o_p(1)\right]
 \sqrt{n}(\wh\bg-\bg_0)\n\\
&& + \left[E\left\{ \frac{\partial}{\partial \ba_0\trans} \left(\left[ \frac{\{A_i - \pi(\X_i,\bg_0)\}\{Y_i - \mu(\X_i,\ba_0)\}}{\pi(\X_i,\bg_0) 
\{1-\pi(\X_i,\bg_0)\}} 
+ \left\{1 - \frac{A_i}{\pi(\X_i,\bg_0)}\right\}\wt Q(\bb_0\trans\X_i,\bb_0,\ba_0,\bg_0) \right] \right.\right. \right.\n\\
&&\left.\left.\left.\times\left\{\X_{Li} -\wh E(\X_{Li} \mid \bb_0\trans\X_i) \right\} \right) \right\}+o_p(1)\right]
 \sqrt{n}(\wh\ba-\ba_0)\n\\
&=& 
\frac{1}{\sqrt{n}}\sumi\bpsi_\bb(\X_i,Y_i,A_i,\bb_0,\ba_0,\bg_0)+\T_1+\T_2\n\\
&& +\{\T_\bb+o_p(1)\}\sqrt{n}(\wh\bb_L-\bb_{L0}) + \{\T_{\bg}+o_p(1)\}\sqrt{n}(\wh\bg-\bg_0) 
+\{\T_{\ba}+o_p(1)\}\sqrt{n}(\wh\ba-\ba_0). 
\ee 
Here $\wt\bb$, $\wt\bg$ and $\wt\ba$ are on the line connecting
$\wh\bb_L$ and $\bb_{L0}$, $\wh\bg$ and $\bg_0$ 
and $\wh\ba$ and $\ba_0$, $\T_\bb+o_p(1)$, $\T_\bg+o_p(1)$, $\T_\ba+o_p(1)$
represent the multipliers in front of $\sqrt{n}(\wh\bb-\bb_0)$, 
$\sqrt{n}(\wh\bg-\bg_0)$, $\sqrt{n}(\wh\ba-\ba_0)$ respectively, and
\bse
\T_1&\equiv&\frac{1}{\sqrt{n}} \sumi
 \left\{1 - \frac{A_i}{\pi(\x_i,\bg_0)}\right\} \left\{ \wt Q(\bb_0\trans\x_i,\bb_0,\ba_0,\bg_0) - Q(\bb_0\trans\x_i) \right\} 
 \left\{\x_{Li} - E(\X_{Li} \mid \bb_0\trans\x_i) \right\},\\
\T_2 &\equiv& \frac{1}{\sqrt{n}} \sumi \left[ \frac{\{A_i - \pi(\x_i,\bg_0)\}\{Y_i - \mu(\x_i,\ba_0)\}}{\pi(\x_i,\bg_0)\{1-\pi(\x_i,\bg_0)\}} 
+ \left\{1 - \frac{A_i}{\pi(\x_i,\bg_0)}\right\}  Q(\bb_0\trans\x_i)\right] \\
&& \times \left\{ E(\X_{Li} \mid \bb_0\trans\x_i) - \wh E(\X_{Li} \mid \bb_0\trans\x_i) \right\}.
\ese
 and we used the fact that
$O_p(n^{1/2}h^4+n^{-1/2}h^{-1})=o_p(1)$ due to 
Condition \ref{con:C5}. Using (\ref{eq:q}),
we further have
\bse
\T_1&=&\frac{1}{\sqrt{n}} \sumi
 \left\{1 - \frac{A_i}{\pi(\x_i,\bg_0)}\right\} \left\{ \wt Q(\bb_0\trans\x_i,\bb_0,\ba_0,\bg_0) - Q(\bb_0\trans\x_i) \right\} 
 \left\{\x_{Li} - E(\X_{Li} \mid \bb_0\trans\x_i) \right\}\\
&=&
\frac{1}{\sqrt{n}} \sumi \left\{\x_{Li} - E(\X_{Li} \mid \bb_0\trans\x_i) \right\}
 \left\{1 - \frac{A_i}{\pi(\x_i,\bg_0)}\right\} 
\\
&&\times\left\{
\frac{\sumj K_h (\bb_0\trans\X_j - \bb_0\trans\X_i) \{A_j - \pi(\X_j,\bg_0)\}\{ Y_j - \mu(\X_j,\ba_0) \} / [ \pi(\X_j,\bg_0)\{ 1- \pi(\X_j,\bg_0)\}  ] }
{\sumj K_h (\bb_0\trans\X_j - \bb_0\trans\X_i) A_j / \pi(\X_j,\bg_0)}\right.\\
&&\left.- Q(\bb_0\trans\x_i) \right\} \\
&=&
\frac{1}{n^{3/2}} \sumi \left\{\x_{Li} - E(\X_{Li} \mid \bb_0\trans\x_i) \right\}
 \left\{1 - \frac{A_i}{\pi(\x_i,\bg_0)}\right\} \\
&&\times\left\{
\frac
{\sumj K_h (\bb_0\trans\X_j - \bb_0\trans\X_i) \{A_j - \pi(\X_j,\bg_0)\}\{ Y_j - \mu(\X_j,\ba_0) \} / [ \pi(\X_j,\bg_0)\{ 1- \pi(\X_j,\bg_0)\}  ] }{f(\bb_0\trans\X_i) E\{\pi_0(\X_i) /
  \pi(\X_i,\bg_0)\mid\bb\trans\X_i)}\right.\\
&&\left.-\frac{
Q(\bb_0\trans\X_i)
(\sumj K_h (\bb_0\trans\X_j - \bb_0\trans\X_i) \{A_j/ \pi(\X_j,\bg_0)\})}{f(\bb_0\trans\X_i) E\{\pi_0(\X_i) /
  \pi(\X_i,\bg_0)\mid\bb\trans\X_i\}}\right\}+o_p(1)\\
&=&
\frac{1}{n^{3/2}} \sumi\sumj \left\{\x_{Li} - E(\X_{Li} \mid \bb_0\trans\x_i) \right\}
 \left\{1 - \frac{A_i}{\pi(\x_i,\bg_0)}\right\}
\frac{K_h (\bb_0\trans\X_j - \bb_0\trans\X_i)}{f(\bb_0\trans\X_i) E\{\pi_0(\X_i) /
  \pi(\X_i,\bg_0)\mid\bb\trans\X_i\}}
 \\
&&\times\left\{\frac{
 \{A_j - \pi(\X_j,\bg_0)\}\{ Y_j - \mu(\X_j,\ba_0) \} }{  \pi(\X_j,\bg_0)\{ 1- \pi(\X_j,\bg_0)\}  }
- \frac{Q(\bb_0\trans\X_i) A_j}{ \pi(\X_j,\bg_0)}\right\}
+o_p(1)\\
&=&
\frac{1}{n^{1/2}} \sumj\frac{
E\left[\left\{\x_{Lj} - E(\X_{Lj} \mid \bb_0\trans\x_j) \right\}
 \left\{1 - \pi_0(\X_j)/\pi(\x_j,\bg_0)\right\}\mid\bb_0\trans\x_j\right]}{E\{\pi_0(\X_j) /
  \pi(\X_j,\bg_0)\mid\bb\trans\x_j\}}
 \\
&&\times\left\{\frac{
 \{A_j - \pi(\X_j,\bg_0)\}\{ Y_j - \mu(\X_j,\ba_0) \} }{  \pi(\X_j,\bg_0)\{ 1- \pi(\X_j,\bg_0)\}  }
- \frac{Q(\bb_0\trans\X_j) A_j}{ \pi(\X_j,\bg_0)}\right\}
+o_p(1).
\ese
Similarly, 
\bse
\T_2
&=&\frac{1}{\sqrt{n}} \sumi \left[ \frac{\{A_i - \pi(\x_i,\bg_0)\}\{Y_i - \mu(\x_i,\ba_0)\}}{\pi(\x_i,\bg_0)\{1-\pi(\x_i,\bg_0)\}} 
+ \left\{1 - \frac{A_i}{\pi(\x_i,\bg_0)}\right\}  Q(\bb_0\trans\x_i)\right] \\
&& \times \left\{ E(\X_{Li} \mid \bb_0\trans\x_i) - \wh E(\X_{Li} \mid
  \bb_0\trans\x_i) \right\}\\
&=&\frac{1}{\sqrt{n}} \sumi \left[ \frac{\{A_i - \pi(\x_i,\bg_0)\}\{Y_i - \mu(\x_i,\ba_0)\}}{\pi(\x_i,\bg_0)\{1-\pi(\x_i,\bg_0)\}} 
+ \left\{1 - \frac{A_i}{\pi(\x_i,\bg_0)}\right\}  Q(\bb_0\trans\x_i)\right] \\
&& \times \left\{ E(\X_{Li} \mid \bb_0\trans\x_i) - 
\frac{\sumj\x_{Lj}K_h(\bb_0\trans\x_j-\bb_0\trans\x_i)}
{\sumj K_h(\bb_0\trans\x_j-\bb_0\trans\x_i)} \right\}\\
&=&\frac{1}{n^{3/2}} \sumi \left[ \frac{\{A_i - \pi(\x_i,\bg_0)\}\{Y_i - \mu(\x_i,\ba_0)\}}{\pi(\x_i,\bg_0)\{1-\pi(\x_i,\bg_0)\}} 
+ \left\{1 - \frac{A_i}{\pi(\x_i,\bg_0)}\right\}  Q(\bb_0\trans\x_i)\right] \\
&& \times \left[- 
\frac{\sumj\x_{Lj}K_h(\bb_0\trans\x_j-\bb_0\trans\x_i)}
{f(\bb_0\trans\x_i)} 
+\frac{E(\X_{Li} \mid \bb_0\trans\x_i)}{f(\bb_0\trans\x_i)}
\{\sumj K_h(\bb_0\trans\x_j-\bb_0\trans\x_i)\}
\right]+o_p(1)\\
&=&\frac{1}{n^{3/2}} \sumi\sumj \left[ \frac{\{A_i - \pi(\x_i,\bg_0)\}\{Y_i - \mu(\x_i,\ba_0)\}}{\pi(\x_i,\bg_0)\{1-\pi(\x_i,\bg_0)\}} 
+ \left\{1 - \frac{A_i}{\pi(\x_i,\bg_0)}\right\}  Q(\bb_0\trans\x_i)\right] \\
&& \times \left\{E(\X_{Li} \mid \bb_0\trans\x_i) -\x_{Lj}
\right\}\frac{K_h(\bb_0\trans\x_j-\bb_0\trans\x_i)}{f(\bb_0\trans\x_i)}+o_p(1)\\
&=&\frac{1}{n^{1/2}} \sumj 
Q(\bb_0\trans\x_j)\left\{E(\X_{Lj} \mid \bb_0\trans\x_j) -\x_{Lj}
\right\}+o_p(1).
\ese
We now derive the expression of $\T_\bb, \T_\ba, \T_\bg$. First,
\bse
&&\frac{\partial}{\partial \bb_L\trans}\wh E(\X_L \mid
\bb\trans\x)\\
&=&
\frac{\partial}{\partial \bb_L\trans}\frac{n^{-1}\sumj \X_{Lj} K_h(\bb\trans\X_j - \bb\trans\x)}{n^{-1} \sumj K_h(\bb\trans\X_j - \bb\trans\x)}\\
&=&\frac{n^{-1}\sumj \X_{Lj}
  K'\{(\bb\trans\X_j - \bb\trans\x)/h\}
(\X_{Lj} - \x_L)\trans
}{h^2f(\bb\trans\x)}\\
&&-
\frac{E(\X_L\mid\bb\trans\x)
n^{-1}\sumj 
  K'\{(\bb\trans\X_j - \bb\trans\x)/h\}
(\X_{Lj} - \x_L)\trans
}{h^2f(\bb\trans\x)}+o_p(1)\\
&=&n^{-1}\sumj \frac{\{\X_{Lj}-E(\X_L\mid\bb\trans\x)\}(\X_{Lj} - \x_L)\trans
  K'\{(\bb\trans\X_j - \bb\trans\x)/h\}
}{h^2f(\bb\trans\x)}+o_p(1)\\
&=&n^{-1}\sumj \frac{E[\{\X_{Lj}-E(\X_L\mid\bb\trans\x)\}(\X_{Lj} - \x_L)\trans\mid\bb\trans\X_j]
  K'\{(\bb\trans\X_j - \bb\trans\x)/h\}
}{h^2f(\bb\trans\x)}+o_p(1)\\
&=&\frac{-1}{f(\bb\trans\x)}\frac{\partial\left( E[\{\X_L-E(\X_L\mid\bb\trans\x)\}^{\otimes2}\mid\bb\trans\x] f(\bb\trans\x)\right)
}{\partial(\bb\trans\x)}+o_p(1).
\ese
In addition, 
\be\label{eq:Qbeta}
&&\frac{\partial}{\partial \bb_L}
\wt Q(\bb\trans\x,\bb,\ba,\bg)\\
&=&\frac{\partial}{\partial \bb_L}\frac{\sumi K_h (\bb\trans\X_i - \bb\trans\x) \{A_i - \pi(\X_i,\bg)\}\{ Y_i - \mu(\X_i,\ba) \} / [ \pi(\X_i,\bg)\{ 1- \pi(\X_i,\bg)\}  ] }
{\sumi K_h (\bb\trans\X_i - \bb\trans\x) A_i / \pi(\X_i,\bg)}\n\\
&=&n^{-1}\sumi \frac{K'\{(\bb\trans\X_i - \bb\trans\x)/h\}
Q(\bb\trans\X_i) E\{\pi_0(\X_i) / \pi(\X_i,\bg)\mid\bb\trans\X_i\} (\X_{Li}-\x_L)}
{h^2f(\bb\trans\x) E\{\pi_0(\X) / \pi(\X,\bg)\mid\bb\trans\x\}}\n\\
&&-n^{-1}\sumi
\frac{Q(\bb\trans\x)\pi_0(\X_i) (\X_{Li} -\x_L) K'\{(\bb\trans\X_i -
  \bb\trans\x)/h\} }{h^2\pi(\X_i,\bg) f(\bb\trans\x) E\{\pi_0(\X) /
  \pi(\X,\bg)\mid\bb\trans\x\}}+o_p(1)\n\\
&=&n^{-1}\sumi 
\frac{K'\{(\bb\trans\X_i - \bb\trans\x)/h\}}{h^2f(\bb\trans\x)
  E\{\pi_0(\X) / \pi(\X,\bg)\mid\bb\trans\x\}}\n\\
&&\times E\left((\X_{Li}-\x_L)
\left[
Q(\bb\trans\X_i) E\left\{\frac{\pi_0(\X_i)}{ \pi(\X_i,\bg)}\mid\bb\trans\X_i\right\} 
-
\frac{Q(\bb\trans\x)\pi_0(\X_i)}{\pi(\X_i,\bg) }\right]\mid\bb\trans\X_i\right)
+o_p(1)\n\\
&=&
\frac{1}{f(\bb\trans\x)
  E\{\pi_0(\X) / \pi(\X,\bg)\mid\bb\trans\x\}}\n\\
&&\times \frac{\partial}{\partial\bb\trans\x}\left\{E\left(\X_L
\left[\frac{\pi_0(\X)}{\pi(\X,\bg) }-
E\left\{\frac{\pi_0(\X)}{ \pi(\X,\bg)}\mid\bb\trans\x\right\} 
\right]\mid\bb\trans\x\right) Q(\bb\trans\x) f(\bb\trans\x)\right\}
+o_p(1),\n
\ee
\be\label{eq:Qalpha}
&&\frac{\partial}{\partial \ba}
\wt Q(\bb\trans\x,\bb,\ba,\bg)\\
&=&\frac{\partial}{\partial \ba}\frac{\sumi K_h (\bb\trans\X_i - \bb\trans\x) \{A_i - \pi(\X_i,\bg)\}\{ Y_i - \mu(\X_i,\ba) \} / [ \pi(\X_i,\bg)\{ 1- \pi(\X_i,\bg)\}  ] }
{\sumi K_h (\bb\trans\X_i - \bb\trans\x) A_i / \pi(\X_i,\bg)}\n\\
&=&-\frac{\sumi K_h (\bb\trans\X_i - \bb\trans\x) \{A_i - \pi(\X_i,\bg)\} \bmu_\ba(\X_i,\ba) / [ \pi(\X_i,\bg)\{ 1- \pi(\X_i,\bg)\}  ] }
{\sumi K_h (\bb\trans\X_i - \bb\trans\x) A_i / \pi(\X_i,\bg)}\n\\
&=&-\frac{E(
\{\pi_0(\X) - \pi(\X,\bg)\} \bmu_\ba(\X,\ba) / [ \pi(\X,\bg)\{ 1- \pi(\X,\bg)\}  ] \mid\bb\trans\X)}
{E\{\pi_0(\X) / \pi(\X,\bg)\mid\bb\trans\x\}}+o_p(1),\n
\ee
and
\be\label{eq:Qgamma}
&&\frac{\partial}{\partial \bg}
\wt Q(\bb\trans\x,\bb,\ba,\bg)\\
&=&\frac{\partial}{\partial \bg}\frac{\sumi K_h (\bb\trans\X_i - \bb\trans\x) \{A_i - \pi(\X_i,\bg)\}\{ Y_i - \mu(\X_i,\ba) \} / [ \pi(\X_i,\bg)\{ 1- \pi(\X_i,\bg)\}  ] }
{\sumi K_h (\bb\trans\X_i - \bb\trans\x) A_i / \pi(\X_i,\bg)}\n\\
&=&-\frac{\sumi K_h (\bb\trans\X_i - \bb\trans\x)\bpi_\bg(\X_i,\bg)\{ Y_i - \mu(\X_i,\ba) \} / [ \pi(\X_i,\bg)\{ 1- \pi(\X_i,\bg)\}  ] }
{\sumi K_h (\bb\trans\X_i - \bb\trans\x) A_i / \pi(\X_i,\bg)}\n\\
&&-\frac{\sumi K_h (\bb\trans\X_i - \bb\trans\x) \bpi_\bg(\X_i,\bg)\{A_i - \pi(\X_i,\bg)\}\{ Y_i - \mu(\X_i,\ba) \} / [ \pi^2(\X_i,\bg)\{ 1- \pi(\X_i,\bg)\}  ] }
{\sumi K_h (\bb\trans\X_i - \bb\trans\x) A_i / \pi(\X_i,\bg)}\n\\
&&+\frac{\sumi K_h (\bb\trans\X_i - \bb\trans\x) \bpi_\bg(\X_i,\bg)\{A_i - \pi(\X_i,\bg)\}\{ Y_i - \mu(\X_i,\ba) \} / [ \pi(\X_i,\bg)\{ 1- \pi(\X_i,\bg)\}^2  ] }
{\sumi K_h (\bb\trans\X_i - \bb\trans\x) A_i / \pi(\X_i,\bg)}\n\\
&&+\frac{\sumi K_h (\bb\trans\X_i - \bb\trans\x) \{A_i - \pi(\X_i,\bg)\}\{ Y_i - \mu(\X_i,\ba) \} / [ \pi(\X_i,\bg)\{ 1- \pi(\X_i,\bg)\}  ] }
{\{\sumi K_h (\bb\trans\X_i - \bb\trans\x) A_i / \pi(\X_i,\bg)\}^2}\n\\
&&\times \{\sumi K_h (\bb\trans\X_i - \bb\trans\x) A_i
\bpi_\bg(\X_i,\bg)/ \pi^2(\X_i,\bg)\}+o_p(1)\n\\
&=&-\frac{E(\bpi_\bg(\X_i,\bg)\{
Q(\bb\trans\x)\pi_0(\X)+\mu_0(\X_i)
 - \mu(\X_i,\ba) \} / [ \pi(\X_i,\bg)\{ 1- \pi(\X_i,\bg)\}  ]\mid\bb\trans\x) }
{E\{\pi_0(\X_i) / \pi(\X_i,\bg)\mid\bb\trans\x\}}\n\\
&&-\frac{Q(\bb\trans\x) E\{\bpi_\bg(\X,\bg)
\pi_0(\X)/\pi^2(\X,\bg) \mid\bb\trans\x\}}
{E\{\pi_0(\X_i) / \pi(\X_i,\bg)\mid\bb\trans\x\}}\n\\
&&+\frac{Q(\bb\trans\x) E(\bpi_\bg(\X_i,\bg)\pi_0(\X)
 / [ \pi(\X_i,\bg)\{ 1- \pi(\X_i,\bg)\}  ]\mid\bb\trans\x) }
{E\{\pi_0(\X_i) / \pi(\X_i,\bg)\mid\bb\trans\x\}}\n\\
&&+\frac{Q(\bb\trans\x) E\left\{\pi_0(\X)
\bpi_\bg(\X,\bg)/ \pi^2(\X,\bg)\mid\bb\trans\x\right\}}
{E\{\pi_0(\X) / \pi(\X,\bg)\mid\bb\trans\x\}} \n\\
&=&\frac{E(\bpi_\bg(\X_i,\bg)\{
 \mu(\X_i,\ba)-\mu_0(\X_i) \} / [ \pi(\X_i,\bg)\{ 1- \pi(\X_i,\bg)\}  ]\mid\bb\trans\x) }
{E\{\pi_0(\X_i) / \pi(\X_i,\bg)\mid\bb\trans\x\}}+o_p(1).\n
\ee
Thus,
\bse
\T_\bb
&=&E\left\{ \frac{\partial}{\partial \bb_{L0}\trans} \left(\left[ \frac{\{A_i - \pi(\X_i,\bg_0)\}\{Y_i - \mu(\X_i,\ba_0)\}}{\pi(\X_i,\bg_0) 
\{1-\pi(\X_i,\bg_0)\}} 
+ \left\{1 - \frac{A_i}{\pi(\X_i,\bg_0)}\right\}\wt Q(\bb_0\trans\X_i,\bb_0,\ba_0,\bg_0) \right] \right.\right. \n\\
&&\left.\left.\times\left\{\X_{Li} -\wh E(\X_{Li} \mid \bb_0\trans\X_i)
    \right\} \right) \right\}\\
&=&E\left[ 
\left\{1 - \frac{A_i}{\pi(\X_i,\bg_0)}\right\}\left\{\X_{Li} - E(\X_{Li} \mid \bb_0\trans\X_i)
    \right\}
\frac{\partial}{\partial \bb_{L0}\trans}
\wt Q(\bb_0\trans\X_i,\bb_0,\ba_0,\bg_0) 
\right]\\
&&-E\left\{ \left(\left[ \frac{\{A_i - \pi(\X_i,\bg_0)\}\{Y_i - \mu(\X_i,\ba_0)\}}{\pi(\X_i,\bg_0) 
\{1-\pi(\X_i,\bg_0)\}} 
+ \left\{1 - \frac{A_i}{\pi(\X_i,\bg_0)}\right\}\wt Q(\bb_0\trans\X_i,\bb_0,\ba_0,\bg_0) \right] \right.\right. \n\\
&&\left.\left.\times \frac{\partial}{\partial \bb_0\trans}\wh E(\X_{Li} \mid \bb_0\trans\X_i)
    \right) \right\}+o_p(1)\\
&=&
E\left[ 
\frac{-\pi_0(\X)\left\{\X_L - E(\X_L \mid \bb_0\trans\X)
    \right\}
}{\pi(\X,\bg_0) f(\bb\trans\x)
  E\{\pi_0(\X) / \pi(\X,\bg)\mid\bb\trans\X\}}\right.\\
&&\left.\times\frac{\partial}{\partial\bb\trans\x}\left\{E\left(\X_L\trans
\left[\frac{\pi_0(\X)}{\pi(\X,\bg) }-
E\left\{\frac{\pi_0(\X)}{ \pi(\X,\bg)}\mid\bb\trans\x\right\} 
\right]\mid\bb\trans\x\right) Q(\bb\trans\x) f(\bb\trans\x)\right\}
\right]\\
&&+E\left\{ 
\frac{Q(\bb_0\trans\X)}{f(\bb\trans\x)}\frac{\partial\left( E[\{\X_L-E(\X_L\mid\bb\trans\x)\}^{\otimes2}\mid\bb\trans\x] f(\bb\trans\x)\right)
}{\partial(\bb\trans\x)}\right\}+o_p(1)\\
&=&\B+o_p(1).
\ese
Similarly,
\bse
\T_\ba&=&
E\left\{ \frac{\partial}{\partial \ba_0\trans} \left(\left[ \frac{\{A_i - \pi(\X_i,\bg_0)\}\{Y_i - \mu(\X_i,\ba_0)\}}{\pi(\X_i,\bg_0) 
\{1-\pi(\X_i,\bg_0)\}} 
+ \left\{1 - \frac{A_i}{\pi(\X_i,\bg_0)}\right\}\wt Q(\bb_0\trans\X_i,\bb_0,\ba_0,\bg_0) \right] \right.\right. \n\\
&&\left.\left.\times\left\{\X_{Li} -\wh E(\X_{Li} \mid
      \bb_0\trans\X_i) \right\} \right) \right\}+o_p(1)\\
&=&E\left(
\left\{\X_{Li} -E(\X_{Li} \mid
      \bb_0\trans\X_i) \right\}
\left[ \frac{-\{A_i - \pi(\X_i,\bg_0)\}\bmu_\ba\trans(\X_i,\ba_0)}{\pi(\X_i,\bg_0) \{1-\pi(\X_i,\bg_0)\}} 
-\left\{1 - \frac{A_i}{\pi(\X_i,\bg_0)}\right\}\right.\right.\\
&&\left.\left.\times\frac{E(\{\pi_0(\X) - \pi(\X,\bg)\} \bmu_\ba\trans(\X,\ba) / [\pi(\X,\bg)\{
  1- \pi(\X,\bg)\}]\mid\bb\trans\X)}
{E\{\pi_0(\X) / \pi(\X,\bg)\mid\bb\trans\x\}} \right] \right)
+o_p(1)\\
&=&E\left(
\left\{\X_{L} -E(\X_{L} \mid
      \bb_0\trans\X) \right\}
\left\{1-\frac{\pi_0(\X)}{\pi(\X,\bg_0)}\right\}\right.\\
&&\left.\times\left[
\frac{\bmu_\ba\trans(\X,\ba_0)}{\{1-\pi(\X,\bg_0)\}} 
-\frac{E(\{\pi_0(\X) - \pi(\X,\bg)\} \bmu_\ba\trans(\X,\ba) / [\pi(\X,\bg)\{
  1- \pi(\X,\bg)\}]\mid\bb\trans\X)}
{E\{\pi_0(\X) / \pi(\X,\bg)\mid\bb\trans\X\}} \right] \right)\\
&&+o_p(1)\\
&=&\B_\ba+o_p(1),
\ese
and
\bse
\T_\bg&=&
E\left\{ \frac{\partial}{\partial \bg_0\trans} \left(\left[ \frac{\{A - \pi(\X,\bg_0)\}\{Y - \mu(\X,\ba_0)\}}{\pi(\X,\bg_0) 
\{1-\pi(\X,\bg_0)\}} 
+ \left\{1 - \frac{A}{\pi(\X,\bg_0)}\right\}\wt Q(\bb_0\trans\X,\bb_0,\ba_0,\bg_0) \right] \right.\right. \n\\
&&\left.\left.\times\left\{\X_{L} - \wh E(\X_{L}  \mid
      \bb_0\trans\X) \right\} \right) \right\}+o_p(1)\\
&=&E\left(\left\{\X_{L} -  E(\X_{L}  \mid
      \bb_0\trans\X) \right\}
E\left[
\frac{\{ - \bpi_\bg\trans(\X,\bg_0)\}\{Y - \mu(\X,\ba_0)\}}{\pi(\X,\bg_0) 
\{1-\pi(\X,\bg_0)\}} \right.\right.\\
&&-\frac{\{A - \pi(\X,\bg_0)\}\bpi_\bg\trans(\X,\bg_0)\{Y - \mu(\X,\ba_0)\}}{\pi^2(\X,\bg_0) 
\{1-\pi(\X,\bg_0)\}} \\
&&+\frac{\{A - \pi(\X,\bg_0)\}\bpi_\bg\trans(\X,\bg_0)\{Y - \mu(\X,\ba_0)\}}{\pi(\X,\bg_0) 
\{1-\pi(\X,\bg_0)\}^2} 
+\frac{\bpi_\bg\trans(\X,\bg_0) A}{\pi^2(\X,\bg_0)}
\wt Q(\bb_0\trans\X,\bb_0,\ba_0,\bg_0) \\
&&\left.\left.
+ \left\{1 - \frac{A}{\pi(\X,\bg_0)}\right\}
\frac{E(\bpi_\bg(\X,\bg)\{
 \mu(\X,\ba)-\mu_0(\X) \} / [ \pi(\X,\bg)\{ 1- \pi(\X,\bg)\}  ]\mid\bb\trans\x) }
{E\{\pi_0(\X) / \pi(\X,\bg)\mid\bb\trans\x\}}\mid\X
 \right]\right)\\
&&+o_p(1)\\
&=&E\left(\left\{\X_{L} -  E(\X_{L}  \mid
      \bb_0\trans\X) \right\}
\left[
\frac{\{ - \bpi_\bg\trans(\X,\bg_0)\}\{\mu_0(\X) - \mu(\X,\ba_0)\}}{\pi(\X,\bg_0) 
\{1-\pi(\X,\bg_0)\}} \right.\right.\\
&&\left.\left.
+ \left\{1 - \frac{\pi_0(\X)}{\pi(\X,\bg_0)}\right\}
\frac{E(\bpi_\bg(\X,\bg)\{
 \mu(\X,\ba)-\mu_0(\X) \} / [ \pi(\X,\bg)\{ 1- \pi(\X,\bg)\}  ]\mid\bb\trans\x) }
{E\{\pi_0(\X) / \pi(\X,\bg)\mid\bb\trans\x\}}
 \right]\right)\\
&&+o_p(1)\\
&=&E\left[
\frac{\left\{\X_{L} -  E(\X_{L}  \mid
      \bb_0\trans\X) \right\}\bpi_\bg\trans(\X,\bg_0)\{\mu(\X,\ba_0)-\mu_0(\X) \}}{\pi(\X,\bg_0) 
\{1-\pi(\X,\bg_0)\}} \right]+o_p(1)\\
&=&\B_\bg+o_p(1).
\ese

Inserting the result regarding $\T_1, \T_2, \T_\bb, \T_\ba,\T_\bg$ in (\ref{eq:expandmain}), we get
\be\label{eq:betaexpan}
\0&=&\frac{1}{n^{1/2}} \sumi\left[
\bpsi_\bb(\X_i,Y_i,A_i,\bb,\ba,\bg)\right.\n\\
&&\left.+
\frac{
E\left[\left\{\x_{Li} - E(\X_{Li} \mid \bb_0\trans\x_i) \right\}
 \left\{1 - \pi_0(\X_i)/\pi(\x_i,\bg_0)\right\}\mid\bb_0\trans\X_i\right]}{E\{\pi_0(\X_i) /
  \pi(\X_i,\bg_0)\mid\bb\trans\X_i\}}\right.
 \n\\
&&\left.\times\left\{\frac{
 \{A_i - \pi(\X_i,\bg_0)\}\{ Y_i - \mu(\X_i,\ba_0) \} }{  \pi(\X_i,\bg_0)\{ 1- \pi(\X_i,\bg_0)\}  }
- \frac{Q(\bb_0\trans\X_i) A_i}{ \pi(\X_i,\bg_0)}\right\}
+
Q(\bb_0\trans\x_i)\left\{E(\X_{Li} \mid \bb_0\trans\x_i) -\x_{Li}
\right\}\right]\n\\
&&+\B\sqrt{n}(\wh\bb-\bb_0)
+\B_\bg n^{-1/2}\sumi\bphi_\bg(\X_i,Y_i,A_i)
+\B_\ba n^{-1/2}\sumi\bphi_\ba(\X_i,Y_i,A_i)
+o_p(1)\n\\
&=&\frac{1}{n^{1/2}} \sumi\left\{
\bphi_\bb(\X_i,Y_i,A_i,\bb,\ba,\bg)
+\B_\bg \bphi_\bg(\X_i,Y_i,A_i)
+\B_\ba \bphi_\ba(\X_i,Y_i,A_i)\right\}
+\B\sqrt{n}(\wh\bb-\bb_0)\n\\
&&+o_p(1).
\ee
This hence proves the results.\qed

\subsection{Proof of Lemma \ref{lem:Q}}
Let the probability density function of 
$\bb_0\trans\X_j$ be $f(\bb_0\trans\X_j)$. Denote 
\bse 
H_1(\X_j,A_j,Y_j,\bg_0,\ba_0) &=& \frac{\{A_j-\pi(\X_j,\bg_0)\}\{Y_j-\mu(\X_j,\ba_0)\}}{\pi(\X_j,\bg_0)\{1-\pi(\X_j,\bg_0)\}}, \\
H_2(\X_j,A_j,\bg_0) &=& \frac{A_j}{\pi(\X_j,\bg_0)},\\
r(\bb_0\trans\x) &=& E\left\{ \pi_0(\X_j)/\pi(\X_j,\bg_0) \mid 
  \bb_0\trans\x \right\}f(\bb_0\trans\x),\\
u_1(\bb_0\trans\x)&=&
\frac{d^2 \{ r(\bb_0\trans\x)Q(\bb_0\trans\x)\}}{ d(\bb_0\trans\x)^2} 
\frac{1}{2} \int z^2 K(z) dz,\\
u_2(\bb_0\trans\x)&=&r''(\bb_0\trans\x)\frac{1}{2} \int z^2 K(z) dz. 
\ese 
Then, from Lemma \ref{lem:order}, (\ref{eq:expectH1}) and (\ref{eq:expectH2}), we get 
\be 
&&E\{H_1(\X_j,A_j,Y_j,\bg_0,\ba_0) \Kh (\bb_0\trans\X_j -
\bb_0\trans\x) \} \n\\
&=& E\{H_1(\X_j,A_j,Y_j,\bg_0,\ba_0)\mid\bb\trans\x\}
f(\bb_0\trans\x) \n\\
&& + \frac{d^2}{ d(\bb_0\trans\x)^2} 
\left[ E\{H_1(\X_j,A_j,Y_j,\bg_0,\ba_0)\mid\bb_0\trans\x\}
f(\bb_0\trans\x)\right]
\frac{h_{\rm opt}^2}{2} \int z^2 K(z) dz + o(h_{\rm opt}^2) \n\\
&=& Q(\bb_0\trans\x)r(\bb_0\trans\x) 
+ u_1(\bb_0\trans\x) h_{\rm opt}^2+ o(h_{\rm opt}^2), \\
&&E\{H_2(\X_j,A_j,\bg_0) \Kh (\bb_0\trans\X_j - \bb_0\trans\x) 
\} \n\\
&=&E\{H_2(\X_j,A_j,\bg_0) \mid\bb\trans\x\}f(\bb_0\trans\x) \n\\
&& + \frac{d^2}{ d(\bb_0\trans\x)^2} 
\left[ E\{H_2(\X_j,A_j,\bg_0) \mid\bb\trans\x\}f(\bb_0\trans\x) \right]
\frac{h_{\rm opt}^2}{2} \int z^2 K(z) dz + o(h_{\rm opt}^2)\n\\
&=& r(\bb_0\trans\x) +u_2(\bb_0\trans\x) h_{\rm opt}^2+ o(h_{\rm opt}^2). 
\ee 
Thus, 
\bse 
&& E \left\{ \wh Q(\bb_0\trans\x) \right\}\\
&=& E\left[ \frac{n^{-1}\sumj H_1(\X_j,A_j,Y_j,\bg_0,\ba_0)\Kh(\bb_0\trans\X_j - \bb_0\trans\x)}{n^{-1}\sumj H_2(\X_j,A_j,\bg_0)\Kh(\bb_0\trans\X_j-\bb_0\trans\x)}
\right] \\
&=& E\bigg(\frac{E\left\{ H_1(\X_j,A_j,Y_j,\bg_0,\ba_0)\Kh(\bb_0\trans\X_j - \bb_0\trans\x) \right\}}{E\left\{ H_2(\X_j,A_j,\bg_0)\Kh(\bb_0\trans\X_j-\bb_0\trans\x) \right\} } + \\
&& \frac{n^{-1}\sumj H_1(\X_j,A_j,Y_j,\bg_0,\ba_0)\Kh(\bb_0\trans\X_j - \bb_0\trans\x) 
-E\left\{ H_1(\X_j,A_j,Y_j,\bg_0,\ba_0)\Kh(\bb_0\trans\X_j - \bb_0\trans\x) \right\}}
{E\left\{ H_2(\X_j,A_j,\bg_0)\Kh(\bb_0\trans\X_j-\bb_0\trans\x) \right\}} \\
&& - \frac{E\left\{ H_1(\X_j,A_j,Y_j,\bg_0,\ba_0)\Kh(\bb_0\trans\X_j - \bb_0\trans\x) \right\}}{\left[E\left\{ H_2(\X_j,A_j,\bg_0)\Kh(\bb_0\trans\X_j-\bb_0\trans\x) \right\}\right]^2}
\big[ n^{-1}\sumj H_2(\X_j,A_j,\bg_0)\Kh(\bb_0\trans\X_j-\bb_0\trans\x)\\
&& - E\left\{ H_2(\X_j,A_j,\bg_0)\Kh(\bb_0\trans\X_j-\bb_0\trans\x) \right\} \big]
\bigg) + o(n^{-1/2}h_{\rm opt}^{-1/2}) \\
&=& \frac{r(\bb_0\trans\x)Q(\bb_0\trans\x) + 
u_1(\bb_0\trans\x)h_{\rm opt}^2 
}{ r(\bb_0\trans\x) +u_2(\bb_0\trans\x)h_{\rm opt}^2}  
+ o(h_{\rm opt}^2 + n^{-1/2}h_{\rm opt}^{-1/2})\\
&=& Q(\bb_0\trans\x) 
+\left\{\frac{u_1(\bb_0\trans\x)}{r(\bb_0\trans\x)}
-\frac{Q(\bb_0\trans\x) u_2(\bb_0\trans\x)}{r(\bb_0\trans\x)}\right\}h_{\rm opt}^2 
+ o(h_{\rm opt}^2 + n^{-1/2}h_{\rm opt}^{-1/2})\\
&=&Q(\bb_0\trans\x)+h_{\rm opt}^2 
\left\{\frac{r'(\bb_0\trans\x) 
Q'(\bb_0\trans\x)}{r(\bb_0\trans\x)} 
+ \frac{Q''(\bb_0\trans\x)}{2}\right\} \int z^2 K(z) dz 
+ o(h_{\rm opt}^2 + n^{-1/2}h_{\rm opt}^{-1/2}). 
\ese 
Because we have shown that $\wh\bb-\bb_0=O_p(n^{-1/2})=o_p(h^2)$ under 
Condition \ref{con:C5}, 
thus, we get the bias 
\bse 
\bias\{\wh  Q(\wh\bb\trans\x) \}
&=& E\{\wh  Q(\wh\bb\trans\x) \} - Q(\bb_0\trans\x)\\
&=&
h_{\rm opt}^2 
\left\{\frac{r'(\bb_0\trans\x) 
Q'(\bb_0\trans\x)}{r(\bb_0\trans\x)} 
+ \frac{Q''(\bb_0\trans\x)}{2}\right\}
\int z^2 K(z) dz 
+ o(h_{\rm opt}^2 + n^{-1/2}h_{\rm opt}^{-1/2}). 
\ese 

To analyze the variance, 
we first note that 
\bse 
&&E\{ H_1^2(\X_j,A_j,Y_j,\bg_0,\ba_0) \mid \bb_0\trans\x\} \\
&=& E \left[ \frac{\{A_j-\pi(\X_j,\bg_0)\}^2\{Y_j-\mu(\X_j,\ba_0)\}^2}{\pi(\X_j,\bg_0)^2\{1-\pi(\X_j,\bg_0)\}^2} \mid \bb_0\trans\x \right]\\
&=& E \left( \frac{\{A_j-\pi(\X_j,\bg_0)\}^2 [ A_j^2\{Y_{1j}-\mu(\X_j,\ba_0)\}^2 + 
2A_j(1-A_j)\{Y_{1j}-\mu(\X_j,\ba_0)\}\{Y_{0j}-\mu(\X_j,\ba_0)\} }{\pi(\X_j,\bg_0)^2\{1-\pi(\X_j,\bg_0)\}^2} \right. \\
&& \left. + \frac{(1-A_j)^2 \{Y_{0j} - \mu(\X_j,\ba_0)\}^2 }{\pi(\X_j,\bg_0)^2\{1 - \pi(\X_j,\bg_0)\}^2} \mid \bb_0\trans\x \right) \\
&=& E\left( \frac{\pi_0(\X_j)}{\pi^2(\X_j,\bg_0)} E[ \{ Y_{1j}-\mu(\X_j,\ba_0)\}^2 \mid \X_j] 
+ \frac{1-\pi_0(\X_j)}{\{1-\pi(\X_j,\bg_0)\}^2} E[ \{ Y_{0j} -
\mu(\X_j,\ba_0) \}^2 \mid \X_j]  \mid \bb_0\trans\x\right), 
\ese 
and 
\bse 
 E\{ H_2^2(\X_j,A_j,\bg_0) \mid \bb_0\trans\x \}
= E\{A_j^2 / \pi^2(\X_j,\bg_0) \mid \bb_0\trans\x \}
= E\{\pi_0(\X_j) / \pi^2(\X_j,\bg_0)\mid \bb_0\trans\x \}. 
\ese 
In addition, 
\bse 
&&E\{ H_1(\X_j,A_j,Y_j,\bg_0,\ba_0)H_2(\X_j,A_j,\bg_0) \mid \bb_0\trans\x \} \\
&=& E\left[ \frac{ \{A_j  - \pi(\X_j,\bg_0) \}\{Y_j - \mu(\X_j,\ba_0) \} }{\pi(\X_j,\bg_0)\{1- \pi(\X_j,\bg_0)\} } 
\frac{A_j}{\pi(\X_j,\bg_0)} \mid \bb_0\trans\x \right]\\
&=&E\left( \frac{ \{A_j  - \pi(\X_j,\bg_0) \}[A_j\{Y_{1j} - \mu(\X_j,\ba_0) \} + (1-A_j) \{Y_{0j} - \mu(\X_j,\ba_0) \} ] }
{\pi(\X_j,\bg_0)\{1- \pi(\X_j,\bg_0)\} } 
\frac{A_j}{\pi(\X_j,\bg_0)} \mid \bb_0\trans\x \right)\\
&=& E\left[ \frac{\pi_0(\X_j)}{\pi^2(\X_j,\bg_0)} \{ Q(\bb_0\trans\X_j)+ \mu_0(\X_j) - \mu(\X_j,\ba_0)\} \mid \bb_0\trans\x \right] . 
\ese 
We define 
\bse 
v_1(\bb_0\trans\x) &=& E\left\{ H_1^2 ( \X_j,A_j,Y_j,\bg_0,\ba_0) \mid 
\bb_0\trans\x \right\} f(\bb_0\trans\x), \\
v_2(\bb_0\trans\x) &=& E\left\{ \frac{\pi_0(\X_j)}{\pi^2(\X_j,\bg_0)} \mid \bb_0\trans\x\right\}f(\bb_0\trans\x),\\
v_3(\bb_0\trans\x) &=& E\left[ \frac{\pi_0(\X_j)}{\pi^2(\X_j,\bg_0)} 
\left\{\mu_0(\X_j)-\mu(\X_j,\ba_0)\right\} \mid \bb_0\trans\x\right]f(\bb_0\trans\x),\\
v_4(\bb_0\trans\x) &=& Q(\bb_0\trans\x)v_2(\bb_0\trans\x). 
\ese 
From Lemma \ref{lem:order}, we get 
\bse 
&&\var \left\{  n^{-1} \sumj H_1(\X_j,A_j,Y_j,\bg_0,\ba_0) \Kh (\bb_0\trans\X_j - \bb_0\trans\x) \right\} \\
&=& (nh_{\rm opt})^{-1} E\left\{ H_1^2 ( \X_j,A_j,Y_j,\bg_0,\ba_0) \mid 
\bb_0\trans\x \right\} f(\bb_0\trans\x) \int K^2(z) dz + O(n^{-1}),\\
&=& (nh_{\rm opt})^{-1} v_1(\bb_0\trans\x) \int K^2(z) dz + O(n^{-1}),\\
&&\var \left\{  n^{-1} \sumj H_2(\X_j,A_j,\bg_0) \Kh(\bb_0\trans\X_j - \bb_0\trans\x) \right\} \\
&=& (nh_{\rm opt})^{-1} E\left\{ H_2^2 ( \X_j,A_j,\bg_0) \mid 
\bb_0\trans\x \right\} f(\bb_0\trans\x) \int K^2(z) dz + O(n^{-1}) \\
&=& (nh_{\rm opt})^{-1} E\{ \pi_0(\X_j)/ \pi^2(\X_j,\bg_0) \mid \bb_0\trans\x\}
f(\bb_0\trans\x) \int K^2(z) dz + O(n^{-1}) \\
&=& (nh_{\rm opt})^{-1} v_2(\bb_0\trans\x) \int K^2(z) dz + O(n^{-1}), 
\ese 
and 
\bse 
&& \cov \left\{ \frac{1}{n} \sumj H_1(\X_j,A_j,Y_j,\bg_0,\ba_0)\Kh(\bb_0\trans\X_j-
\bb_0\trans\x), \frac{1}{n}\sumj H_2(\X_j,A_j,\bg_0)\Kh(\bb_0\trans\X_j-
\bb_0\trans\x) \right\} \\
&=& n^{-1} E\{ H_1(\X_j,A_j,Y_j,\bg_0,\ba_0)H_2(\X_j,A_j,\bg_0)\Kh^2(\bb_0\trans\X_j - \bb_0\trans\x)\} +O(n^{-1})\\
&=& n^{-1} \int E\{ H_1(\X_j,A_j,Y_j,\bg_0,\ba_0)H_2(\X_j,A_j,\bg_0) \mid \bb_0\trans\x_j \} 
\Kh^2(\bb_0\trans\x_j - \bb_0\trans\x) f(\bb_0\trans\x_j)d\mu(\bb_0\trans\x_j) \\
&& + O(n^{-1})\\
&=& \frac{1}{n} \int E\left[ \frac{\pi_0(\X_j)}{\pi^2(\X_j,\bg_0)} \{ Q(\bb_0\trans\X_j) + \mu_0(\X_j) - \mu(\X_j,\ba_0)\}  \mid \bb_0\trans\x_j \right]
\Kh^2(\bb_0\trans\x_j - \bb_0\trans\x) f(\bb_0\trans\x_j)d\mu(\bb_0\trans\x_j) \\
&&+ O(n^{-1})\\
&=& n^{-1}  \int E\left\{ \frac{\pi_0(\X_j)}{\pi^2(\X_j,\bg_0)} \mid \bb_0\trans\x_j\right\}Q(\bb_0\trans\x_j) 
\Kh^2(\bb_0\trans\x_j - \bb_0\trans\x) f(\bb_0\trans\x_j)d\mu(\bb_0\trans\x_j) \\
&& +n^{-1} \int E\left[ \frac{\pi_0(\X_j)}{\pi^2(\X_j,\bg_0)} \{ \mu_0(\X_j) - \mu(\X_j,\ba_0)\}  \mid \bb_0\trans\x_j \right]
\Kh^2(\bb_0\trans\x_j - \bb_0\trans\x) f(\bb_0\trans\x_j)d\mu(\bb_0\trans\x_j) \\
&&+O(n^{-1})\\
&=& n^{-1} \int E\left\{ \frac{\pi_0(\X_j)}{\pi^2(\X_j,\bg_0)} \mid \bb_0\trans\x+zh_{\rm opt} \right\}Q(\bb_0\trans\x + zh_{\rm opt}) 
h_{\rm opt}^{-1}
K^2(z) f(\bb_0\trans\x+zh_{\rm opt})dz \\
&& +n^{-1} \int E\left[ \frac{\pi_0(\X_j)}{\pi^2(\X_j,\bg_0)} \{ \mu_0(\X_j) - \mu(\X_j,\ba_0)\}  \mid \bb_0\trans\x+zh_{\rm opt} \right]
h_{\rm opt}^{-1}K^2(z) f(\bb_0\trans\x+zh_{\rm opt})dz \\
&&+O(n^{-1})\\
&=& n^{-1} Q(\bb_0\trans\x)  E\left\{ \frac{\pi_0(\X_j)}{\pi^2(\X_j,\bg_0)} \mid \bb_0\trans\x \right\}f(\bb_0\trans\x) h_{\rm opt}^{-1} 
\int K^2(z) dz \\
&& +n^{-1}   E\left[ \frac{\pi_0(\X_j)}{\pi^2(\X_j,\bg_0)} \{ \mu_0(\X_j) - \mu(\X_j,\ba_0)\}  \mid \bb_0\trans\x \right]
f(\bb_0\trans\x) h_{\rm opt}^{-1}\int K^2(z) dz 
+ O(n^{-1})\\
&=& (nh_{\rm opt})^{-1} \{v_4(\bb_0\trans\x)+v_3(\bb_0\trans\x)\} \int K^2(z) dz 
 + O(n^{-1}). 
\ese 
Define 
\bse 
\Delta &=& 
\begin{bmatrix}
\{r(\bb_0\trans\x)\}^{-1} \\
\\
 -r(\bb_0\trans\x)Q(\bb_0\trans\x)/\{r(\bb_0\trans\x)\}^2 
\end{bmatrix} = \frac{1}{r(\bb_0\trans\x)}
\begin{Bmatrix}
1 \\
\\
-Q(\bb_0\trans\x) 
\end{Bmatrix}. 
\ese 
\bse 
&&\var\{ \wh Q(\bb_0\trans\X) \} \\
&=& 
\Delta\trans\left(
\frac{\int K^2(z) dz }{nh_{\rm opt}}
\begin{bmatrix}
v_1(\bb_0\trans\x)  & 
\{v_4(\bb_0\trans\x)+v_3(\bb_0\trans\x)\} \\
\{v_4(\bb_0\trans\x)+v_3(\bb_0\trans\x)\} 
 & 
v_2(\bb_0\trans\x) 
\end{bmatrix}
+O(n^{-1})\right) 
\Delta \\
&=&(nh_{\rm opt})^{-1}r^{-2}(\bb_0\trans\x) 
\{ v_1(\bb_0\trans\x) 
- 2 Q(\bb_0\trans\x)v_3(\bb_0\trans\x)  - Q(\bb_0\trans\x)v_4(\bb_0\trans\x) \} \int K^2(z) dz + O(n^{-1}). 
\ese 
As we already proved that $\wh\bb-\bb_0=O_p(n^{-1/2})=o_p(n^{-1/2}h^{-1/2})$ 
under 
Condition \ref{con:C5}, we get 
\bse 
&&\var\{ \wh Q(\wh\bb\trans\X) \} \\
&=&
(nh_{\rm opt})^{-1}r^{-2}(\bb_0\trans\x) 
\{ v_1(\bb_0\trans\x) 
- 2 Q(\bb_0\trans\x)v_3(\bb_0\trans\x)  - Q(\bb_0\trans\x)v_4(\bb_0\trans\x) \} \int K^2(z) dz + O(n^{-1}). 
\ese 
\qed 

\subsection{Proof of Theorem \ref{th:root}}

First, 
\bse
0&=&\wh Q(\wh z)-Q(z_0)\\
&=&\wh Q(\wh z)- Q(\wh z)+Q(\wh z)-Q(z_0)\\
&=&\wh Q(\wh z)- Q(\wh z)+\{Q'(z_0)+O_p(\wh z-z_0)\}(\wh z-z_0),
\ese
hence.
\bse
\wh z-z_0&=&-\{Q'(z_0)^{-1}+O_p(\wh z-z_0)\}\{\wh Q(\wh z)- Q(\wh z)\}\\
&=&-Q'(z_0)^{-1}\{\wh Q(\wh z)- Q(\wh z)\}\{1+o_p(1)\}.
\ese
Using the bias and variance results of $\wh Q(\cdot)$ in Lemma 
\ref{lem:Q}, we get that the leading term of the bias of $\wh Q(\wh z)- Q(\wh z)$
is
\bse
h_{\rm opt}^2
\left\{\frac{Q'(\wh z)
d[E\{\pi_0(\X)/\pi(\X,\bg_0)\mid \bb_0\trans\X=\wh z\}f(\wh z)]/d(\wh z)
}{
E\{\pi_0(\X)/\pi(\X,\bg_0)\mid \bb_0\trans\X=\wh z\}
f(\wh z)} 
+ \frac{Q''(\wh z)}{2}\right\}
\int z^2 K(z) dz,
\ese
which equals
\bse
h_{\rm opt}^2
\left\{\frac{Q'(z_0)
d[E\{\pi_0(\X)/\pi(\X,\bg_0)\mid \bb_0\trans\X=z_0\}f(z_0)]/d(z_0)
}{
E\{\pi_0(\X)/\pi(\X,\bg_0)\mid \bb_0\trans\X=z_0\}
f(z_0)} 
+ \frac{Q''(z_0)}{2}\right\}\int z^2 K(z) dz
\ese
to the leading order. Similarly, 
and the leading term of the variance  of $\wh Q(\wh z)- Q(\wh z)$ is
\bse
&&\frac{1}{nh_{\rm opt}} \bigg( 
E\left[ \frac{\pi_0(\X)}{\pi^2(\X,\bg_0)} \{
  Y_{1}-\mu(\X,\ba_0)\}^2  \mid \bb_0\trans\X=z_0\right]\\
&&+ E\left[\frac{1-\pi_0(\X)}{\{1-\pi(\X,\bg_0)\}^2}  \{ Y_{0} -
\mu(\X,\ba_0) \}^2   \mid \bb_0\trans\X=z_0\right]
\bigg)  \\
&& \times \frac{1}{f(z_0)}\left[E\left\{ \frac{\pi_0(\X)}{\pi(\X,\bg_0)} \mid \bb_0\trans\X=z_0\right\} \right]^{-2}\int K^2(z) dz.
\ese
These results lead to Theorem \ref{th:root}.
\qed

\subsection{Proof of Theorem \ref{th:value}}

It is obvious that $\wh V\{\wh Q(\cdot),\wh\bb,\wh\ba,\wh\bg\}\to
V\{Q(\cdot),\bb_0\}$. We first define 
\bse
J_a(t)\equiv I(t\le-a)+[1+\sin\{-\pi t/(2a)\}]/2I(-a<t< a)
\ese
for any $a>0$.
Then $|J_a(t)- I(t\le0)|=0$ if $|t|>a$ and $|J_a(t)- I(t\le0)|\le1$
if $|t|\le a$, and  $J_a(t)$ is a differentiable function. 
Define
\bse
&&\wt V\{\wh Q(\cdot),\wh\bb,\wh\ba,\wh\bg\}\\
&\equiv&n^{-1} \sumi \left(
\frac{ [ A_i 
+ (1-2A_i) J_a\{\wh Q(\wh\bb\trans\X_i,\wh\bb,\wh\ba,\wh\bg)  \}  ]Y_i }
{\pi(\X_i,\wh\bg) 
+ \{1-2\pi(\X_i,\wh\bg)\} J_a\{\wh
Q(\wh\bb\trans\X_i,\wh\bb,\wh\ba,\wh\bg)  \}}
+\{\pi(\X_i,\wh\bg)-A_i\}\right.\n\\
&&\left.\times
\frac{
[\mu(\X_i,\wh\ba)+\wh Q(\wh\bb\trans\X_i,\wh\bb,\wh\ba,\wh\bg)
-\{2 \mu(\X_i,\wh\ba)+
\wh Q(\wh\bb\trans\X_i,\wh\bb,\wh\ba,\wh\bg)\}
J_a\{\wh
Q(\wh\bb\trans\X_i,\wh\bb,\wh\ba,\wh\bg)\}]}{
\pi(\X_i,\wh\bg) 
+ \{1-2\pi(\X_i,\wh\bg)\} J_a\{\wh
Q(\wh\bb\trans\X_i,\wh\bb,\wh\ba,\wh\bg)  \}}\right).
\ese
then
\bse
&&n^{1/2}|\wt V\{\wh Q(\cdot),\wh\bb,\wh\ba,\wh\bg\}-
\wh V\{\wh Q(\cdot),\wh\bb,\wh\ba,\wh\bg\}|\\
&\le& n^{-1/2} \sumi \left|
\frac{ [ A_i 
+ (1-2A_i) I\{\wh Q(\wh\bb\trans\X_i,\wh\bb,\wh\ba,\wh\bg) \leq 0 \}  ]Y_i }
{\pi(\X_i,\wh\bg) 
+ \{1-2\pi(\X_i,\wh\bg)\} I\{\wh
Q(\wh\bb\trans\X_i,\wh\bb,\wh\ba,\wh\bg) \leq 0 \}}\right.\\
&&\left.-\frac{[ A_i 
+ (1-2A_i) J_a\{\wh Q(\wh\bb\trans\X_i,\wh\bb,\wh\ba,\wh\bg)  \}  ]Y_i }
{\pi(\X_i,\wh\bg) 
+ \{1-2\pi(\X_i,\wh\bg)\} J_a\{\wh
Q(\wh\bb\trans\X_i,\wh\bb,\wh\ba,\wh\bg)  \}}\right|+n^{-1/2} \sumi\\
&&\left|
\frac{
[\mu(\X_i,\wh\ba)+\wh Q(\wh\bb\trans\X_i,\wh\bb,\wh\ba,\wh\bg)
-\{2 \mu(\X_i,\wh\ba)+
\wh Q(\wh\bb\trans\X_i,\wh\bb,\wh\ba,\wh\bg)\}
I\{\wh
Q(\wh\bb\trans\X_i,\wh\bb,\wh\ba,\wh\bg)\le0 \}]}{
\pi(\X_i,\wh\bg) 
+ \{1-2\pi(\X_i,\wh\bg)\} I\{\wh
Q(\wh\bb\trans\X_i,\wh\bb,\wh\ba,\wh\bg) \leq 0 \}}\right.\\
&&\left.-
\frac{
[\mu(\X_i,\wh\ba)+\wh Q(\wh\bb\trans\X_i,\wh\bb,\wh\ba,\wh\bg)
-\{2 \mu(\X_i,\wh\ba)+
\wh Q(\wh\bb\trans\X_i,\wh\bb,\wh\ba,\wh\bg)\}
J_a\{\wh
Q(\wh\bb\trans\X_i,\wh\bb,\wh\ba,\wh\bg)\}]}{
\pi(\X_i,\wh\bg) 
+ \{1-2\pi(\X_i,\wh\bg)\} J_a\{\wh
Q(\wh\bb\trans\X_i,\wh\bb,\wh\ba,\wh\bg)  \}}
\right|\\
&=& n^{-1/2} \sumi \left(\left|
 \{A_i-\pi(\X_i,\wh\bg)\}[ J_a\{\wh
Q(\wh\bb\trans\X_i,\wh\bb,\wh\ba,\wh\bg)  \}- I\{\wh
Q(\wh\bb\trans\X_i,\wh\bb,\wh\ba,\wh\bg) \leq 0 \}]Y_i\right|\right.\\
&&+\left|[\mu(\X_i,\wh\ba)+\wh
    Q(\wh\bb\trans\X_i,\wh\bb,\wh\ba,\wh\bg)
\{1-\pi(\X_i,\wh\bg)\}]\right.\\
&&\left.\left.\times[ J_a\{\wh
Q(\wh\bb\trans\X_i,\wh\bb,\wh\ba,\wh\bg)  \}- I\{\wh
Q(\wh\bb\trans\X_i,\wh\bb,\wh\ba,\wh\bg) \leq 0 \}]\right|\right)\\
&&/\left([\pi(\X_i,\wh\bg) 
+ \{1-2\pi(\X_i,\wh\bg)\} I\{\wh
Q(\wh\bb\trans\X_i,\wh\bb,\wh\ba,\wh\bg) \leq0\}]\right.\\
&&\left.\times[\pi(\X_i,\wh\bg) 
+ \{1-2\pi(\X_i,\wh\bg)\} J_a\{\wh
Q(\wh\bb\trans\X_i,\wh\bb,\wh\ba,\wh\bg)  \}]\right)\\
&\le&n^{-1/2} \sumi \left|
[ J_a\{\wh
Q(\wh\bb\trans\X_i,\wh\bb,\wh\ba,\wh\bg)  \}- I\{\wh
Q(\wh\bb\trans\X_i,\wh\bb,\wh\ba,\wh\bg) \leq 0 \}]\right|Cc^{-2}\\
&=&n^{-1/2} \sumi 
I\{|\wh
Q(\wh\bb\trans\X_i,\wh\bb,\wh\ba,\wh\bg)|<a\}
\left[1-\sin\left\{\frac{\pi|
\wh
Q(\wh\bb\trans\X_i,\wh\bb,\wh\ba,\wh\bg)|
}{2a}\right\}\right]
\frac{C}{2c^2}\\
&\le&n^{-1/2} \sumi 
I\{|\wh
Q(\wh\bb\trans\X_i,\wh\bb,\wh\ba,\wh\bg)|<a\}
C/(2c^2)\\
&\le&n^{-1/2} \sumi 
[I\{|Q(\bb\trans\X_i)|<2a\}
+I\{|\wh
Q(\wh\bb\trans\X_i,\wh\bb,\wh\ba,\wh\bg)-Q(\bb\trans\X_i)|>a\}]
C/(2c^2).
\ese
Now let $a=n^{-1/5}$. Then
$n^{-1/2}\sumi I\{|Q(\bb\trans\X_i)|<2a\}$ converges in distribution to a normal
random variable with mean $\pr\{|Q(\bb\trans\X_i)|<2n^{-1/5}\}\to0$ and
variance
$\pr\{|Q(\bb\trans\X_i)|<2n^{-1/5}\}\pr\{|Q(\bb\trans\X_i)|>2n^{-1/5}\}\to0$.
Hence, 
$n^{-1/2}\sumi I\{|Q(\bb\trans\X_i)|<2a\}$ converges in probability to
0. On the other hand, 
\bse
&&\pr[\sumi I\{|\wh
Q(\wh\bb\trans\X_i,\wh\bb,\wh\ba,\wh\bg)-Q(\bb\trans\X_i)|>a\}>0]\\
&=&\pr\{\sup_i |\wh
Q(\wh\bb\trans\X_i,\wh\bb,\wh\ba,\wh\bg)-Q(\bb\trans\X_i)|>a\}\\
&\to&0
\ese
at the bandwidth $h=O(n^{-1/5})$ by Lemma \ref{lem:Q}. Thus,
\bse
n^{-1/2} \sumi 
I\{|\wh
Q(\wh\bb\trans\X_i,\wh\bb,\wh\ba,\wh\bg)-Q(\bb\trans\X_i)|>a\}\to0
\ese
in probability as well. Therefore, 
$n^{1/2}|\wt V\{\wh Q(\cdot),\wh\bb,\wh\ba,\wh\bg\}-
\wh V\{\wh Q(\cdot),\wh\bb,\wh\ba,\wh\bg\}|\to0$ in probability.

We also have
\be\label{eq:expandvalue}
&&n^{1/2}[\wt V\{\wh Q(\cdot),\wh\bb,\wh\ba,\wh\bg\}
-V\{Q(\cdot),\bb_0\}]\n\\
&=&
n^{1/2}[\wt V\{\wh Q(\cdot),\wh\bb,\wh\ba,\wh\bg\}-
\wt V\{\wh Q(\cdot),\bb_0,\ba_0,\bg_0\}]
+n^{1/2}[\wt V\{\wh Q(\cdot),\bb_0,\ba_0,\bg_0\}-V\{Q(\cdot),\bb_0\}]\n\\
&=&E\left[
\frac{\partial\wt V\{
  \wh Q(\bb_0\trans\X),\bb_0,\ba_0,\bg_0\}}{\partial\bb_{0L}\trans}\right]n^{1/2}(\wh\bb_L-\bb_{0L})
+
E\left[\frac{\partial\wt V\{
  \wh Q(\bb_0\trans\X),\bb_0,\ba_0,\bg_0\}}{\partial\ba_0\trans}\right]n^{1/2}(\wh\ba-\ba_0)\n\\
&&+E\left[
\frac{\partial \wt V\{
  \wh
  Q(\bb_0\trans\X),\bb_0,\ba_0,\bg_0\}}{\partial\bg_0\trans}\right]n^{1/2}(\wh\bg-\bg_0)\n\\
&&+n^{-1/2} \sumi \frac{\partial}{\partial \wh Q(\bb_0\trans\X_i,\bb_0,\ba_0,\bg_0)}
\left(\frac{ [ A_i 
+ (1-2A_i) J_a\{\wh Q(\bb_0\trans\X_i,\bb_0,\ba_0,\bg_0)\}  ]Y_i }
{\pi(\X_i,\bg_0) 
+ \{1-2\pi(\X_i,\bg_0)\} J_a\{\wh
Q(\bb_0\trans\X_i,\bb_0,\ba_0,\bg_0)\}}\right.\n\\
&&\left.+[\mu(\X_i,\ba_0)+\wh
  Q(\bb_0\trans\X_i,\bb_0,\ba_0,\bg_0)\right.\n\\
&&\left.-
\{2 \mu(\X_i,\ba_0)+\wh Q(\bb_0\trans\X_i,\bb_0,\ba_0,\bg_0) \}
J_a\{ \wh
Q(\bb_0\trans\X_i,\bb_0,\ba_0,\bg_0) \}]\right.\n\\
&&\left.\times
\frac{\{\pi(\X_i,\bg_0)-A_i\}
}{\pi(\X_i,\bg_0) 
+\{1-2\pi(\X_i,\bg_0)\}J_a\{\wh
Q(\bb_0\trans\X_i,\bb_0,\ba_0,\bg_0) \}}\right)\n\\
&&\times\left\{
\frac{\sumj K_h (\bb_0\trans\X_j - \bb_0\trans\X_i) \{A_j - \pi(\X_j,\bg_0)\}\{ Y_j - \mu(\X_j,\ba_0) \} / [ \pi(\X_j,\bg_0)\{ 1- \pi(\X_j,\bg_0)\}  ] }
{\sumj K_h (\bb_0\trans\X_j - \bb\trans\X_i) A_j /
  \pi(\X_j,\bg_0)}\right.\n\\
&&\left.-
Q(\bb_0\trans\X_i)\right\}
+n^{1/2}
[\wt V\{Q(\cdot),\bb_0,\ba_0,\bg_0\}-V\{Q(\cdot),\bb_0\}]
+o_p(1)\n\\
&=&E\left[
\frac{\partial\wt V\{
  \wh Q(\bb_0\trans\X),\bb_0,\ba_0,\bg_0\}}{\partial\bb_{0L}\trans}\right]n^{1/2}(\wh\bb_L-\bb_{0L})
+
E\left[\frac{\partial\wt V\{
  \wh Q(\bb_0\trans\X),\bb_0,\ba_0,\bg_0\}}{\partial\ba_0\trans}\right]n^{1/2}(\wh\ba-\ba_0)\n\\
&&+E\left[
\frac{\partial \wt V\{
  \wh
  Q(\bb_0\trans\X),\bb_0,\ba_0,\bg_0\}}{\partial\bg_0\trans}\right]n^{1/2}(\wh\bg-\bg_0)\n\\
&&+n^{-1/2} \sumi \frac{\partial}{\partial Q(\bb_0\trans\X_i)}
\left(
\frac{ [ A_i 
+ (1-2A_i) J_a\{Q(\bb_0\trans\X_i)\}  ]Y_i }
{\pi(\X_i,\bg_0) 
+ \{1-2\pi(\X_i,\bg_0)\} J_a\{
Q(\bb_0\trans\X_i)\}}+\{\pi(\X_i,\bg_0)-A_i\}\right.\n\\
&&\left.
\times\frac{
[\mu(\X_i,\ba_0)+Q(\bb_0\trans\X_i)-
 \{2\mu(\X_i,\ba_0)+Q(\bb_0\trans\X_i)\} J_a\{ 
Q(\bb_0\trans\X_i) \}]}{\pi(\X_i,\bg_0) 
+\{1-2\pi(\X_i,\bg_0)\}J_a\{
Q(\bb_0\trans\X_i) \}}\right)\{1+o_p(1)\}\n\\
&&\times\left\{
\frac{\sumj K_h (\bb_0\trans\X_j - \bb_0\trans\X_i) \{A_j - \pi(\X_j,\bg_0)\}\{ Y_j - \mu(\X_j,\ba_0) \} / [ \pi(\X_j,\bg_0)\{ 1- \pi(\X_j,\bg_0)\}  ] }
{\sumj K_h (\bb_0\trans\X_j - \bb_0\trans\X_i) A_j /
  \pi(\X_j,\bg_0)}\right.\n\\
&&\left.-
Q(\bb_0\trans\X_i)\right\}
+n^{1/2}
[\wt V\{Q(\cdot),\bb_0,\ba_0,\bg_0\}-V\{Q(\cdot),\bb_0\}]
+o_p(1).
\ee
For convenience, for $a=n^{-1/5}$, denote
\bse
&&T(Y,A,\X)\\
&=&\frac{\partial}{\partial Q(\bb_0\trans\X)}
\left(
\frac{ [ A 
+ (1-2A) J_a\{Q(\bb_0\trans\X)\}  ]Y }
{\pi(\X,\bg_0) 
+ \{1-2\pi(\X,\bg_0)\} J_a\{
Q(\bb_0\trans\X)\}}+\{\pi(\X,\bg_0)-A\}\right.\n\\
&&\left.
\times\frac{
[\mu(\X,\ba_0)+Q(\bb_0\trans\X)-
 \{2\mu(\X,\ba_0)+Q(\bb_0\trans\X)\} J_a\{ 
Q(\bb_0\trans\X) \}]}{\pi(\X,\bg_0) 
+\{1-2\pi(\X,\bg_0)\}J_a\{
Q(\bb_0\trans\X) \}}\right)\\
&=&
\frac{J_a'\{Q(\bb_0\trans\X)\}
[Y-\mu(\X,\ba_0)-
Q(\bb_0\trans\X)\{1-\pi(\X,\bg_0)\}]
 \{\pi(\X,\bg_0)-A\}}{[\pi(\X,\bg_0)+\{1-2\pi(\X,\bg_0)\} J_a\{
Q(\bb_0\trans\X)\}]^2}\n\\
&&+\frac{
[1-J_a\{Q(\bb_0\trans\X)\}]
 \{\pi(\X,\bg_0)-A\}}{\pi(\X,\bg_0)+\{1-2\pi(\X,\bg_0)\} J_a\{
Q(\bb_0\trans\X)\}}\\
&=&
\frac{J_a'\{Q(\bb_0\trans\X)\}
[Y-\mu(\X,\ba_0)-
Q(\bb_0\trans\X)\{1-\pi(\X,\bg_0)\}]
 \{\pi(\X,\bg_0)-A\}}{[\pi(\X,\bg_0)+\{1-2\pi(\X,\bg_0)\} J_a\{
Q(\bb_0\trans\X)\}]^2}\n\\
&&+\frac{
[1-J_a\{Q(\bb_0\trans\X)\}]
 \{\pi(\X,\bg_0)-A\}}{\pi(\X,\bg_0)}+o_p(1),
\ese
then
\bse
&&E\{T(Y,A,\X)\mid \X\}\\
&=&
\frac{J_a'\{Q(\bb_0\trans\X)\}
 \{\pi(\X,\bg_0)-1\}\pi(\X,\bg_0)Q(\bb_0\trans\X)}{[\pi(\X,\bg_0)+\{1-2\pi(\X,\bg_0)\} J_a\{
Q(\bb_0\trans\X)\}]^2}
+
\frac{
[1-J_a\{Q(\bb_0\trans\X)\}]
 \{\pi(\X,\bg_0)-\pi_0(\X)\}}{\pi(\X,\bg_0)}\\
&&+o_p(1),
\ese
and for any $\b(\X)$, 
\bse
E\{T(Y,A,\X)\b(\X)\}
=E\left(
\frac{\{\pi(\X,\bg_0)-\pi_0(\X)\}
[1-J_a\{Q(\bb_0\trans\X)\}]
 }{\pi(\X,\bg_0)}\b(\X)\right)+o_p(1).
\ese
Now, incorporating (\ref{eq:Qbeta}), for $a=n^{-1/5}$, 
\be\label{eq:vbeta}
&&E\left[
\frac{\partial\wt V\{
  \wh
  Q(\bb_0\trans\X),\bb_0,\ba_0,\bg_0\}}{\partial\bb_{0L}}\right]\n\\
&=&E\left[
T(Y,A,\X)
\frac{\partial \wh
Q(\bb_0\trans\X,\bb_0,\ba_0,\bg_0)}{\partial\bb_{0L}}\right]\n\\
&=&
E\left(
\frac{\{\pi(\X,\bg_0)-\pi_0(\X)\}
[1-J_a\{Q(\bb_0\trans\X)\}]}
{\pi(\X,\bg_0)}\frac{\partial \wh
Q(\bb_0\trans\X,\bb_0,\ba_0,\bg_0)}{\partial\bb_{0L}}\right)+o(1)\n\\
&=&
E\left[\frac{[1-E\{\pi_0(\X)/\pi(\X,\bg_0)\mid\bb_0\trans\X\}]
[1-J_a\{Q(\bb_0\trans\X)\}]}{f(\bb_0\trans\X)
  E\{\pi_0(\X) / \pi(\X,\bg_0)\mid\bb_0\trans\X\}}\right.\n\\
&&\left.\times \frac{\partial}{\partial\bb_0\trans\X}\left\{E\left(\X_L
\left[\frac{\pi_0(\X)}{\pi(\X,\bg_0) }-
E\left\{\frac{\pi_0(\X)}{ \pi(\X,\bg_0)}\mid\bb_0\trans\X\right\} 
\right]\mid\bb_0\trans\X\right) Q(\bb_0\trans\X)
f(\bb_0\trans\X)\right\}\right]+o(1)\n\\
&=&
E\left(\frac{[1-E\{\pi_0(\X)/\pi(\X,\bg_0)\mid\bb_0\trans\X\}]
I\{Q(\bb_0\trans\X)>0\}}{f(\bb_0\trans\X)
  E\{\pi_0(\X) / \pi(\X,\bg_0)\mid\bb_0\trans\X\}}\right.\n\\
&&\left.\times \frac{\partial}{\partial\bb_0\trans\X}
\left[
\cov\left\{\X_L,\frac{\pi_0(\X)}{\pi(\X,\bg_0) }\mid\bb_0\trans\X\right\}
Q(\bb_0\trans\X)
f(\bb_0\trans\X)\right]\right)+o(1)\n\\
&=&\U_\bb+o_p(1).
\ee
Similarly, incorporating  (\ref{eq:Qalpha}), for $a=n^{-1/5}$, 
\be\label{eq:valpha}
&&E\left[
\frac{\partial\wt V\{
  \wh Q(\bb_0\trans\X),\bb_0,\ba_0,\bg_0\}}{\partial\ba_0}\right]\n\\
&=&E\left(
E\{T(Y,A,\X)\mid\X\}\frac{\partial \wh
Q(\bb_0\trans\X,\bb_0,\ba_0,\bg_0)}{\partial\ba_0}\right.\n\\
&&\left.+\{\pi(\X,\bg_0)-\pi_0(\X)\}\frac{\bmu_\ba(\X,\ba_0)[1-2J_a\{Q(\bb_0\trans\X)\}]}
{\pi(\X,\bg_0)+\{1-2\pi(\X,\bg_0)\}J_a\{Q(\bb_0\trans\X)\}}\right)+o(1)\n\\
&=&
E\left(
\frac{\{\pi(\X,\bg_0)-\pi_0(\X)\}
[1-J_a\{Q(\bb_0\trans\X)\}]}
{\pi(\X,\bg_0)}\frac{\partial \wh
Q(\bb_0\trans\X,\bb_0,\ba_0,\bg_0)}{\partial\ba_0}\right.\n\\
&&\left.+\{\pi(\X,\bg_0)-\pi_0(\X)\}\frac{\bmu_\ba(\X,\ba_0)[1-2J_a\{Q(\bb_0\trans\X)\}]}
{\pi(\X,\bg_0)+\{1-2\pi(\X,\bg_0)\}J_a\{Q(\bb_0\trans\X)\}}\right)+o(1)\n\\
&=&E\left(
[1-E\{\pi_0(\X)/\pi(\X,\bg_0)\mid\bb_0\trans\X\}]
I\{Q(\bb_0\trans\X)>0\}
\right.\n\\
&&\left.\times\frac{E(
\{\pi(\X,\bg_0)-\pi_0(\X)\} \bmu_\ba(\X,\ba_0) / [ \pi(\X,\bg_0)\{ 1- \pi(\X,\bg_0)\}  ] \mid\bb_0\trans\X)}
{E\{\pi_0(\X) / \pi(\X,\bg_0)\mid\bb_0\trans\X\}}\right.\n\\
&&\left.+\frac{\{\pi(\X,\bg_0)-\pi_0(\X)\}\bmu_\ba(\X,\ba_0)[1-2I\{Q(\bb_0\trans\X)\le0\}]}
{\pi(\X,\bg_0)+\{1-2\pi(\X,\bg_0)\}I\{Q(\bb_0\trans\X)\le0\}}\right)+o(1)\n\\
&=&\U_\ba+o_p(1).
\ee
Incorporating  (\ref{eq:Qgamma}), for $a=n^{-1/5}$, 
\be\label{eq:vgamma}
&&E\left[
\frac{\partial\wt V\{
  \wh Q(\bb_0\trans\X),\bb_0,\ba_0,\bg_0\}}{\partial\bg_0}\right]\n\\
&=&
E\left(
\frac{\{\pi(\X,\bg_0)-\pi_0(\X)\}
[1-J_a\{Q(\bb_0\trans\X)\}]}
{\pi(\X,\bg_0)}\frac{\partial \wh
Q(\bb_0\trans\X,\bb_0,\ba_0,\bg_0)}{\partial\bg_0}\right)\n\\
&&+E\left(
\frac{[ A+(1-2A)J_a\{Q(\bb_0\trans\X)\}][1-2 J_a\{Q(\bb_0\trans\X)\}]
Y\bpi_\bg(\X,\bg_0)
}
{-[\pi(\X,\bg_0)+\{1-2\pi(\X,\bg_0)\}J_a\{Q(\bb_0\trans\X)\}]^2}\right)\n\\
&&+E\left(
\frac{\bpi_\bg(\X,\bg_0)
[\mu(\X,\ba_0)+Q(\bb_0\trans\X)
-\{2 \mu(\X,\ba_0)+
Q(\bb_0\trans\X)\}
J_a\{Q(\bb_0\trans\X)\}]}{
\pi(\X_i,\bg_0) 
+ \{1-2\pi(\X_i,\bg_0)\} J_a\{
Q(\bb_0\trans\X)  \}}\right)\n\\
&&+
E\left(
\frac{\{\pi(\X,\bg_0)-A\}
[\mu(\X,\ba_0)+Q(\bb_0\trans\X)
-\{2 \mu(\X,\ba_0)+
Q(\bb_0\trans\X)\}
J_a\{Q(\bb_0\trans\X)\}]
\bpi_\bg(\X_i,\bg_0) 
}{
-[\pi(\X_i,\bg_0) 
+ \{1-2\pi(\X_i,\bg_0)\} J_a\{
Q(\bb_0\trans\X)  \}]^2}\right.\n\\
&&\left.\times [1-2 J_a\{Q(\bb_0\trans\X)\}]
\right)+o(1)\n\\
&=&
E\left(
[1-E\{\pi_0(\X)/\pi(\X,\bg_0)\mid\bb_0\trans\X\}]
I\{Q(\bb_0\trans\X)>0\}\right.\n\\
&&\left.\times\frac{E(\bpi_\bg(\X,\bg_0)\{
 \mu(\X,\ba_0)-\mu_0(\X) \} / [ \pi(\X,\bg_0)\{ 1- \pi(\X,\bg_0)\}  ]\mid\bb_0\trans\X) }
{E\{\pi_0(\X) / \pi(\X,\bg_0)\mid\bb_0\trans\X\}}\right)
\n\\
&&+E\left(\left[
\frac{
I\{
Q(\bb_0\trans\X) \leq 0 \}\{1-\pi_0(\X)\}}{
\{1-\pi(\X,\bg_0)\}^2}
-\frac{I\{
Q(\bb_0\trans\X)>0 \}\pi_0(\X)}{
\pi(\X,\bg_0)^2}\right]\right.\n\\
&&\left.\times \{\mu_0(\X)-\mu(\X,\ba_0)\}\bpi_\bg(\X,\bg_0)
\right)+o(1)\n\\
&=&\U_\bg+o(1).
\ee
Further, we have
\be\label{eq:vQ}
&&n^{-1/2} \sumi \frac{\partial}{\partial Q(\bb_0\trans\X_i)}
\left(
\frac{ [ A_i 
+ (1-2A_i) J_a\{Q(\bb_0\trans\X_i)\}  ]Y_i }
{\pi(\X_i,\bg_0) 
+ \{1-2\pi(\X_i,\bg_0)\} J_a\{
Q(\bb_0\trans\X_i)\}}+\{\pi(\X_i,\bg_0)-A_i\}\right.\n\\
&&\left.
\times\frac{
[\mu(\X_i,\ba_0)+Q(\bb_0\trans\X_i)-
 \{2\mu(\X_i,\ba_0)+Q(\bb_0\trans\X_i)\} J_a\{ 
Q(\bb_0\trans\X_i) \}]}{\pi(\X_i,\bg_0) 
+\{1-2\pi(\X_i,\bg_0)\}J_a\{
Q(\bb_0\trans\X_i) \}}\right)\n\\
&&\times\left\{
\frac{\sumj K_h (\bb_0\trans\X_j - \bb_0\trans\X_i) \{A_j - \pi(\X_j,\bg_0)\}\{ Y_j - \mu(\X_j,\ba_0) \} / [ \pi(\X_j,\bg_0)\{ 1- \pi(\X_j,\bg_0)\}  ] }
{\sumj K_h (\bb_0\trans\X_j - \bb\trans\X_i) A_j /
  \pi(\X_j,\bg_0)}\right.\n\\
&&\left.-
Q(\bb_0\trans\X_i)\right\}\n\\
&=&n^{-1/2} \sumi T(Y_i,\X_i,A_i)\n\\
&&\times\left\{
\frac{\sumj K_h (\bb_0\trans\X_j - \bb_0\trans\X_i) \{A_j - \pi(\X_j,\bg_0)\}\{ Y_j - \mu(\X_j,\ba_0) \} / [ \pi(\X_j,\bg_0)\{ 1- \pi(\X_j,\bg_0)\}  ] }
{\sumj K_h (\bb_0\trans\X_j - \bb\trans\X_i) A_j /
  \pi(\X_j,\bg_0)}\right.\n\\
&&\left.-
Q(\bb_0\trans\X_i)\right\}\n\\
&=&
n^{-3/2} \sumi\sumj T(Y_i,\X_i,A_i)K_h (\bb_0\trans\X_j - \bb_0\trans\X_i) \n\\
&&\left(
\frac{ \{A_j - \pi(\X_j,\bg_0)\}\{ Y_j - \mu(\X_j,\ba_0) \} / [ \pi(\X_j,\bg_0)\{ 1- \pi(\X_j,\bg_0)\}  ] }
{f(\bb_0\trans\X_i) E\{\pi_0(\X_i) /
  \pi(\X_i,\bg_0)\mid\bb_0\trans\X_i\}}\right.\n\\
&&\left.-
\frac{A_j/\pi(\X_j,\bg_0)}{f(\bb_0\trans\X_i)E\{\pi_0(\X_i)
/\pi(\X_i,\bg_0)\mid\bb_0\trans\X_i\}}
Q(\bb_0\trans\X_i)
\right)+o_p(1)\n\\
&=&
n^{-1/2}\sumj 
E\{T(Y_j,\X_j,A_j)\mid\bb_0\trans\X_j\} \n\\
&&\left(
\frac{ \{A_j - \pi(\X_j,\bg_0)\}\{ Y_j - \mu(\X_j,\ba_0) \} / [ \pi(\X_j,\bg_0)\{ 1- \pi(\X_j,\bg_0)\}  ] }
{E\{\pi_0(\X_j) /
  \pi(\X_j,\bg_0)\mid\bb_0\trans\X_j\}}
\right.\n\\
&&\left.-
\frac{A_j Q(\bb_0\trans\X_j)/\pi(\X_j,\bg_0)}{E\{\pi_0(\X_j) /
  \pi(\X_j,\bg_0)\mid\bb_0\trans\X_j\}}\right)
+o_p(1)\n\\
&=&
n^{-1/2}\sumi
\left\{
E\left(\frac{J_a'\{Q(\bb_0\trans\X)\}
 \{\pi(\X,\bg_0)-1\}\pi(\X,\bg_0)Q(\bb_0\trans\X)}{[\pi(\X,\bg_0)+\{1-2\pi(\X,\bg_0)\} J_a\{
Q(\bb_0\trans\X)\}]^2}\mid\bb_0\trans\X_i\right)\right.\n\\
&&\left.+
E\left(\frac{
I\{Q(\bb_0\trans\X)>0\}
 \{\pi(\X,\bg_0)-\pi_0(\X)\}}{\pi(\X,\bg_0)}\mid\bb_0\trans\X_i\right)\right\}\n\\
&&\left(
\frac{ \{A_i - \pi(\X_i,\bg_0)\}\{ Y_i - \mu(\X_i,\ba_0) \} / \{ 1- \pi(\X_i,\bg_0)\}  
-A_i Q(\bb_0\trans\X_i)
}
{\pi(\X_i,\bg_0) E\{\pi_0(\X_i) /
  \pi(\X_i,\bg_0)\mid\bb_0\trans\X_i\}}
\right)+o_p(1)\n\\
&=&n^{-1/2}\sumi \wt v_Q\{\X_i,A_i,Y_i,\bb_0,\ba_0,\bg_0,Q(\cdot)\}+o_p(1).
\ee
Finally, we have
\be\label{eq:v0}
&&\wt V\{Q(\cdot),\bb_0,\ba_0,\bg_0\}
-V\{Q(\cdot),\bb_0\}\\
&\equiv&n^{-1} \sumi \left(
\frac{ [ A_i 
+ (1-2A_i) J_a\{Q(\bb_0\trans\X_i)  \}  ]Y_i }
{\pi(\X_i,\bg_0) 
+ \{1-2\pi(\X_i,\bg_0)\} J_a\{
Q(\bb_0\trans\X_i)  \}}
+\{\pi(\X_i,\bg_0)-A_i\}\right.\n\\
&&\left.\times
\frac{
[\mu(\X_i,\ba_0)+Q(\bb_0\trans\X_i)
-\{2 \mu(\X_i,\ba_0)+
Q(\bb_0\trans\X_i)\}
J_a\{
Q(\bb_0\trans\X_i)\}]}{
\pi(\X_i,\bg_0) 
+ \{1-2\pi(\X_i,\bg_0)\} J_a\{
Q(\bb_0\trans\X_i)  \}}
-V\{Q(\cdot),\bb_0\}
\right)\n\\
&=&n^{-1} \sumi \left(
\frac{ [ A_i 
+ (1-2A_i) I\{Q(\bb_0\trans\X_i) \le0 \}  ]Y_i }
{\pi(\X_i,\bg_0) 
+ \{1-2\pi(\X_i,\bg_0)\} I\{
Q(\bb_0\trans\X_i) \le0 \}}
+\{\pi(\X_i,\bg_0)-A_i\}\right.\n\\
&&\left.\times
\frac{
[\mu(\X_i,\ba_0)+Q(\bb_0\trans\X_i)
-\{2 \mu(\X_i,\ba_0)+
Q(\bb_0\trans\X_i)\}
I\{Q(\bb_0\trans\X_i)\le0\}]}{
\pi(\X_i,\bg_0) 
+ \{1-2\pi(\X_i,\bg_0)\} I\{
Q(\bb_0\trans\X_i) \le0 \}}
-V\{Q(\cdot),\bb_0\}
\right)\n\\
&=&n^{-1} \sumi v_0(\X_i,A_i,Y_i).\n
\ee
Inserting the results of (\ref{eq:vbeta}), (\ref{eq:valpha}),
(\ref{eq:vgamma}), (\ref{eq:vQ}) and (\ref{eq:v0}) into
(\ref{eq:expandvalue}), taking into account (\ref{eq:betaexpan}), the
results of Propositions \ref{pro:gamma} and \ref{pro:alpha}, we get
\bse
&&n^{1/2}[\wt V\{\wh Q(\cdot),\wh\bb,\wh\ba,\wh\bg\}
-V\{Q(\cdot),\bb_0\}]\n\\
&=&n^{-1/2}\sumi \left[-\U_\bb\trans
\B^{-1}\left\{\bphi_\bb(\X_i,Y_i,A_i,\bb_0,\ba_0,\bg_0)
 + \B_{\bg}\bphi_\bg(\X_i,A_i,\bg_0)
 +\B_{\ba} \bphi_\ba(\X_i,A_i,Y_i,\ba_0)
\right\}\right.\\
&&\left.+\U_\ba\trans \bphi_\ba(\X_i,A_i,Y_i,\ba_0)
+\U_\bg\trans \bphi_\bg(\X_i,A_i,\bg_0)
+\wt v_Q\{\X_i,A_i,Y_i,\bb_0,\ba_0,\bg_0,Q(\cdot)\}
+v_0(\X_i,A_i,Y_i)
\right]\\
&&+o_p(1).
\ese
Thus, the variance of the estimated value function has leading order
\bse
\sigma^2
&=&E
\left[-\U_\bb\trans
\B^{-1}\left\{\bphi_\bb(\X_i,Y_i,A_i,\bb_0,\ba_0,\bg_0)
 + \B_{\bg}\bphi_\bg(\X_i,A_i,\bg_0)
 +\B_{\ba} \bphi_\ba(\X_i,A_i,Y_i,\ba_0)
\right\}\right.\\
&&\left.+\U_\ba\trans \bphi_\ba(\X_i,A_i,Y_i,\ba_0)
+\U_\bg\trans \bphi_\bg(\X_i,A_i,\bg_0)
+\wt v_Q\{\X_i,A_i,Y_i,\bb_0,\ba_0,\bg_0,Q(\cdot)\}
+v_0(\X_i,A_i,Y_i)
\right]^2\\
&=&
E
\left[-\U_\bb\trans
\B^{-1}\left\{\bphi_\bb(\X_i,Y_i,A_i,\bb_0,\ba_0,\bg_0)
 + \B_{\bg}\bphi_\bg(\X_i,A_i,\bg_0)
 +\B_{\ba} \bphi_\ba(\X_i,A_i,Y_i,\ba_0)
\right\}\right.\\
&&\left.+\U_\ba\trans \bphi_\ba(\X_i,A_i,Y_i,\ba_0)
+\U_\bg\trans \bphi_\bg(\X_i,A_i,\bg_0)
+v_Q\{\X_i,A_i,Y_i,\bb_0,\ba_0,\bg_0,Q(\cdot)\}
+v_0(\X_i,A_i,Y_i)
\right]^2,
\ese
which is the result in Theorem \ref{th:value}.\qed

\clearpage

\subsection{Results of the simulations based on the method of \cite{fanetal2017} with constant propensity score model}

	\begin{table}[htbp]
		\centering 
		\begin{tabular}{ccccccclc} 
			\hline 
			Case &parameters& True  & Estimate & sd & $\hat{\rm sd}$ & cvg & MSE  \\ 
			\hline 
			\hline 
			\multicolumn{8}{c}{ CAL estimates for $\bb$, value function $V$ and the root of $Q_0(t)=2t$.}\\
			\hline 
			\multirow{5}{*}{I}&$\beta_2$& 1 & 1.0041 &0.1048 &0.0917 &   90.8\%  & 0.0110 \\
			&$\beta_3$      & -1&-0.9977&0.1021&0.0907& 90.5\%  &0.0104\\
			&$\beta_4$      & 1&1.0058&0.1030 & 0.0918&  92.6\%   & 0.0106 \\ 
			&V& 2.5958 &2.5577 &0.2181& 0.2264&  95.2\%   & 0.0490\\
			&root& 0 &0.0016 &  0.0073&    -   &  -   &   0.0001   \\ 
			\hline 
			
			\multirow{5}{*}{II}&$\beta_2$& 1 &1.0541&0.2754&0.2444&   93.6\%  &0.0788\\
			&$\beta_3$      & -1    &-1.0131&0.2101&0.2027&   94.6\%  &0.0443\\
			&$\beta_4$      & 1     &1.0395&0.2762&0.2442&  93.5\%   & 0.0779\\ 
			&V& 2.5958 &2.9728&0.2030 &0.2098&  55.9\%   & 0.1834\\
			&root& 0 &0.0041&0.0069&    -   &  -   &   0.0001   \\ 
			\hline 
			
			\multirow{5}{*}{III}&$\beta_2$&1&1.0784&0.3371&0.2627&90.1\%&0.1198\\
			&$\beta_3$&-1&-1.0177&0.2364&0.1999&91.4\%&0.0562\\
			&$\beta_4$&1&1.0650&0.3238&0.2539&90.4\%&0.1091\\ 
			&V& 2.5958 &3.2963 &0.2516&0.2500&  18.6\%   & 0.5541\\
			&root& 0 &  0.0009 &  0.0067&    -     &  -   &   0.00004   \\ \hline 
			
			\multirow{5}{*}{IV}&$\beta_2$& 1&0.9670&0.2054&0.1860&   88.3\%  & 0.0433\\
			&$\beta_3$      & -1    &-0.9452&0.1700&0.1617& 88.8\%  & 0.0319\\
			&$\beta_4$      & 1  &1.0241& 0.2117& 0.1938&  93.5\%   &   0.0454\\ 
			&V& 2.5958 &3.3160&0.2501&0.2480&  15.8\%   & 0.5813\\
			&root& 0 & 0.0003&  0.0068 &    -   &  -   &   0.0001   \\ 
			\hline 
			
			\multicolumn{8}{c}{ CAL-DR estimates for $\bb$, value function $V$ and the root of $Q_0(t)=2t$.}\\ \hline
			\multirow{5}{*}{I}&$\beta_2$&1&0.9991&0.0605&0.0850&99.0\%&0.0037\\
			&$\beta_3$&-1&-0.9958&0.0615&0.0783&97.8\%&0.0038\\
			&$\beta_4$&1&0.9989&0.0603&0.0845&98.6\%&0.0036\\ 
			& V& 2.5958 &2.5628& 0.2186&0.2259&  95.3\%   &  0.0489 \\
			&root& 0 &0.0012 & 0.0073 &    -    &  -   &   0.0001   \\ 
			\hline 
			
			\multirow{5}{*}{II}&$\beta_2$& 1 &1.0572 & 0.2504&0.2688&   96.9\%  & 0.0660\\
			&$\beta_3$      & -1 &-1.0136&0.1894&0.2150&   97.6\%  &0.0360\\
			&$\beta_4$      & 1 &1.0413&0.2556&0.2682&  96.7\%   &0.0671\\ 
			&V& 2.5958 &2.9761&0.2031& 0.2100 &  55.6\%   & 0.1859\\
			&root& 0 &0.0027 & 0.0067&    -   &  -   &   0.0001   \\ 
			\hline 
			
			\multirow{5}{*}{III}&$\beta_2$&1&1.0714&0.2919&0.3289&97.1\%&0.0903\\
			&$\beta_3$&-1&-1.0121&0.2104&0.2451&98.2\%&0.0444\\
			&$\beta_4$&1&1.0665&0.2906&0.3234&97.1\%&0.0888\\ 
			&V& 2.5958 &3.3091&0.2510&0.2482&  17.6\%   & 0.5718\\
			&root& 0 &0.0014 &  0.0076 &    -      &  -   &   0.0001   \\
			\hline 
			
			\multirow{5}{*}{IV}&$\beta_2$& 1 &0.9747& 0.1926&0.2097&   93.3\%  &0.0377\\
			&$\beta_3$      & -1 &-0.9461&0.1548&0.1738&   93.2\%  &0.0269\\
			&$\beta_4$      & 1  &1.0298&0.2050& 0.2163&  96.4\%   &0.0429\\ 
			&V& 2.5958 &3.3246&0.2502&0.2475&  14.5\%   & 0.5938\\
			&root& 0 &-0.0010& 0.0065 &    -   &  -   &   0.00004   \\ 
			\hline 
		\end{tabular}
\caption{Simulation 1
	competitor. See also the caption of Table \ref{table:mono1}. 
}
	\label{table: CAL-DR1}
\end{table}

\begin{table}[htbp]
	\centering 
	\begin{tabular}{ccccccclc} 
		\hline 
		Case &parameters& True  & Estimate & sd & $\hat{\rm sd}$ & cvg & MSE  \\ 
		\hline 
		\hline 
		\multicolumn{8}{c}{ CAL estimates for $\bb$, value function $V$ and the root of $Q_0(t)=t+\sin(t)$.}\\
		\hline 
		\multirow{5}{*}{I}&$\beta_2$& 1 &  1.0697& 0.3125&0.2340 & 93.1\%  & 0.1025\\
		&$\beta_3$      & -1 & -1.0179& 0.2165&0.3699& 95.2\%  & 0.0472 \\
		&$\beta_4$      & 1     & 1.0801&0.3093&0.2351&92.9\%   & 0.1021 \\ 
		&V&3.0533 & 2.9731 & 0.1949 &0.2091& 93.6\%   &0.0444 \\
		&root& 0 &0.0048 & 0.0063 &    -   &  -   &  0.0001\\ 
		\hline 
		
		\multirow{5}{*}{II}&$\beta_2$& 1 & 1.0587&0.2737&0.2455& 93.9\%  & 0.0784\\
		&$\beta_3$      & -1    & -1.0247 &0.2249&0.2064 & 94.4\%  & 0.0512\\
		&$\beta_4$      & 1  & 1.0584&0.2808& 0.2425& 91.7\%   & 0.0823\\ 
		&V& 3.0533 &2.9843&0.2025&0.2098& 93.8\%   &0.0458\\
		&root& 0  &0.0042 &0.0071&    -   &  -   &0.0001\\ 
		\hline 
		
		\multirow{5}{*}{III}&$\beta_2$&1&1.0815&0.3119&0.2572&91.0\%&0.1040\\
		&$\beta_3$&-1&-1.0314&0.2303&0.2016&92.1\%&0.0540\\
		&$\beta_4$&1&1.0578&0.3111&0.2483&90.1\%&0.1001\\ 
		&V& 3.0533 & 3.3050& 0.2481 &0.2492& 83.9\%   & 0.1249\\
		&root& 0  & 0.0007 & 0.0066&    -     &  -   &  0.00004 \\ \hline 
		
		\multirow{5}{*}{IV}&$\beta_2$& 1&1.0505&0.2704&0.2138&  89.0\%  & 0.0757\\
		&$\beta_3$      & -1    &-1.0168& 0.2219&0.1776&   89.8\%  & 0.0495\\
		&$\beta_4$      & 1 &1.0499&0.2696& 0.2070& 88.4\%   & 0.0752\\ 
		&V&3.0533&3.3180& 0.2417&0.2483&84.0\%   &0.1285\\
		&root& 0 &0.0002&0.0069 &    -   &  -   & 0.0001  \\ 
		\hline 
		
		\multicolumn{8}{c}{ CAL-DR estimates for $\bb$, value function $V$ and the root of $Q_0(t)=t+\sin(t)$.}\\ \hline
		\multirow{5}{*}{I}&$\beta_2$&1&1.0775&0.3351&0.3221&96.6\%&0.1183\\
		&$\beta_3$&-1&-1.0225&0.2185&0.2369&97.4\%&0.0482\\
		&$\beta_4$&1&1.0902&0.3409&0.3214&96.8\%&0.1244\\ 
		& V&3.0533 & 2.9782 & 0.1948&0.2083& 93.7\%   & 0.0436\\
		&root& 0  & 0.0034 &  0.0062 &    -    &  -   &  0.0001 \\ 
		\hline 
		
		\multirow{5}{*}{II}&$\beta_2$& 1&1.0544 &0.2580 & 0.2608& 96.4\%  &0.0695\\
		&$\beta_3$      & -1 &-1.0133&0.1979&0.2123& 96.8\%  &0.0393\\
		&$\beta_4$      & 1 & 1.0561&0.2688 & 0.2563& 96.0\%   &0.0754\\ 
		&V&3.0533& 2.9888&0.2025&0.2089&  93.8\%   & 0.0452 \\
		&root& 0 & 0.0029&  0.0069 &    -   &  -   &  0.0001 \\ 
		\hline 
		
		\multirow{5}{*}{III}&$\beta_2$&1&1.0723&0.2738&0.2828&96.9\%&0.0802\\
		&$\beta_3$&-1&-1.0228&0.2032&0.2153&97.6\%&0.0418\\
		&$\beta_4$&1&1.0537&0.2728&0.2806&96.2\%&0.0773\\ 
		&V& 3.0533& 3.3168& 0.2476 & 0.2480& 83.0\%   & 0.1307\\
		&root& 0 &-0.0006 & 0.0064&    -      &  -   & 0.00004  \\
		\hline 
		
		\multirow{5}{*}{IV}&$\beta_2$& 1 &1.0511&0.2820&0.2324&   93.3\%  & 0.0822\\
		&$\beta_3$      & -1 &-1.0109&0.1962&0.1869& 94.5\%  &0.0386\\
		&$\beta_4$      & 1  &1.0459& 0.2664& 0.2243& 92.6\%   & 0.0731\\ 
		&V& 3.0533 &3.3271&0.2412&0.2473&83.0\%   & 0.1332\\
		&root& 0  &-0.0010 & 0.0067 &    -   &  -   &   0.0001 \\ 
		\hline 
	\end{tabular}
\caption{Simulation 2 competitor. See also the caption of Table \ref{table:mono1}. 
}
	\label{table:CAL-DR2}
\end{table}

\begin{table}[htbp]
	\centering 
	\begin{tabular}{ccccccclc} 
		\hline 
		Case &parameters& True  & Estimate & sd & $\hat{\rm sd}$ & cvg & MSE  \\ 
		\hline 
		\hline 
		\multicolumn{8}{c}{ CAL estimates for $\bb$, value function $V$ and the root of $Q_0(t)=t^2-2$.}\\
		\hline 
		\multirow{5}{*}{I}&$\beta_2$& 1 & 0.1896  &16.065&180.53& 77.2\%  &258.75\\
		&$\beta_3$      & -1 &-0.3297&16.740&284.51& 79.1\%  & 280.66\\
		&$\beta_4$      & 1     & 0.2828& 12.534&93.518&75.6\%   & 157.62\\ 
		&V& 4.7166 & 3.6907 & 0.4367&0.4574 & 38.4\%   &1.2432\\
		&root& $\pm$1.4142 &-0.1031 &  0.0283 &    -   &  -   &  -\\ 
		\hline 
		
		\multirow{5}{*}{II}&$\beta_2$& 1 & -6.1005&207.21&66816& 78.0\%  & 42987\\
		&$\beta_3$      & -1    & -46.056&1558.4&502935& 76.5\%  & 2430757\\
		&$\beta_4$      & 1  & 3.8245& 137.01& 43241& 75.4\%   & 18781\\ 
		&V& 4.7166 &3.7069& 0.4528&0.4602& 41.4\%   &1.2245\\
		&root& $\pm$1.4142  &-0.1070 &0.0267&    -   &  -   &-\\ 
		\hline 
		
		\multirow{5}{*}{III}&$\beta_2$&1&6.0469&184.58&75607&73.9\%&34095\\
		&$\beta_3$&-1&-3.9209&140.46&57214&79.9\%&19739\\
		&$\beta_4$&1&-1.6121&52.100&20912&73.2\%&2721.2\\ 
		&V& 4.7166 & 3.6993&0.4305 & 0.4476& 37.8\%   & 1.2202\\
		&root& $\pm$1.4142  &-0.0970 &  0.0300&    -     &  -   &  - \\ \hline 
		
		\multirow{5}{*}{IV}&$\beta_2$& 1&-1.4839&48.991&3160.1 &  75.0\%  & 2406.3 \\
		&$\beta_3$      & -1    & 0.0049 & 8.6964& 472.80&   79.9\%  & 76.637\\
		&$\beta_4$      & 1 &-2.3429 &72.514& 4663.7& 74.7\%   & 5269.6\\ 
		&V& 4.7166&3.8064& 0.4614&0.4825&51.7\%   & 1.0414\\
		&root& $\pm$1.4142  &-0.1003&  0.0281  &    -   &  -   & -  \\ 
		\hline 
		
		\multicolumn{8}{c}{ CAL-DR estimates for $\bb$, value function $V$ and the root of $Q_0(t)=t^2-2$.}\\ \hline
		\multirow{5}{*}{I}&$\beta_2$&1&0.6729&24.883&499.86&83.3\%&619.29\\
		&$\beta_3$&-1&-0.7673&29.866&595.41&84.3\%&892.05\\
		&$\beta_4$&1&0.8199&50.211&1023.4&81.6\%&2521.2\\ 
		& V&4.7166 & 3.7061&.4342&0.4525& 39.0\%   &1.2096 \\
		&root& $\pm$1.4142  &-0.1056 &0.0275&    -    &  -   &  -  \\ 
		\hline 
		
		\multirow{5}{*}{II}&$\beta_2$& 1&-3.8486 &93.644& 5309.0& 83.1\%  & 8792.7\\
		&$\beta_3$      & -1 &-3.6347&75.200&4264.8 & 83.2\%  &5661.9\\
		&$\beta_4$      & 1 & 3.8347 &96.155 & 5400.7 & 83.4\%   & 9253.8 \\ 
		&V&4.7166& 3.7188&0.4509&0.4547&  41.5\%   & 1.1989\\
		&root& $\pm$1.4142  &-0.1092 &  0.0260 &    -   &  -   &  - \\ 
		\hline 
		
		\multirow{5}{*}{III}&$\beta_2$&1&1.2712&26.446&520.67&80.1\%&699.48\\
		&$\beta_3$&-1&0.2419&21.320&243.83&87.0\%&456.06\\
		&$\beta_4$&1&1.7351&36.291&765.29&80.6\%&1317.6\\ 
		&V& 4.7166& 3.7113&  0.4276& 0.4441& 38.2\%   & 1.1935\\
		&root& $\pm$1.4142  &-0.0995 & 0.0300 &    -      &  -   & -  \\
		\hline 
		
		\multirow{5}{*}{IV}&$\beta_2$& 1 &0.0050&13.226 &117.41&   84.1\%  & 175.91\\
		&$\beta_3$      & -1 &-0.9763&11.797&102.71& 85.9\%  & 139.18\\
		&$\beta_4$      & 1  &-0.1352& 18.7417& 164.01& 82.9\%   & 352.54 \\ 
		&V& 4.7166 &3.8198&0.4595&0.4762&51.9\%   & 1.0154 \\
		&root& $\pm$1.4142  & -0.1025 & 0.02802 &    -   &  -   &   - \\ 
		\hline 
	\end{tabular}
\caption{Simulation 3 competitor. Similar to Table
	\ref{table:mono1}.}
	\label{table:CAL-DR3}
\end{table}

\newpage
\subsection{Results of the simulations based on the method of \cite{fanetal2017} with non-constant propensity score model}

\begin{table}[htbp]
	\centering 
	\begin{tabular}{ccccccclc} 
		\hline 
		Case &parameters& True  & Estimate & sd & $\hat{\rm sd}$ & cvg & MSE  \\ 
		\hline 
		\hline 
		\multicolumn{8}{c}{ CAL estimates for $\bb$, value function $V$ and the root of $Q_0(t)=2t$.}\\
		\hline 
		\multirow{5}{*}{I}&$\beta_2$& 1     &    1.0064     &  0.1061  &      0.0927        &   90.6\%  &   0.0113   \\
		&$\beta_3$      & -1    &    -1.0043     &  0.1020  &    0.0925        &   92.2\%  &   0.0104   \\
		&$\beta_4$      & 1     &    1.0101      &  0.1055  &     0.0929        &  92.3\%   &   0.0112   \\ 
		&V& 2.5958 &2.5490 &0.2023 &0.2082&  94.5\%   &0.0431\\
		&root& 0 &    0.0021      &  0.0072  &    -   &  -   &   0.0001   \\ 
		\hline 
		
		\multirow{5}{*}{II}&$\beta_2$& 1     & 1.0397 &  0.2439 &0.2378&   92.9\%  & 0.0611 \\
		&$\beta_3$      & -1    &    -1.0254&  0.2299&    0.2356&   94.7\%  &   0.0535\\
		&$\beta_4$      & 1     &    1.0104  &  0.1829 & 0.1810&  94.6\%   &  0.0336\\ 
		&V& 2.5958 &2.5401&0.1949&0.2092&  95.6\%   &0.0411\\
		&root& 0 &    0.0017 &  0.0065 &    -   &  -   &   0.00004   \\ 
		\hline 
		
		\multirow{5}{*}{III}&$\beta_2$&1&1.0067&0.1081&0.0917&90.1\%&0.0117\\
		&$\beta_3$&-1&-0.9981&0.1053&0.0906&89.6\%&0.0111\\
		&$\beta_4$&1&1.0069&0.1058&0.0917&91.2\%&0.0112\\ 
		&V& 2.5958 &2.4206 &0.2064& 0.2217&  87.5\%   & 0.0733\\
		&root& 0 &  0.0019 &  0.0076&    -     &  -   &   0.0001   \\ \hline 
		
		\multirow{5}{*}{IV}&$\beta_2$& 1&1.0398&  0.2142& 0.2307 &   94.6\%  &   0.0475  \\
		&$\beta_3$      & -1    & -1.0339&  0.2020&    0.2316&   95.6\%  &   0.0419 \\
		&$\beta_4$      & 1  & 1.0195&  0.1867 &  0.1800 &  94.5\%   &   0.0352\\ 
		&V& 2.5958 & 2.3971&0.2122& 0.2230&  85.1\%   & 0.0845\\
		&root& 0 &    0.0018 &  0.0069 &    -   &  -   &   0.0001   \\ 
		\hline 
		
		\multicolumn{8}{c}{ CAL-DR estimates for $\bb$, value function $V$ and the root of $Q_0(t)=2t$.}\\ \hline
		\multirow{5}{*}{I}&$\beta_2$&1&0.9999&0.0618&0.0899&99.3\%&0.0038\\
		&$\beta_3$&-1&-1.0013&0.0610&0.0833&98.2\%&0.0037\\
		&$\beta_4$&1&1.0019&0.0582&0.0903&99.5\%&0.0034\\ 
		& V& 2.5958 &2.5547 &0.2023 &0.2076 &  94.6\%   &0.0426\\
		&root& 0 &    0.0014&  0.0073  &    -    &  -   &   0.0001   \\ 
		\hline 
		
		\multirow{5}{*}{II}&$\beta_2$& 1 & 1.0433 &  0.2412& 0.2260&   95.4\%  &   0.0600\\
		&$\beta_3$      & -1    & -1.0236&  0.2208&    0.1989&   94.3\%  &   0.0493 \\
		&$\beta_4$      & 1 &    1.0040 &  0.1711&     0.1697&  96.7\%   &   0.0293 \\ 
		&V& 2.5958 & 2.5464&0.1936&0.2096 &  95.5\%   &   0.0399 \\
		&root& 0 &    0.0012 &  0.0065 &    -   &  -   &   0.00004   \\ 
		\hline 
		
		\multirow{5}{*}{III}&$\beta_2$&1&1.0004&0.0618&0.0855&98.9\%&0.0038\\
		&$\beta_3$&-1&-0.9960&0.0602&0.0786&98.3\%&0.0036\\
		&$\beta_4$&1&0.9986&0.0603&0.0850&98.7\%&0.0036\\ 
		&V& 2.5958 &2.4258&0.2073&0.2213&  87.7\%   &0.0719 \\
		&root& 0 &0.0015 &  0.0076 &    -      &  -   &   0.0001   \\
		\hline 
		
		\multirow{5}{*}{IV}&$\beta_2$& 1 & 1.0424& 0.1988 &0.2050&   96.6\%  &   0.0413 \\
		&$\beta_3$      & -1 & -1.0322&  0.1826& 0.1819&   95.1\%  &   0.0344 \\
		&$\beta_4$      & 1  &1.0189&  0.1663& 0.1746&  98.1\%   &   0.0280   \\ 
		&V& 2.5958 &2.4022 &0.2117&0.2234&  85.2\%   & 0.0823\\
		&root& 0 &    0.0013 &  0.0068 &    -   &  -   &   0.0001   \\ 
		\hline 
	\end{tabular}
	\caption{Simulation 4
		competitor. See also the caption of Table \ref{table:mono1}. 
	}
	\label{table: CAL-DR4}
\end{table}

\begin{table}[htbp]
	\centering 
	\begin{tabular}{ccccccclc} 
		\hline 
		Case &parameters& True  & Estimate & sd & $\hat{\rm sd}$ & cvg & MSE  \\ 
		\hline 
		\hline 
		\multicolumn{8}{c}{ CAL estimates for $\bb$, value function $V$ and the root of $Q_0(t)=t+\sin(t)$.}\\
		\hline 
		\multirow{5}{*}{I}&$\beta_2$& 1 & 1.0442&0.2489&0.2740 & 95.7\%  & 0.0639\\
		&$\beta_3$      & -1 &-1.0121&0.2123&0.21779& 95.3\%  & 0.0452\\
		&$\beta_4$      & 1     & 1.0563&0.2573&0.2723&96.7\%   & 0.0694\\ 
		&V&3.0533 & 2.9202 & 0.1647 & 0.2092& 95.0\%   &0.0448\\
		&root& 0 &0.0048 & 0.0077 &    -   &  -   &  0.0001\\ 
		\hline 
		
		\multirow{5}{*}{II}&$\beta_2$& 1 & 1.0433&0.2414&0.3039& 94.5\%  & 0.0601\\
		&$\beta_3$      & -1    & -1.0143 &0.2032 &0.2200& 94.4\%  &0.0415\\
		&$\beta_4$      & 1  & 1.0405& 0.2385& 0.3002& 94.1\%   & 0.0585\\ 
		&V& 3.0533 &2.9695& 0.1531&0.1768& 94.5\%   &0.0305\\
		&root& 0  &0.0051&0.0079&    -   &  -   &0.0001\\ 
		\hline 
		
		\multirow{5}{*}{III}&$\beta_2$&1&1.3530&0.6259&0.5921&98.0\%&0.5164\\
		&$\beta_3$&-1&-1.0364&0.3592&0.3546&95.0\%&0.1303\\
		&$\beta_4$&1&1.3669&0.6571&0.6025&98.5\%&0.5664\\ 
		&V& 3.0533 & 2.6987&0.2031& 0.2465& 72.0\%   & 0.1670\\
		&root& 0  &0.0060 &  0.0076&    -     &  -   &  0.0001\\ \hline 
		
		\multirow{5}{*}{IV}&$\beta_2$& 1&1.2291&0.8755&0.4244 &  96.4\%  & 0.8190\\
		&$\beta_3$      & -1 &-1.0359&0.3283&0.2669& 94.2\%  & 0.1090\\
		&$\beta_4$      & 1  &1.2162& 0.4616& 0.3624& 97.0\%   & 0.2598 \\ 
		&V& 3.0533 &2.6891&0.1992&0.2472&71.6\%   &0.1724\\
		&root& 0  & 0.0063 & 0.0077 &    -   &  -   &   0.0001\\ 
		\hline 
		
		\multicolumn{8}{c}{ CAL-DR estimates for $\bb$, value function $V$ and the root of $Q_0(t)=t+\sin(t)$.}\\ \hline
		\multirow{5}{*}{I}&$\beta_2$&1&1.0479&0.3106&0.2694&96.5\%&0.0988\\
		&$\beta_3$&-1&-1.0053&0.2062&0.2221&96.7\%&0.0425\\
		&$\beta_4$&1&1.0559&0.3092&0.2652&97.4\%&0.0987\\ 
		& V&3.0533 & 2.9283 &0.1629&0.2065& 95.0\%   & 0.0422\\
		&root& 0  & 0.0033 &  0.0074 &    -    &  -   &  0.0001 \\ 
		\hline 
		
		\multirow{5}{*}{II}&$\beta_2$& 1&1.0374 &0.2257& 0.2409 & 96.9\%  & 0.0523\\
		&$\beta_3$      & -1 &-1.0084&0.1920&0.2092 & 97.1\%  &0.0370\\
		&$\beta_4$      & 1 & 1.0316 &0.2175 & 0.2348& 96.6\%   & 0.0483\\ 
		&V&3.0533&2.9755&0.1518&0.1761&  94.9\%   &0.0291 \\
		&root& 0 & 0.0035&  0.0077 &    -   &  -   &  0.0001 \\ 
		\hline 
		
		\multirow{5}{*}{III}&$\beta_2$&1&1.3946&1.1256&0.8694&99.6\%&1.4225\\
		&$\beta_3$&-1&-1.0406&0.5801&0.4573&96.6\%&0.3382\\
		&$\beta_4$&1&1.4214&1.4059&0.9926&99.7\%&2.1541\\ 
		&V& 3.0533& 2.7091& 0.2032 & 0.2434& 73.9\%   & 0.1598\\
		&root& 0 &0.0043 & 0.0075 &    -      &  -   & 0.0001  \\
		\hline 
		
		\multirow{5}{*}{IV}&$\beta_2$& 1&1.2293&0.7231&0.4598&  98.7\%  & 0.5755\\
		&$\beta_3$      & -1 &-1.0199&0.2838&0.2786& 97.0\%  & 0.0809\\
		&$\beta_4$      & 1  &1.2231& 0.4830&0.4059& 99.2\%   &0.2830\\ 
		&V& 3.0533 &2.6994&0.1986&0.2443&72.7\%   &0.1647\\
		&root& 0  & 0.0046 & 0.0075 &    -   &  -   &   0.0001\\ 
		\hline 
	\end{tabular}
	\caption{Simulation 5 competitor. See also the caption of Table \ref{table:mono1}. 
	}
	\label{table:CAL-DR5}
\end{table}

\begin{table}[htbp]
	\centering 
	\begin{tabular}{ccccccclc} 
		\hline 
		Case &parameters& True  & Estimate & sd & $\hat{\rm sd}$ & cvg & MSE  \\ 
		\hline 
		\hline 
		\multicolumn{8}{c}{ CAL estimates for $\bb$, value function $V$ and the root of $Q_0(t)=t^2-2$.}\\
		\hline 
		\multirow{5}{*}{I}&$\beta_2$& 1 &  -0.5431  & 20.501&272.11 & 78.1\%  & 422.65\\
		&$\beta_3$      & -1 & 0.8042& 21.955&276.85& 75.9\%  & 485.27 \\
		&$\beta_4$      & 1     & 0.6835 & 23.240&271.48&77.0\%   & 540.21 \\ 
		&V& 4.7166 & 3.7004 & 0.4882&0.4566 & 41.1\%   &1.2708\\
		&root& $\pm$1.4142 &-0.1059 & 0.0293 &    -   &  -   &  -\\ 
		\hline 
		
		\multirow{5}{*}{II}&$\beta_2$& 1 & -1.4798&103.69&3385.1& 75.3\%  & 10758 \\
		&$\beta_3$      & -1    & -5.1392 &109.59 &4186.9 & 76.7\%  & 12028\\
		&$\beta_4$      & 1  & 1.4680& 60.410& 2261.4& 75.8\%   & 3649.6\\ 
		&V& 4.7166 &3.7164&0.5095&0.4578& 43.0\%   &1.2600 \\
		&root& $\pm$1.4142  &-0.1082 &0.0279&    -   &  -   &-\\ 
		\hline 
		
		\multirow{5}{*}{III}&$\beta_2$&1&6.0412&184.58&75607&73.8\%&34095\\
		&$\beta_3$&-1&-3.8956&140.46&57214&79.6\%&19739\\
		&$\beta_4$&1&-1.5895&52.102&20913&72.8\%&2721.4\\ 
		&V& 4.7166 & 3.7002& 0.4279 & 0.4475& 37.3\%   &1.2161\\
		&root& $\pm$1.4142  & -0.0973& 0.0299&    -     &  -   &  - \\ \hline 
		
		\multirow{5}{*}{IV}&$\beta_2$& 1&-1.4839&48.991&3160.1 &  75.0\%  & 2406.3 \\
		&$\beta_3$      & -1    & 0.0049 & 8.6964& 472.80&   79.9\%  & 76.637\\
		&$\beta_4$      & 1 &-2.3429 &72.514& 4663.7& 74.7\%   & 5269.6\\ 
		&V& 4.7166&3.6965& 0.4299&0.4470&37.9\%   & 1.2255\\
		&root& $\pm$1.4142  &-0.1004&  0.0284 &    -   &  -   & -  \\ 
		\hline 
		
		\multicolumn{8}{c}{ CAL-DR estimates for $\bb$, value function $V$ and the root of $Q_0(t)=t^2-2$.}\\ \hline
		\multirow{5}{*}{I}&$\beta_2$&1&4.6108&110.98&4618.8&78.7\%&12330\\
		&$\beta_3$&-1&-0.0221&76.440&3430.7&79.2\%&5844.0\\
		&$\beta_4$&1&-0.5504&25.206&898.73&78.8\%&637.74\\ 
		& V&4.7166 & 3.7029 &0.4856&0.4549& 41.6\%   & 1.2635\\
		&root& $\pm$1.4142  & -0.1060 &0.0282&    -    &  -   &  -  \\ 
		\hline 
		
		\multirow{5}{*}{II}&$\beta_2$& 1&-12.704 &410.17 & 158724 & 79.1\%  & 168424\\
		&$\beta_3$      & -1 &22.623&692.39&268071 & 79.8\%  &479969\\
		&$\beta_4$      & 1 & -4.5086 &125.95  & 47522 & 77.1\%   & 15894 \\ 
		&V&4.7166& 3.7173&0.5057&0.4565&  42.9\%   & 1.2544\\
		&root& $\pm$1.4142  & -0.1080 &  0.0268 &    -   &  -   &  - \\ 
		\hline 
		
		\multirow{5}{*}{III}&$\beta_2$&1&1.1753&26.467&522.08&79.9\%&700.54\\
		&$\beta_3$&-1&0.2243&21.388&252.18&87.7\%&458.94\\
		&$\beta_4$&1&1.8103&36.281&761.45&79.7\%&1317\\ 
		&V& 4.7166&3.7124&0.4249 & 0.4438& 37.7\%   & 1.1891\\
		&root& $\pm$1.4142  &-0.0997 & 0.0299 &    -      &  -   & -  \\
		\hline 
		
		\multirow{5}{*}{IV}&$\beta_2$& 1 &0.0050&13.226 &117.41&   84.1\%  & 175.91\\
		&$\beta_3$      & -1 &-0.9763&11.797&102.71& 85.9\%  & 139.18\\
		&$\beta_4$      & 1  &-0.1352& 18.7417& 164.01& 82.9\%   & 352.54 \\ 
		&V& 4.7166 &3.7085&0.4270&0.4433&38.0\%   & 1.1986 \\
		&root& $\pm$1.4142  &-0.1025 & 0.0283 &    -   &  -   &   - \\ 
		\hline 
	\end{tabular}
	\caption{Simulation 6
		competitor. See also the caption of Table \ref{table:mono1}. 
	}
	\label{table:CAL-DR6}
\end{table}

\clearpage
\baselineskip=14pt
\bibliographystyle{agsm}
\bibliography{notrank}